\documentclass[useAMS,usenatbib,usegraphicx]{mn2e}

\usepackage{amssymb} 
\usepackage{textcomp} 
\usepackage{appendix}
\usepackage{amsmath} 
\usepackage{fixltx2e} 
\usepackage{multirow} 
\usepackage{enumerate} 
\usepackage{mathptmx} 
\usepackage{hyperref} 

\newcommand{\kms}{\rm km\ s^{-1}}
\newcommand{\ergs}{\rm erg\ s^{-1}}
\newcommand{\flux}{\rm erg\ s^{-1}\ cm^{-2}}
\newcommand{\mic}{\mbox{$\mu$m}}
\newcommand{\Hz}{\rm{Hz}}
\newcommand{\kev}{\rm keV}

\let\AAold\AA
\renewcommand{\AA}{\text{\AAold}}

\newcommand{\lbol}{L_{\rm bol}}
\newcommand{\lledd}{L / L_{\rm{Edd}}}
\newcommand{\mbh}{M_{\rm BH}}
\newcommand{\an}{\alpha_{\rm opt}}
\newcommand{\aox}{\alpha_{\rm ox}}
\newcommand{\auv}{\alpha_{\rm UV}}
\newcommand{\aoxb}{\alpha_{\rm {FUV,x}}}
\newcommand{\nln}{\nu L_{\nu}}
\newcommand{\nlnl}{\nu L_{\nu}}
\newcommand{\Ks}{K$_{\rm s}$}
\newcommand{\ebv}{E(B-V)}
\newcommand{\hc}{h^3}
\newcommand{\hd}{h^4}
\newcommand{\dbreak}{D_{4000}}
\newcommand{\msun}{{\rm M_{\odot}}}
\newcommand{\chisq}{\chi^2}
\newcommand{\hii}{\text{H~{\sc ii}}}
\newcommand{\nii}{\text{[N~{\sc ii}]}}
\newcommand{\sii}{\text{[S~{\sc ii}]}}

\newcommand{\oi}{\text{[O~{\sc i}]}}
\newcommand{\oii}{\text{[O~{\sc ii}]}}
\newcommand{\oiii}{\text{[O~{\sc iii}]}}
\newcommand{\feii}{\text{Fe~{\sc ii}}}
\newcommand{\caii}{\text{Ca~{\sc ii}}}
\newcommand{\neiii}{\text{[Ne~{\sc iii}]}}
\newcommand{\mgii}{\text{Mg~{\sc ii}}}
\newcommand{\civ}{\text{C~{\sc iv}}}
\newcommand{\La}{\text{L$\alpha$}}
\newcommand{\Ha}{\text{H$\alpha$}}
\newcommand{\bHa}{\text{bH$\alpha$}}
\newcommand{\nHa}{\text{nH$\alpha$}}
\newcommand{\Hb}{\text{H$\beta$}}
\newcommand{\Hg}{\text{H$\gamma$}}
\newcommand{\Hd}{\text{H$\delta$}}
\newcommand{\He}{\text{H$\epsilon$}}

\newcommand{\tio}{\text{Ti}{\sc O}}
\newcommand{\mi}{M_{\rm{i}}}

\newcommand{\lbha}{L_{\rmn{bH\alpha}}}
\newcommand{\fbha}{F_{\rmn{bH\alpha}}}

\newcommand{\ewha}{\rm{EW}_{\rmn{bH\alpha}}}
\newcommand{\dv}{\Delta {\rm v}_{1000}}
\newcommand{\dva}{\Delta {\rm v}}
\newcommand{\lagn}{L_{\rm AGN}}
\newcommand{\lhost}{L_{\rm host}}

\newcommand{\sone}{\sigma_{\rm data}}
\newcommand{\normdf}{\Delta F / \sone}
\renewcommand{\v}{\text{v}}
\newcommand{\normhieght}{F_{\lambda, \rm{b\Ha}} / \epsilon_{\lambda}}

\newcommand{\nd}{$^{\rm nd}$}

\renewcommand{\th}{$^{\rm th}$}

\newcommand{\Brinchmannoffset}{-0.05}

\newcommand{\NbadlyfitBLRs}{35}
\newcommand{\NeyeInspectThrows}{93}
\newcommand{\NhighChisq}{209}
\newcommand{\NinitiallyPassed}{3\,602}
\newcommand{\Npasshighscore}{80}
\newcommand{\NprcntHighchisq}{3}
\newcommand{\Ntotalprofilehighfwhm}{45}

\newcommand{\badpixeloiiiprcnt}{4}
\newcommand{\expectfalseXmatches}{7}
\newcommand{\extendedParentSample}{230\,413}
\newcommand{\falseRatePrcnt}{3}
\newcommand{\highBLR}{14}

\newcommand{\insignificantoiiiprcnt}{0.6}
\newcommand{\lbolcoeff}{130}
\newcommand{\luvscatter}{2.4}

\newcommand{\midsmpsz}{6\,986}
\newcommand{\minHafluxdensity}{2}
\newcommand{\minfluxinwingsA}{2.5}
\newcommand{\minfluxinwingsB}{4}
\newcommand{\nInPG}{26}
\newcommand{\nRejected}{3\,407}
\newcommand{\nResolved}{2\,766}
\newcommand{\nirdetectionprcnt}{96.5}
\newcommand{\nlowfwhm}{149}
\newcommand{\nlowfwhmp}{4}

\newcommand{\notInCompleteSamplesPrcnt}{0.4}
\newcommand{\observedbyGalex}{89}

\newcommand{\offsetShen}{0.13}

\newcommand{\oursunderfortyone}{297}
\newcommand{\oursunderfortytwo}{1\,878}
\newcommand{\prcntOflargezQCV}{24}
\newcommand{\prcntbasedonoiii}{14}
\newcommand{\prcntnearthres}{4}
\newcommand{\prcntsbtrcthighchisq}{1}

\newcommand{\psmpsz}{232\,837}

\newcommand{\selectionNearThresA}{6.4}
\newcommand{\selectionNearThresB}{3.6}
\newcommand{\selectionNearThresC}{17.1}
\newcommand{\sigOffsetShen}{0.10}
\newcommand{\smpsz}{3\,579}
\newcommand{\szHighZ}{502}
\newcommand{\szHighZinQCVprcnt}{88}
\newcommand{\szMidZ}{3\,077}
\newcommand{\szMidZinQCV}{423}
\newcommand{\szMidZinQCVprcnt}{14}

\newcommand{\xraydetection}{1\,545}
\newcommand{\xraydetectionprcnt}{43}
\newcommand{\xraydetectionprcnthighf}{61}
\newcommand{\xraylowbindetectionprcnt}{20}
\newcommand{\xslope}{1.5}
\newcommand{\xtolbhaIndex}{0.79}
\newcommand{\xtolbhaIndexError}{0.02}
\newcommand{\xtolbhaOffset}{0.45}
\newcommand{\xtolbhaOffsetError}{0.01}

\newcommand{\minsn}{10}
\newcommand{\slopebinmingroupsize}{10}

\newcommand{\maxfwhm}{25}
\newcommand{\pureAGNbHa}{44.2}

\title[Type 1 low $z$ AGN. I. Emission properties]{Type 1 AGN at low $z$. I. Emission properties} 
\author[Jonathan Stern and Ari Laor]
{Jonathan Stern\thanks{E-mail: \href{mailto:stern@physics.technion.ac.il}{stern@physics.technion.ac.il} (JS);\newline \href{mailto:laor@physics.technion.ac.il}{laor@physics.technion.ac.il} (AL)}
and Ari Laor\footnotemark[1]\\
Department of Physics, Technion -- Israel Institute of Technology, Haifa 32000, Israel}

\begin{document}
\maketitle

\begin{abstract}

We analyze the emission properties of a new sample of \smpsz\ type 1 AGN, selected from the SDSS DR7 based on the detection of broad \Ha\ emission. The sample extends over a broad \Ha\ luminosity $\lbha$ of $10^{40} - 10^{44}\ \ergs$ and a broad \Ha\ FWHM of $1\,000 - 25\,000\ \kms$, which covers the range of black hole mass $10^6<\mbh/\msun<10^{9.5}$ and luminosity in Eddington units $10^{-3} < \lledd < 1$.
We combine ROSAT, GALEX and 2MASS observations to form the SED from 2.2 \mic\ to 2 \kev.
We find the following: 
1. The distribution of the \Ha\ FWHM values is independent of luminosity.
2. The observed mean optical-UV SED is well matched by a fixed shape SED of luminous quasars, which scales linearly with $\lbha$, and a host galaxy contribution. 
3. The host galaxy $r$-band (fibre) luminosity function follows well the luminosity function of inactive non-emission line galaxies (NEG), consistent with a fixed fraction of $\sim 3$\% of NEG hosting an AGN, regardless of the host luminosity. 
4. The hosts of lower luminosity AGN have a mean $z$ band luminosity and $u-z$ colour which are identical to NEG with the same redshift distribution. With increasing $\lbha$ the AGN hosts become bluer and less luminous than NEG. The implied increasing star formation rate with $\lbha$ is consistent with the relation for SDSS type 2 AGN of similar bolometric luminosity.
5. The optical-UV SED of the more luminous AGN shows a small dispersion, consistent with dust reddening of a blue SED, as expected for thermal thin accretion disc emission. 
6. There is a rather tight relation of $\nln(2 \kev)$ and $\lbha$, which provides a useful probe for unobscured (true) type 2 AGN.
7. The primary parameter which drives the X-ray to UV emission ratio is the luminosity, rather than $\mbh$ or $\lledd$. 
\end{abstract}
\section{INTRODUCTION}\label{sec:Introduction}

The first systematic study, based on optical spectroscopy, of a complete and well defined sample of Broad Line AGN was conducted by Boroson \& Green (1992, hereafter BG92). It included 87 $z<0.5$ AGN from the Bright Quasar Survey (BQS, Green et al. 1986), with spectra taken over the 4100\AA--5900\AA\ range, at a spectral resolution of $\sim 700$. This survey, and additional systematic studies at other wavelengths, led to major new understandings of the emission and absorption properties of AGN (BG92, and citations thereafter). The BQS survey was followed by the Large Bright Quasar Survey, which yielded spectra of 1055 quasars (Hewett et al. 1995), and by the 2dF QSO Redshift Survey which yielded spectra for 23\,338 quasars (Croom et al. 2004). The next major step in AGN optical spectroscopy came with the Sloan Digital Sky Survey (SDSS) Quasar Catalog (latest release QCV, Schneider et al. 2010), which produced a sample of quasars a factor of 1000 larger, compared to BG92, with a factor of 10 more spectroscopic data per object (a factor of three larger wavelength coverage and a factor of three larger spectral resolution). This sample is now a prime resource for studies of AGN (see Schneider et al. 2002, and citations thereafter). The QCV sample is mainly based on objects selected for spectroscopy due to their non-stellar colours, supplemented by SDSS spectra selected based on other surveys. The QCV sample includes luminous objects ($\mi < -22.0$) which exhibit at least one emission line with FWHM $>1000\ \kms$.

Various studies used the SDSS to produce samples of lower luminosity AGN. At $\mi>-22.0$ the AGN are generally dominated by the host light, which needs to be subtracted to detect the AGN continuum and broad line emission. Hao et al. (2005a), who used the SDSS second data release (DR2), subtracted the host light based on a principal component analysis method. They then modeled the \Ha\ profile using one or two Gaussians, and derived 
the FWHM of the broader component. Their type 1 AGN sample includes a total of 1317 objects with 
FWHM $>$ 1200~$\kms$, and was used to study the extension of the luminosity function of type 1 AGN to low luminosities. The later study of Vanden Berk et al. (2006), based on DR3, extended the earlier SDSS quasar sample (Schneider et al. 2005) to lower luminosity by removing the absolute magnitude criterion, which yielded 4\,666 low-luminosity type 1 AGN, used to study the relation between the AGN luminosity and the host properties. The definition of a type 1 AGN in Vanden Berk et al. (2006), and also in QCV, requires the detection of at least one emission line with FWHM $> 1000\ \kms$. This criterion works well for luminous AGN, as the emission line width is dominated by the Broad Line Region (BLR) in type 1 AGN, and it excludes
type 2 AGN where the emission lines are from the Narrow Line Region (NLR), where the FWHM $< 1000\ \kms$.
However, as we show below 
lower luminosity type 1 AGN show an increasing relative contribution from the NLR, and would fail the FWHM $>1000\ \kms$ criterion. Thus, the Vanden Berk et al. (2006) sample becomes increasingly incomplete with decreasing AGN luminosity. The latest SDSS sample of low luminosity type 1 AGN is of Green \& Ho (2007, hereafter GH07), which is based on DR4. This sample is based on the detection of a broad component for the \Ha\ emission line in the $6400\AA-6700\AA$ wavelength region. It includes 8435 objects, and was used to study the AGN Black Hole mass ($\mbh$) function. We note that broad \Ha\ with FWHM $>10,000~\kms$ are selected 
against in the GH07 sample. due to their restricted search region.

Here we present a new SDSS based sample of type 1 AGN which extends to low luminosity, with the purpose of providing a useful and well defined database which extends the QCV sample to low luminosity, using DR7 (Abazajian et al. 2009). We combine this data set with observations in the UV, IR, and X-ray, based on GALEX (Martin et al. 2005), 2MASS (Skrutskie et al. 2006) and ROSAT (Voges et al. 1999, 2000), in order to study the NIR to X-ray Spectral Energy Distribution (SED) of low luminosity AGN, and its dependence on various AGN properties. We also derive the mean properties of low luminosity AGN host galaxies, and compare them to the mean properties of inactive galaxies. In following papers we study some of the emission line properties of this sample.

The first study of the IR to X-ray SED of a complete and well defined sample of AGN was made by Sanders et al. (1989) for the BQS optically selected quasar sample. It showed the overall similarity of the SED of AGN, composed of an IR bump, an optical-UV bump, and an X-ray power-law continuum. A follow up study was made by Richards et al. (2006) for a small subsample of 259 quasars from the SDSS, which determined the wavelength dependence of the bolometric correction factor. However, little is known about the dependence of the AGN SED on luminosity, $\mbh$, and luminosity in Eddington units $\lledd$ (Vasudevan \& Fabian 2009; Grupe et al. 2010). This dependence may provide important clues to the nature of the accretion process (e.g. Davis \& Laor 2011).

Here we provide an order of magnitude larger sample of type 1 AGN with nearly complete detections in most bands.
These allow us to form an unbiased SED of the sample which extends to low luminosity AGN, and to study its relation with various AGN properties. A significant limitation is that the SED of low luminosity AGN is often dominated by the host galaxy emission. High angular resolution, not available here, is required to isolate the low luminosity AGN emission from the host galaxy light (e.g. Maoz et al. 2005). 

In addition, we explore how broad the broad \Ha\ gets. Laor (2003) noted that type 1 AGN with BLR FWHM$>25\,000~\kms$ are extremely rare. Various theoretical models of the BLR and the ionizing continuum predict an upper limit on the FWHM, beyond which the BLR does not exist (Nicastro 2000, Elitzur and Shlosman 2006, Laor \& Davis 2011). The models differ in the dependence of the maximal FWHM on luminosity. Here, we use the new sample to explore the FWHM distribution and its dependence on luminosity. 

The paper is organized as follows. In \S2.1 -- \S2.5 we describe our type 1 sample selection. In \S2.6 we explore the selection effects which arise from our detection procedure, and in \S2.7 we add photometry from IR, UV, and X-ray surveys. In \S3.1 the distribution of the type 1 sample in various emission properties is explored, and compared to previous AGN samples, and to samples of inactive galaxies. In \S3.2 -- \S3.4, we decompose the mean observed NIR -- UV SED into the net-AGN and host SEDs. The implied host properties are examined in \S3.5.
In \S 3.6 we explore the relation between $\nln$(X-ray) and $\lbha$. The origin of the SED dispersion around the mean SED is explored in \S3.7. 
In \S3.8 -- \S3.9 we investigate the dependence of the AGN SED on $\mbh$ and $\lledd$. We discuss our results in \S4, and summarize them in \S5. 

Throughout the paper, we assume a FRW cosmology with $\Omega$ = 0.3, $\Lambda$ = 0.7 and $H_0 = 70\ \kms$ Mpc$^{-1}$.

\section{SAMPLE CREATION}\label{sec:sample creation}

\subsection{parent sample}\label{subsec:SDSS}
Our purpose is to derive a complete and well defined sample of low luminosity broad line AGN, based on the SDSS DR7 (Abazajian et al. 2009). The SDSS obtained imaging of a quarter of the sky in five bands ($ugriz$; Fukugita et al. 1996) with a CCD mosaic camera (Gunn et al. 1998) coupled to a 2.5-m telescope. The 95\% completeness limit in the $r$ band is 22.2 mag. Spectroscopic follow-ups of the photometric catalog were performed using spectrographs fed by 3\arcsec-aperture fibres. Targets were selected for spectroscopic followup if one of the following criteria applied to the photometry: an extended morphology, which creates the $\sim 10^6$ objects of the main galaxy sample (Strauss et al. 2002);  non-stellar colours or a nearby FIRST (Becker et al. 1995) source, which is the base of the $\sim 10^5$ objects of QCV. Based on the availability of fibers, targets with a nearby ROSAT (Voges et al. 1999, 2000) source might have also been selected for spectroscopy. The spectrographs cover the wavelength range between 3800\AA\ and 9200\AA\ at a resolution of $\sim 150\ \kms$, with a typical S/N of 10 pixel$^{-1}$ for a galaxy near the flux limit. The flux of the spectra is calibrated by matching the spectra of simultaneously observed standard stars to their PSF magnitude (Adelman-McCarthy et al. 2008). On each spectrum, the SDSS pipeline measures the emission lines, and then determines the redshift $z$ and the classification of the spectrum, according to the locations of emission lines, or according to correlations with template spectra.

We apply several initial filters to the DR7 database:
\begin{enumerate}
 \item The spectrum is classified as non-stellar. 
\item $z < 0.31$, to ensure the continuum level can be measured at least $20\,000\ \kms$ to the red away from \Ha, so that very broad \Ha\ lines with FWHM $\lesssim 25\,000~\kms$ are detectable.
\item $z > 0.005$, to ensure reliable distance estimates based on the assumed $H_0$, without significant errors due to peculiar motions.
\item A S/N $>10$ in the spectrum\footnote{Based on median S/N of spectrum pixels covered by the $g$, $r$, $i$ bands (available in the SDSS database).}, to improve the accuracy of the NLR/BLR decomposition. 
\item A percentage of pixels near \Ha\ indicated as `bad' in the pixel mask (Stoughton et al. 2002) 
which is no more than 0\%, 20\% or 50\%, depending on the distance from \Ha\ (see appendix A1). The main source of bad pixels is poor sky subtraction, which degrades the spectrum mainly at $8000\AA < \lambda$.
\end{enumerate}

A total of \psmpsz\ spectra, out of the 1.6 million spectra in DR7, pass these five criteria. We dub them as the `parent sample'.
The spectra are then corrected for foreground dust, using the maps of Schlegel et al. (1998) and the extinction law of Cardelli et al. (1989). We now need to determine which of these objects host a broad line AGN.

\subsection{Host subtraction}\label{subsec: host-galaxy subtraction}

The rather large 3\arcsec\  fibre aperture of the SDSS spectra implies that a significant fraction of the host light is typically included, which needs to be removed in order to identify low luminosity AGN.
As discussed below (\S 2.3), the AGN are identified through the detection of a broad \Ha\ component. To detect this component we need to subtract the broad stellar absorption features, and derive a smooth continuum. In particular, in the vicinity of the \Ha\ feature, there are flanking \tio\ bands (Bica \& Alloin 1986, and see appendix A2), which need to be properly subtracted off. 

To roughly model the absorption features, we follow an approach similar in principle to that described in Hao et al. (2005a). We utilize the galaxy eigenspectra (ESa) presented in Yip et al. (2004; hereafter Yip04), which were derived from a principle component analysis of SDSS galaxies. A linear sum of the first three ESa and a power law component with $L_{\lambda} \propto \lambda^{-1.5}$, representing the AGN continuum, is fit to wavelength regions which lack the strongest galaxy and AGN emission lines (see appendix A2).  Note that the four components of the fit are non-orthogonal, in contrast with the original ESa of Yip04.

In practice, light emitted by hot stars near \Ha\ is degenerate with the power-law component, as can be seen in the third ES of Yip04 (fig. 5 there). This degeneracy produces an unrealistic result that the mean fraction of continuum emission emitted by AGN in all galaxies is $\approx 5\%$. In the more thorough PCA decomposition performed by Vanden Berk et al. (2006), in which they use also eigenvectors derived from a sample of quasars, they found that host-AGN separation is unreliable if the contribution of either component is $<10\%$. This problem probably also arises from the non-orthogonality of the base spectra, as we find here. Note that if one excludes objects in which one of the components is weak, a correlation between the host and AGN luminosity is introduced. To avoid these complications and uncertainties, we use our PCA results only to subtract the stellar features and search for a broad \Ha, and not to derive the host and AGN emission.

The linear sum of the three ESa is subtracted from the SDSS spectra in the wavelength ranges 4800\AA --5250\AA\ and 6125\AA --7000\AA, to derive a featureless continuum near the \Ha, \Hb\ and \oiii\ lines. Note that the Yip04 ESa include narrow emission lines from \hii\ regions, with typical FWHM of $\sim 250\ \kms$, which should be avoided. We interpolate over all the strong narrow features in the ESa prior to their subtraction (note the narrow lines do not participate in the fit). Since \Hb\ in ES 1 and 2, and \Ha\ in ES 1, reside in wider stellar absorption features (FWHM $\sim 1300\ \kms$), there may be additional narrower stellar absorption features, which are lost by the interpolation, and not properly subtracted off with the other stellar absorption features. The possible amplitude of this effect is small, and is discussed in following papers on the narrow emission line properties.

In \prcntsbtrcthighchisq\% of the spectra in the parent sample only poor fits (reduced $\chisq > 2$) are obtained. The algorithm described below might falsely detect a broad \Ha\ feature when the fit is poor, and such false detections are filtered in the final stage of the pipeline (\S 2.5).

\subsection{Candidate broad \Ha\ objects}\label{subsec: Excess Search}

A broad emission line is a unique signature of AGN emission\footnote{Excluding the rare examples of broad \Ha\ due to supernovae (see figures 13 \& 14 in Filippenko 1997). Only a handful of such objects reached the final sample (\S 2.5).}. We search for a broad \Ha\ line, as \Ha\ is the strongest AGN line in the visible region. 
Identifying AGN through the detection of a broad \Ha\ line is easier than the detection of a power-law component, as the average flux density of the broad \Ha\ feature is $\sim 5$ times larger than the power-law continuum flux density (Fig. 1). Also, as discussed above, the power-law component is degenerate with the continuum of young stars.
To identify a broad \Ha\ line we search for excess flux above the continuum in the vicinity of \Ha, which cannot be attributed to be part of the narrow emission line profiles. 

After subtracting the stellar emission, the featureless continuum is derived by linearly interpolating the mean continuum level at the continuum windows at 6125\AA --6250\AA\ and 6880\AA --7000\AA, which are $\sim 300$\AA, or $\sim 14\,000\ \kms$ away from \Ha. The continuum is subtracted, and a broad \Ha\ is looked for in the residual spectrum. The residual flux at 6250\AA--6880\AA\ is summed, excluding regions 20-pixel wide (1\,380 $\kms$) centred on the \oi, \sii, \nii\ and \Ha\ narrow emission lines. Also, we exclude excess flux detected away from \Ha, with a region at least ten pixels wide with null mean residual flux, as it is implausible that such distinct emission features originate from a broad \Ha\ emission feature. Bad pixels are interpolated over. The total excess flux is denoted by $\Delta F$. 

The significance of $\Delta F$ is assessed by comparing it to $\sone$, the dispersion in the flux density at 6125\AA --6250\AA\ and 6880\AA --7000\AA, 
which incorporates both the S/N of the spectrum and residual stellar features which were not fully subtracted in \S 2.2. A total of \midsmpsz\ (3\%) objects with $\normdf > \minfluxinwingsA$, from
the parent sample of \psmpsz\ objects, are found as potential broad-line AGN, and were passed on to the next stage of the pipeline. A depiction of this process, and the reasoning behind the 2.5 threshold, appear in appendix A3.

\subsection{Broad \Ha\ and narrow lines fit}\label{subsec: GHfits} 
The purpose of this stage is to fit the excess flux near \Ha\ by a broad line profile.
We measure the profiles of various forbidden lines, \oiii\ $\lambda 5007$, \oi\ $\lambda\lambda 6300,6363$, 
\nii\ $\lambda\lambda 6548,6583$ and \sii\ $\lambda\lambda 6716,6730$, which provide the narrow lines profile, and help us separate the broad and narrow components of \Ha. We use 4\th -order Gauss-Hermite functions (GHs; van der Marel \& Franx 1993) to fit the narrow line profiles, and an up to 10\th -order GH to fit the broad \Ha\ profile, which can have a complex shape, such as a double peak. The Levenberg-Marquardt best-fit algorithm (Press et al. 1992) is used in all fits. The fit is used to measure the luminosities and mean velocities of all the lines mentioned above, and the higher moments are measured only for the broad \Ha\ and for \oiii. 

An overview of the fitting process is given below, and further details appear in appendix A4. The narrow line results are used in following papers, so a more elaborate analysis of their fit procedure and its implications is given there. 

The wavelength region near \oiii, 4967\AA --5250\AA, is fit with three components: an iron template derived from observations of I~Zw~1 (kindly provided by T. Boroson), a GH fit for the \oiii\ line, and a continuum component. The iron template can broaden and shift during the fit. 
Following that, all the lines between 6250\AA --6880\AA\ noted above are fit, using a 4\th-order GH for the broad \Ha. To lower the degrees of freedom, the widths and higher moments of all the narrow lines in this region are kept equal. In the cases where all these narrow lines are too weak to significantly constrain their fit (\prcntbasedonoiii\% of final sample, see \S 2.5), 
we tie their widths and higher moments to that of \oiii, which is detected in $99\%$ of the objects in the final sample. Then, the broad \Ha\ is fit with increasingly higher order GHs, until an acceptable fit is achieved. 
Finally, the narrow \Hb\ line is also fit, keeping its centroid, width and shape equal to that of the narrow \Ha. Fitting the broad \Hb\ profile is beyond the scope of this paper.

Following the above process, \NprcntHighchisq\% of the objects have an unacceptable reduced $\chisq > 2$ in the 6250\AA --6880\AA\ region. The treatment of these objects is detailed in appendix A4.

Figure 1 shows various examples of fits. The three upper panels show objects with a range of broad to narrow \Ha\ peak flux density ratio. The upper panel shows a `standard' Seyfert 1.0, where the broad \Ha\ dominates, with a flux density ratio at the line core of $F_{\lambda; \bHa}/F_{\lambda; \nHa}\simeq 3$. Note the extended wings of the broad \Ha\  profile, beyond a simple Gaussian, which are fit by higher order GH components. The second panel shows a Seyfert 1.5 (Osterbrock 1981), where the narrow \Ha\ peak is higher than the broad \Ha\ peak $F_{\lambda; \bHa}/F_{\lambda; \nHa}\simeq 0.5$. The FWHM of the total \Ha\ profile is dominated by the narrow component. Such an object is therefore excluded from QCV and the Vanden Berk et al. (2006) sample of broad line AGN, although it is clearly a broad line AGN. The third panel shows a Seyfert 1.9 ($F_{\lambda; \bHa}/F_{\lambda; \nHa}\simeq 0.05$), where a very broad \Ha\ is still detectable. The broad \Hb\ is unobservable, as expected. Such an object will be defined as a Seyfert 2, if only \Hb\ is available, although the EW of \Ha\ (before the host subtraction) is 28\AA, which may indicate this is actually an unobscured broad line AGN. A full analysis of narrow to broad flux ratios and their dependence on AGN characteristics is deferred to a following paper. 

The bottom panel is an example of an object with no clear transition between the Balmer narrow and broad line profiles. Additionally, all narrow lines near \Ha\ are undetectable. In such objects, we use the $\oiii\ \lambda 5007$ profile as a template profile for the narrow Balmer lines.

\begin{figure}
\includegraphics{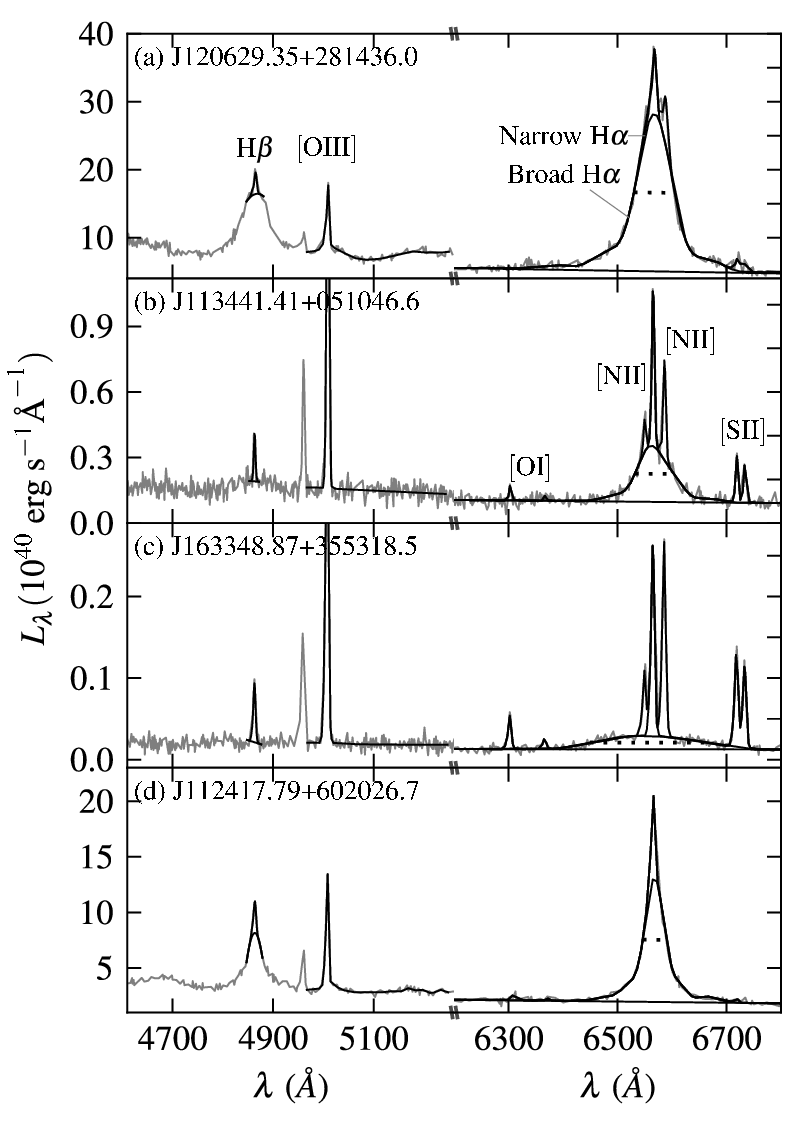}
\caption{Examples of the range of \Ha\ and \Hb\ profiles observed in the host-subtracted spectra. The black lines are the fits to the broad \Ha\ (8\th --10\th\ order GH functions) and to the narrow lines (4\th-order GHs). Dotted lines mark the FWHM of the Broad \Ha.  {\bf (a)} Seyfert 1.0, $F_{\lambda; \bHa}/F_{\lambda; \nHa}\simeq 3$: note the extended wings of the BL profile compared to a Gaussian, fit by high-order GH components. {\bf (b)} Seyfert 1.5, $F_{\lambda; \bHa}/F_{\lambda; \nHa}\simeq 0.5$: the FWHM of the total \Ha\ profile is dominated by the narrow \Ha, therefore this object is excluded from the SDSS quasar catalog and the Vanden Berk et al. (2006) sample.  {\bf (c)} Seyfert 1.9, $F_{\lambda; \bHa}/F_{\lambda; \nHa}\simeq 0.05$: such an object will be defined as a Seyfert 2 if only \Hb\ is available, despite an EW$_{\rm{b\Ha}}=28\AA$, which may indicate this is an unobscured broad line AGN. {\bf (d)} A type 1 AGN with no clear transition in the profile between $F_{\lambda; \bHa}$ and $F_{\lambda; \nHa}$. The Balmer NL profiles are based on \oiii, since \nii, \sii\ and \oi\ are undetectable.}
\label{fig: fits}
\end{figure}

\subsection{Final selection criteria}\label{subsec: final criteria}
The purpose of this stage is to determine if the excess flux found above in \midsmpsz\ objects is produced by a broad \Ha.
As the $\normdf = \minfluxinwingsA$ threshold is chosen to minimize the rejection of broad line AGN (appendix A3), many objects with larger $\normdf$ lack a clear broad \Ha\ feature. To filter out such cases, we employ the additional following selection criteria:

\begin{enumerate}
\item A line width $1 < \dv < \maxfwhm$, where $\dv=\dva/1000$ and $\dva$ is the \Ha\ FWHM in $\kms$. The upper limit filters out cases where a very broad GH component essentially fits the continuum. The lower limit filters out a very narrow GH component which fits the narrow \Ha\ profile.
\item A normalized excess flux density at the line centre $\normhieght > \minHafluxdensity$, where $\epsilon_{\lambda}$ is the SDSS flux density error. 
This ensures significant flux density excess at the expected centre of \Ha, in addition to the above requirement of significant total flux. It filters out significant excess flux placed away from the position of \Ha. Also, it excludes broad and very shallow emission features produced when the continuum is not properly placed (e.g. due to residual stellar absorption features, \S 2.2).
\item A revised excess flux criterion ${\Delta F' / \sone} > \minfluxinwingsB$, 
rather than $\normdf > \minfluxinwingsA$ used earlier (\S 2.3), where
$\Delta F'$ is obtained from the broad \Ha\ profile fit.
Flux at $\lambda$ less than $3 \sigma$ away from the peak positions of the narrow lines is excluded in the summation of $\Delta F'$ (using the $\sigma$ and peak fit to the narrow lines). This filters out objects in which the original excess flux may come from a broad base of the forbidden lines, rather than from the BLR.
\end{enumerate}

A total of \NinitiallyPassed\ objects with acceptable fits (reduced $\chisq < 2$) pass these criteria. The selection effects produced by the above criteria are discussed below. To verify the reality of the broad \Ha\ detection, we also inspected by eye all the \NinitiallyPassed\ spectra, and excluded further \NeyeInspectThrows\ objects where there was no clear BLR, i.e a false-positives rate of $\sim \falseRatePrcnt\%$ for our broad \Ha\ detection algorithm. Also, \Npasshighscore\ objects with high $\chisq$ have an acceptable fit, as described in appendix A4. Finally, we inspected by eye the images of all the $z<0.03$ objects and removed three objects which were offset from the centre of their host galaxies. The size of the final sample, hereafter the `T1 sample', is \smpsz\ objects. 

Table 1 lists the broad \Ha\ luminosity ($\lbha$), and $\dv$ for each object in the T1 sample. Our rejection criteria may have been too restrictive, and potentially excluded objects with an unusual broad line profile. We therefore provide in appendix B a table which lists the other \nRejected\ excess flux objects that were rejected from the T1 sample, along with the reason for rejection. This list may be useful for future studies.

\begin{table*}
\begin{tabular}{l|c|c|c|c|c|c|c|c|c}
SDSS Name & $z$ & $\lbha$ & $\dva$ & \Ks & H & J & NUV & FUV & 2~\kev \\ 
\hline
J000202.95-103037.9  &  0.103  &  41.91  &  2320  &  44.13  &  44.19  &  44.23  &  43.83  &  43.83  &  42.58  \\
J000338.94+160220.6  &  0.116  &  42.19  &  2080  &  44.14  &  44.23  &  44.16  &  43.59  &  43.49  &  42.49  \\
J000410.80-104527.2  &  0.240  &  42.63  &  1370  &  44.55  &  44.63  &  44.47  &  44.65  &  44.69  &  43.22  \\
J000611.55+145357.2  &  0.119  &  42.12  &  3300  &  44.18  &  44.24  &  44.30  &  43.94  &  44.03  &  42.60  \\
J000614.36-010847.2  &  0.090  &  41.60  &  3960  &  43.77  &  43.86  &  43.89  &  43.30  &  43.26  &     0.  \\
\end{tabular}
\caption{The broad \Ha\ luminosities (in $\log\ \ergs$), widths (in $\kms$) and 2MASS, GALEX and ROSAT $\nln$ luminosities (in $\log\ \ergs$) of the T1 sample. GALEX and 2MASS luminosities are in observed frame. Non detections are denoted by `0'. For 2MASS and GALEX, a detection in all bands of the survey is required. An object that was not observed is denoted by `-1' (GALEX only). The electronic version of the paper includes all \smpsz\ objects.}
\label{table: sample}
\end{table*}

\subsection{Selection effects}\label{subsec: selection effects}
Since the T1 sample is selected based on the detection of a broad \Ha, it is important to understand how the sample selection criteria (\S 2.3 and \S 2.5) affect the range of detectable line widths, as a function of line flux. 

Figure 2 presents three panels with the distribution of the T1 sample objects in the broad \Ha\ flux ($\fbha$) vs. $\dv$ plane. In each panel the coloured objects mark the \prcntnearthres\% of the objects which are just above a given selection threshold. The top panel marks the positions of \prcntnearthres\% of the objects where  $\minfluxinwingsA < \normdf < \selectionNearThresA$. The selection of $\normdf > \minfluxinwingsA$ affects the lowest detectable $\dv$.
As $\fbha$ decreases below $10^{-14}\ \flux$, the minimal detectable $\dv$ increases from 1 to 3. As the excess flux becomes lower, the line needs to be broader, otherwise it is lost in the glare of the NLR \Ha\ component. Selection criterion (iii) in \S 2.5 of ${\Delta F' / \sone} > \minfluxinwingsB$ (not plotted), produces a similar effect.

\begin{figure*}
\includegraphics{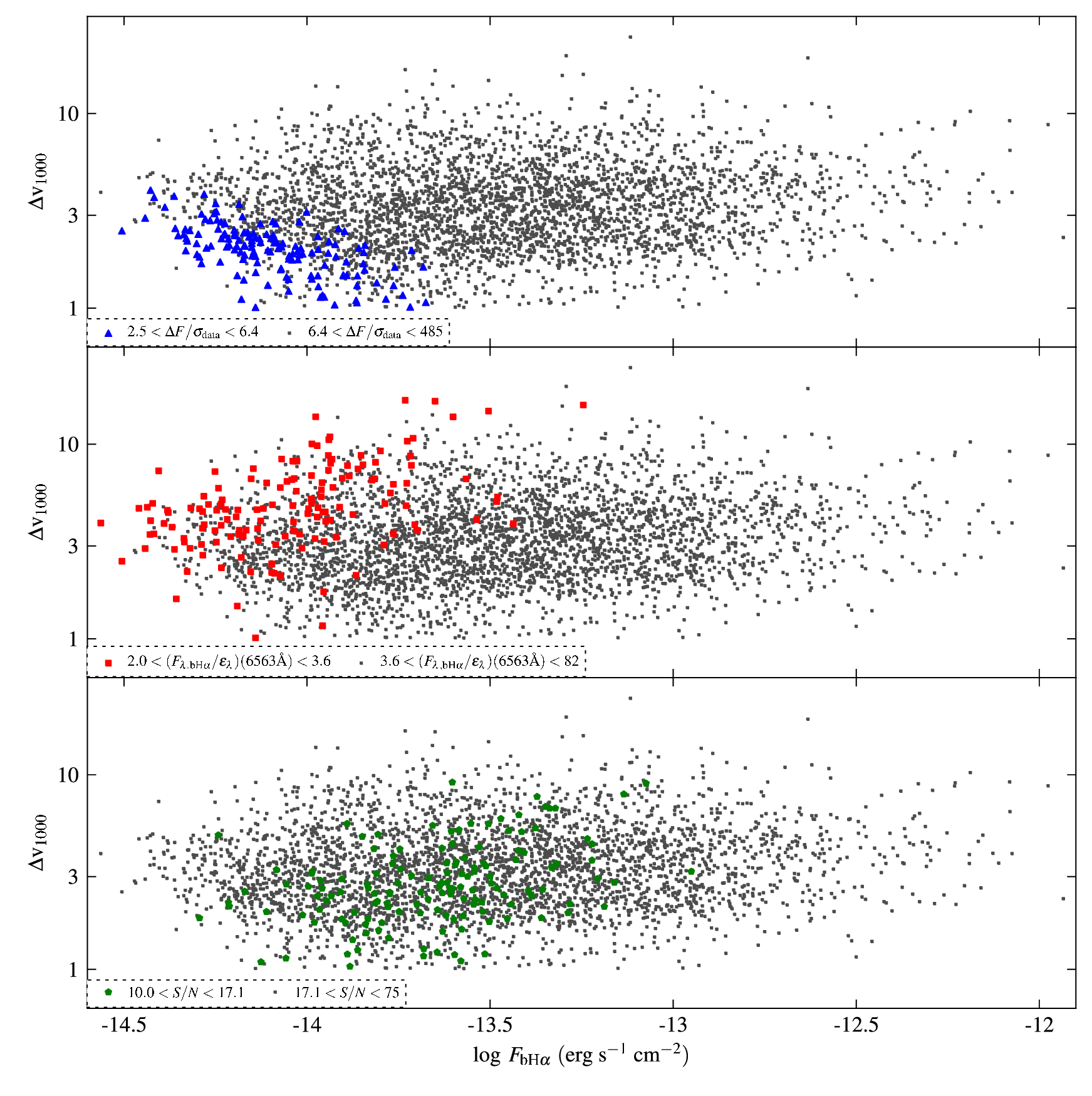}
\caption{The selection criteria effects on the 
distribution of the T1 objects (small gray dots) in the broad \Ha\ flux vs. FWHM plane. Large coloured markers are used to plot the \prcntnearthres\% of the T1 objects which are just above the following three selection thresholds. {\bf (Upper panel)} The $\normdf > \minfluxinwingsA$ selection criterion. At $\fbha<10^{-14}\ \flux$, this selection increases the minimal detectable $\dv$ from 1 to 3. {\bf (Middle panel)} The $\normhieght > \minHafluxdensity$ selection criterion. At $\fbha<10^{-14}\ \flux$, this selection decreases the maximal detectable $\dv$ from 10 to 3. The {\bf (Lower panel)} S/N $>10$ selection criterion. This selection criterion does not form any boundary in the $\fbha$ vs. $\dv$ plane.
The T1 objects extend down to $\dv=1$, in particular at $\fbha<10^{-13.5}\ \flux$, and objects with $\dv<1$ are excluded ($\sim \nlowfwhmp\%$, see text). 
The $\dv < \maxfwhm$ criterion has practically no effect, as objects with $\dv>15$ are extremely rare. Our $\fbha$ detection limit is lowest for $\dv \sim 3$ objects. 
} 
\label{fig: FWHM vs F}
\end{figure*}

The middle panel marks the positions of objects where $\minHafluxdensity<\normhieght<\selectionNearThresB$. This selection produces a reverse effect. As $\fbha$ decreases below $10^{-14}\ \flux$, the maximum detectable $\dv$ decreases from 10 to 3. As the excess flux density becomes lower, the line needs to be narrower, otherwise it is lost in the continuum noise.

The lower panel marks the positions of objects with $\minsn < \rm{S/N} < \selectionNearThresC$. Clearly, this selection criterion does not form any of the boundaries in the $\fbha$ vs. $\dv$ plane. Of the \NeyeInspectThrows\ objects removed by eye (\S 2.5), 85\% have $\fbha$ and $\dv$ values which coincide with the $\minfluxinwingsA < \normdf < \selectionNearThresA$ objects (top panel), indicating the manual filtering does not introduce an additional boundary in this plane. 

A similar analysis in the $\ewha$ vs. $\dv$ plane gives similar results. The minimum detectable $\dva$ rises from 1 to 3 for $\ewha<40\AA$, and the maximum detectable $\dv$ decreases from 10 to 3 for $\ewha<20\AA$.

The objects in Fig. 2 extend towards $\dv=1$, in particular at $\fbha<10^{-13.5}\ \flux$. Thus, the selection criterion $\dv>1$ may exclude some type 1 AGN which have $\dv<1$. What is the possible missed fraction of AGN? We have excluded \nlowfwhm\ ($= \smpsz \times \nlowfwhmp\%$) objects that were fit with $0 < \dv < 1$, due to possible confusion with emission from \nii, some of which may be true broad line AGN with $\dv<1$.
The expected fraction of $\dv<1$ AGN can be also estimated from the GH07 sample, where also 3\% of the objects have $\dv<1$. Since a primary motivation of GH07 was to find broad line AGN with the lowest $\dv$ (low $\mbh$, see Greene \& Ho 2004), 
it is likely to be as complete as possible, and implies the T1 sample likely misses the 3\% of AGN in the SDSS with $\dv<1$, which may constitute part of the objects with an apparent $\dv<1$ excluded here.

The $\dv < \maxfwhm$ selection criterion has practically no effect as very few objects reside at $\dv>10$
(only one object with $\dv > 25$ is detected here\footnote{SDSS J094215.12+090015.8; see Wang et al. (2005).}).

\subsection{Additional surveys}\label{subsec: other surveys}
We supplement the optical SDSS spectra in the T1 sample with photometric measurements in the UV, Near IR and X-ray. In order to include all AGN light, and to be consistent with the luminosity calibration of SDSS spectra (Adelman-McCarthy et al. 2008), we use luminosities from fixed apertures (different for each survey), corrected for the expected full-to-aperture light ratio of a point source. Note that for extended objects, the deduced luminosity is different than the luminosity obtained by a PSF fit to the image. This procedure is preferred over using galaxy-model fits, which perform poorly and differently for different AGN/host light ratios (e.g. \S4.4.5 in Stoughton et al. 2002). Thus, comparing the AGN luminosity of our T1 objects to PSF luminosities of distant quasars (\S3.3) is justified. Host luminosities are compared only with objects that were observed in the same way (\S3.3 -- \S3.5). The specific details of each survey are described below, and the photometric luminosities of each object are listed in Table 1.

\subsubsection{GALEX}\label{subsubsec: uv}
The GALEX mission (Martin et al. 2005) imaged 2/3 of the sky with two broadband filters: far-UV (FUV; effective wavelength 1528\AA) and near-UV (NUV; 2271\AA). We use the GR6 data release, described in Morrissey et al. (2007) and in the MAST website\footnote{http://galex.stsci.edu/GR6}. 

We search for GALEX detections in both bands, within 5\arcsec\  of the T1 objects. In the case of multiple GALEX detections per object, we co-add the observations, weighted by exposure time. Figure 3 presents the detection fractions of the \observedbyGalex\% of the T1 objects observed by GALEX, as a function of broad \Ha\ flux ($\fbha$) for point sources and for extended sources. The point sources in the sample are effectively complete in the UV. A trend of decreasing detection fraction with decreasing $\fbha$ can be seen in the extended objects, but the detected fraction remains $>82$\%. If we include NUV-only detections, 98\% of the T1 objects observed by GALEX are detected.

We use the UV fluxes of the 6\arcsec-radius `aperture 4', so that the residual correction for the PSF is $<10\%$ (Morrissey et al. 2007). We correct for the PSF and for Galactic reddening using $A_{\rm{NUV}}/\ebv = 8.2$ and $A_{\rm{FUV}}/\ebv = 8.24$ (Wyder et al. 2007). When a specific rest-frame UV wavelength is used, we interpolate (or extrapolate) a power-law from the measurements of the two bands. 

\begin{figure}
\includegraphics{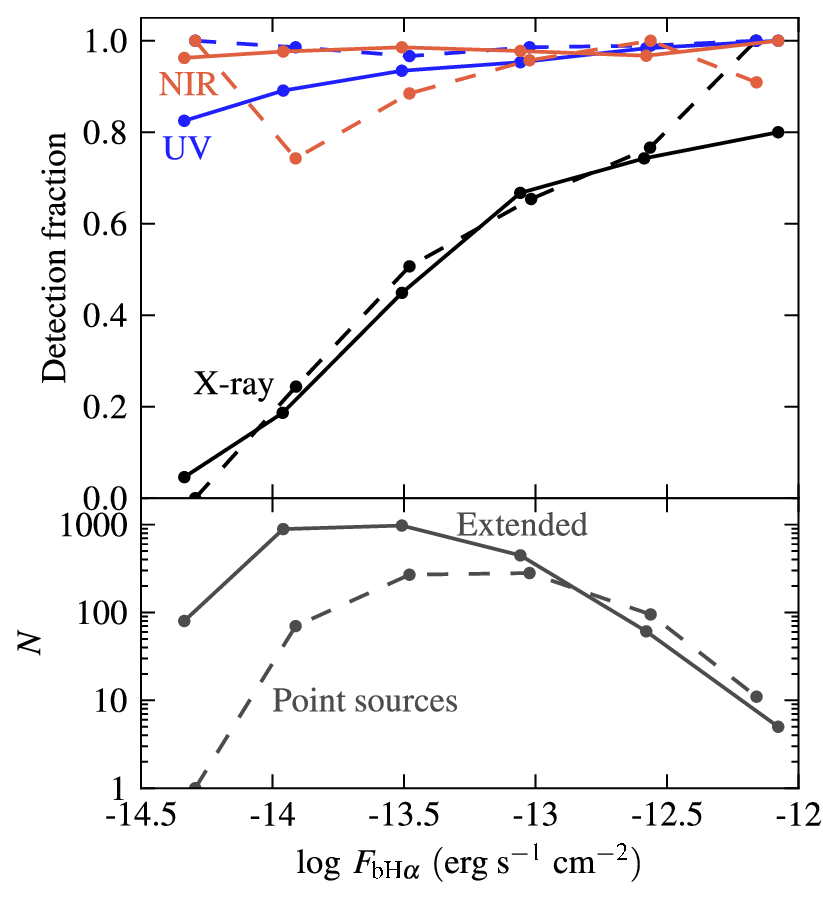}
\caption{The detection fractions of the T1 objects in the GALEX (UV, blue), 2MASS (NIR, red) and ROSAT (X-ray, black) surveys, as a function of $\fbha$, for extended and point like sources. Detections in all bands of each survey is required. Solid/dashed lines represent extended/point source objects. The lower panel shows the number of T1 point-like and extended objects per 0.5 decade $\fbha$ bin. The GALEX detection fraction is relative to the \observedbyGalex\% of T1 objects observed. Note the nearly complete 2MASS and GALEX detections. When analyzing the X-ray luminosities (\S 3.6 and \S 3.9), we utilize only $\fbha>10^{-13.5}\ \flux$ objects, where the detection fraction is $>50\%$. Note that most of the lowest $\fbha$ T1 AGN reside in extended objects. Their bright host enables them to pass the SDSS flux limit, despite their low $\fbha$.} 
\label{fig: band detection fraction}
\end{figure}

\subsubsection{2MASS}\label{subsubsec: nir}
Near infrared photometric information was obtained from the point source catalog of the Two Micron All Sky Survey (2MASS; Skrutskie et al. 2006), which imaged 99.998\% of the sky in the J (1.25 \mic), H (1.65 \mic), and \Ks\ (2.16 \mic) bands to a sensitivity limit of 15 mag in the H-band (for point sources with S/N $\sim 10$). We use the published luminosity from a 4\arcsec-radius aperture, fixed for the residual of the PSF.

For each object in the T1 sample, we search for a 2MASS counterpart using a matching radius of 2\arcsec. We filter 2MASS objects with a risk of nearby star contribution (`cc\_flag' = s,p; Cutri et al. 2003) or an upper limit in either of the three bands. We find that \nirdetectionprcnt\% of the T1 sample have such matches (Fig. 3). We correct for Galactic reddening using the  $A/\ebv$ values in Schlegel et al. (1998). When a specific rest-frame wavelength in the NIR is needed, a power-law is interpolated (or extrapolated) from the two nearest 2MASS bands.

\subsubsection{ROSAT}\label{subsubsec: xray}
The ROSAT All-Sky Survey (RASS; Voges et al. 1999, 2000) covers the entire celestial sphere in the $0.1 - 2.4\ \kev$ range with the Position Sensitive Proportional Counter (Pfeffermann et al. 1987) to a typical limiting sensitivity of $\sim 10^{-13}\ \ergs\ cm^{-2}$. Typical positional uncertainties are $10\arcsec - 30\arcsec$, and the detected source density is $\sim 3$ deg$^{-2}$.
Only \notInCompleteSamplesPrcnt\% of the T1 sample was targeted for spectroscopy by the SDSS solely for being near a ROSAT source, and not as part of the complete SDSS Galaxy and Quasar surveys (see \S 2.1 and Stoughton et al. 2002). Therefore, we do not expect the T1 sample to be biased towards X-ray bright AGN. 

We search for a RASS source within 50\arcsec\ of each T1 object, expecting \expectfalseXmatches\ objects to have false matches. We find matches to \xraydetection\ (\xraydetectionprcnt\%) of the T1 sample. The detection fraction vs. $\fbha$ is shown in Figure 3. The strong trend of detection rate with $\fbha$ implies that most of the objects without an X-ray detection are not a distinct class (`X-ray weak'; Laor et al. 1997), but are rather AGN faint in all bands. 

We use the PIMMS software to convert RASS count rates into X-ray fluxes, assuming a power-law X-ray spectrum with $\alpha_{\rm{x}} = \xslope$ (Laor et al. 1994, Schartel et al. 1996). Then, using the $z$ of the optical match, $\alpha_{\rm{x}} = \xslope$, and the Galactic $N_{\rm{H}}$ measurements of Stark et al. (1992), we derive the rest-frame $\nln(2\ \kev)$ of the matched objects,
corrected for Galactic absorption.

\section{RESULTS}\label{sec: Results}

\subsection{The final sample}\label{subsec: final sample} 
Figure 4 presents $\lbha$ vs. $z$ of the final T1 sample. The figure displays the \szMidZ\ objects at $z<0.2$. 
Empty circles mark the \szMidZinQCV\ objects (\szMidZinQCVprcnt\%) that also appear in the QCV sample. The QCV sample objects mostly reside at $\log \lbha \approx 42.5-43$, while the T1 sample extends further down to $\log \lbha \approx 40$. The T1 sample includes additional \szHighZ\ objects at $0.2<z<0.31$ not displayed in Fig. 4. In this range there is an overlap of \szHighZinQCVprcnt\% of our objects with the QCV sample,
but our sample includes only \prcntOflargezQCV\% of the QCV objects in this range. This small fraction is due to our requirement of high quality data near \Ha, which is satisfied by a decreasing fraction of the objects as \Ha\ shifts into the noisy red end of the SDSS spectra (\S 2.1). The total T1 sample extends over the range $\log \lbha \sim 40-44$, and becomes distinct from the QCV sample at $z<0.2$ .

\begin{figure}
\includegraphics{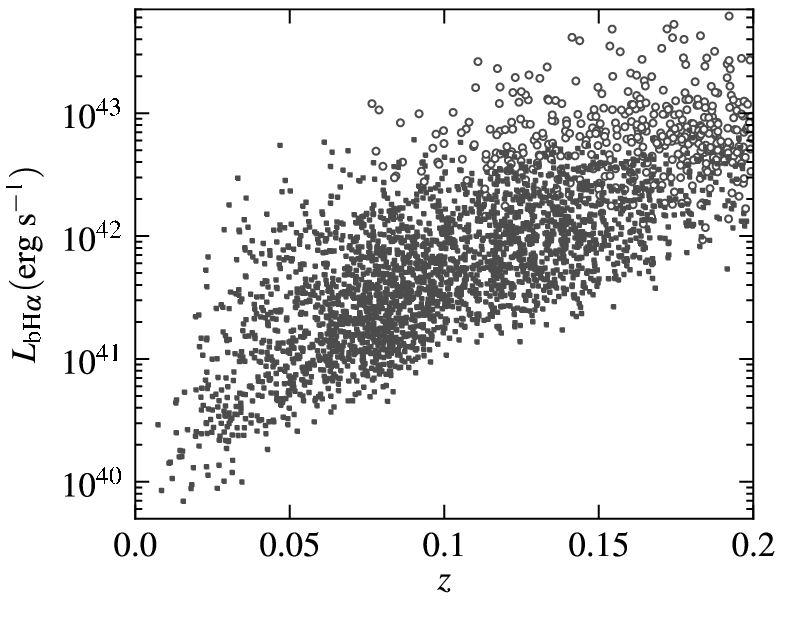}
\caption{The $\lbha$ versus $z$ distribution for the T1 sample at $z<0.2$ (\szMidZ\ objects). Empty circles mark the \szMidZinQCV\ objects (\szMidZinQCVprcnt\%) that also appear in the SDSS quasar catalog (QCV). Note that the QCV sample terminates at $\lbha$ of a few $10^{42}\ \ergs$, while the T1 sample extends further down to $10^{40}\ \ergs$.}
\label{fig: L_bHa_vs_redshift}
\end{figure}

Figure 5 compares the ranges of $\lbha$ spanned by the T1 sample with the $z<0.35$ QCV sample (where \Ha\ is observable). Our sample extends down by nearly two orders in magnitude in $\lbha$, compared to the QCV sample. Specifically, the T1 sample includes \oursunderfortytwo\ objects at $\lbha<10^{42}\ \ergs$, compared to 207 in QCV, and \oursunderfortyone\ objects at $\lbha<10^{41}\ \ergs$, compared to 5 in QCV. We also show the Ho et al. (1997a) sample (44 objects, updated values in Ho 2003), which extends below $10^{38.5}\ \ergs$, while the T1 sample terminates just below $10^{40}\ \ergs$.

\begin{figure}
\includegraphics{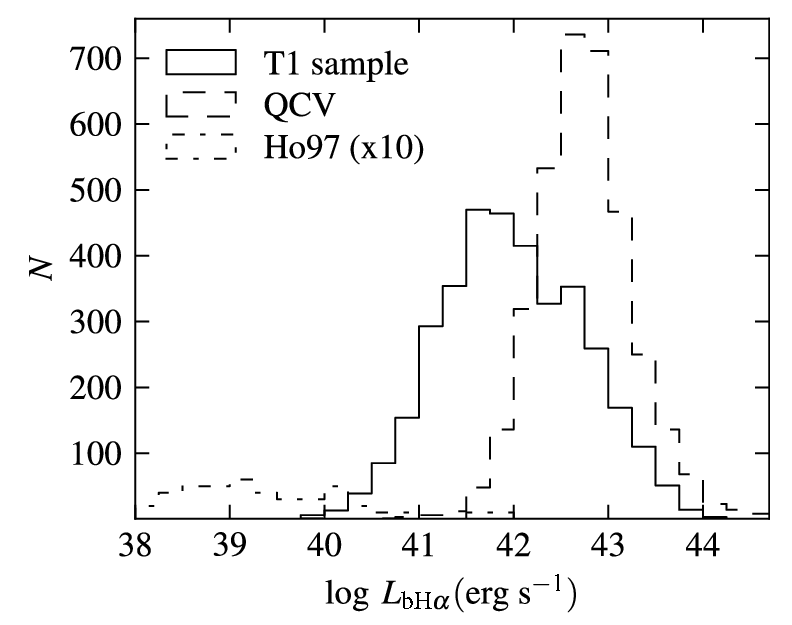}
\caption{Comparison of $\lbha$ ranges of different broad line AGN samples. We include only $z<0.35$ objects of the QCV sample, where \Ha\ is observable. Although the T1 sample extends down by two orders of magnitude in $\lbha$, compared to QCV, it does not extend as low as the Ho et al. (1997a) sample, which reaches $\lbha<10^{38.5}\ \ergs$ (magnified ten-fold for clarity).}
\label{fig: L_bHa hist}
\end{figure}

In this paper we explore the emission properties of the T1 sample as a function of various parameters, including $\mbh$ and $\lledd$. We derive these parameters from \Ha, following Greene \& Ho (2005) and Kaspi et al. (2005). We use the relation between the BLR radius and UV luminosity in Kaspi et al. (2005),
\begin{eqnarray}
\lefteqn{ R_{\rm{b\Hb}} =  \notag  } \\
 && (17.6 \pm 0.26) \times \nln(1450\AA)_{44}^{0.545 \pm 0.06}\ \rm{light}\ \rm{days} ~~,
\label{eq: kaspi}
\end{eqnarray}
We then use the relation $\nln(1450\AA) = 33 \times \lbha^{1.0}$ (eq. 5 below), and Greene \& Ho (2005, eq. 3 there) to convert FWHM(broad \Hb) to FWHM(broad \Ha), which gives
\begin{eqnarray}
\label{eq: mbh}
m  \equiv  \log \frac{\mbh}{\msun} = 7.4 + 2.06 \log \dv + 0.545\log L_{\rm{b\Ha}, 44} .
\end{eqnarray}
We use $\lbha$ to obtain $\lbol$ using the relation $\lbol = \lbolcoeff \times \lbha^{1.0}$ (eq. 6 below), which gives
\begin{eqnarray}
\label{eq: lledd}
l  \equiv  \log\ \frac{L}{L_{\rm Edd}} = 0.6 - 2.06 \log \dv + 0.455\log L_{\rm{b\Ha}, 44}
\end{eqnarray}
Note that eq. 2 is equivalent to eq. 6 in Greene \& Ho (2005), but is slightly different as it uses the Kaspi et al. (2005) relation, and the relations between $\lbha$ and continuum luminosity obtained below. 

It is important to understand how the sample selection criteria affect the observed range of various AGN parameters. Figure 6 shows the observed distribution of the T1 sample objects in the $\lbha$ vs. $\dva$ plane. In the top panel, the larger (coloured) dots in this figure present the objects close to our selection cutoffs (see Fig. 2). The $m$ and $l$ diagonal lines are derived from eqs. 2 and 3. It is clear that $m \lesssim 6$ objects are heavily selected against due to the $\normdf > \minfluxinwingsA$ criterion, while the $\normhieght > \minHafluxdensity$ criterion selects against $l \lesssim -2.8$ objects. 

\begin{figure*}
\includegraphics{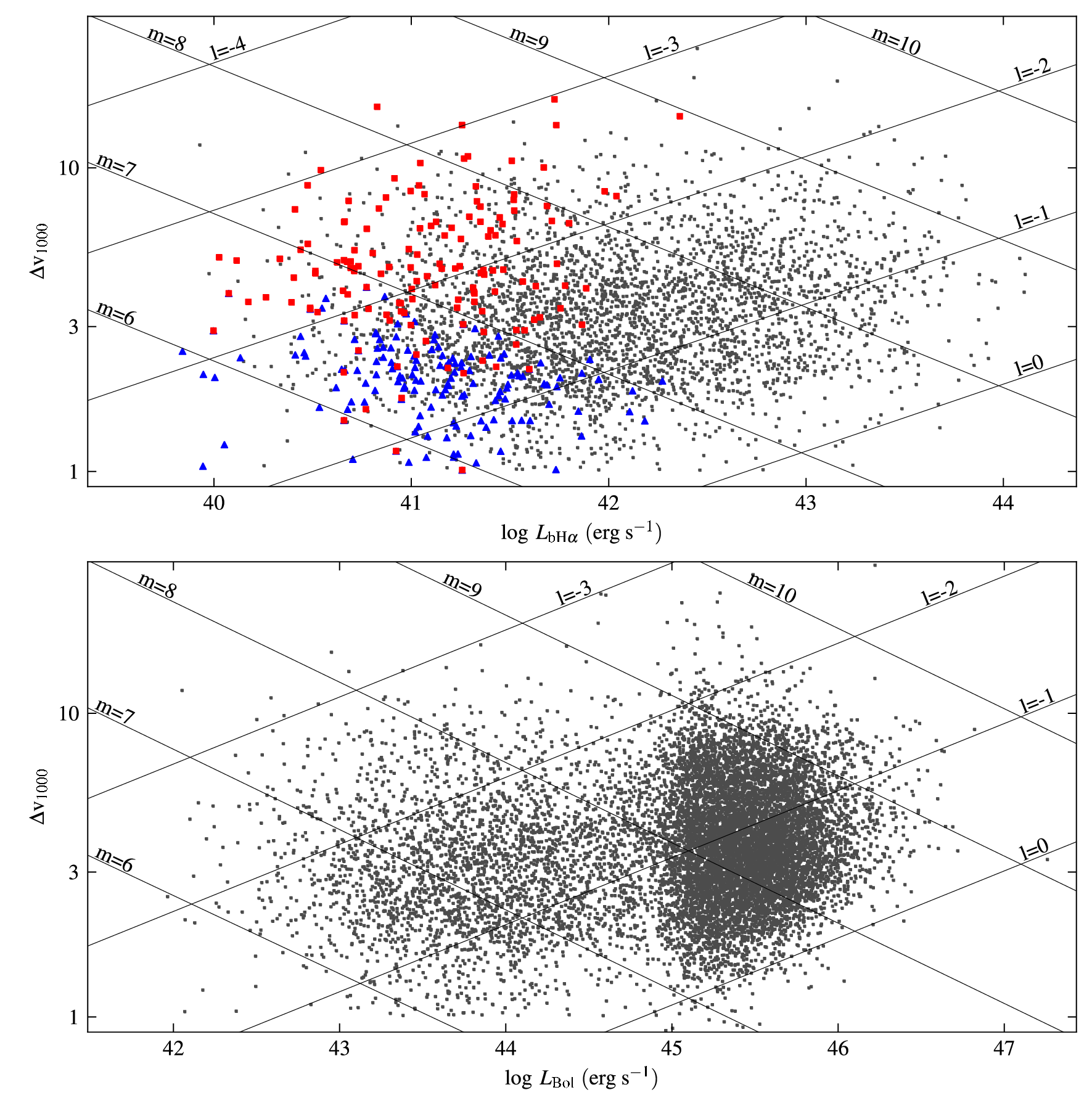}
\caption{{\bf (Upper panel)} The distribution of the T1 sample objects in the $\lbha$ vs. $\dva$ plane. Large coloured markers are as in Fig. 2.  Objects with $m \lesssim 6$ or $l \lesssim -2.8$ are selected against. The Eddington limit ($l=0$) sets a minimum $\dva$ with increasing $\lbha$ for $\lbha > 10^{42.5}\ \ergs$. This, together with the rarity of AGN with $m>9$, leads to a decrease in the range of observed $\dva$ values with increasing $\lbha$.
{\bf (Lower panel)} The distribution of the T1 sample + QCIV sample (values from Shen et al. 2008, see text) in the $\lbol$ vs. $\dva$ plane ($\lbol$ is from eq. 6). The Eddington limit is clearly seen. Note there is a similar steep decline in the number of objects with $\dva>10\,000\ \kms$, at all luminosities. The decline is not set by our selection criteria. }
\label{fig: FWHM and L}
\end{figure*}

Also, we can now identify the effect of the rarity of active very massive black holes ($m \gtrsim 9$) in the local universe, which manifests as a lack of high-$\dva$, high-$\lbha$ objects. Another limit on the distribution is due to the Eddington luminosity, which leads to an increasing minimum $\dva$ with $\lbha$ for $\lbha > 10^{42.5}\ \ergs$. 

The lower panel shows $\dva$ as a function of $\lbol$.
Here we add the 8\,185 QCIV objects from Shen et al. (2008), where the broad \Hb\ is available ($z<0.66$). We fix for a mean offset of \offsetShen\ dec (with a dispersion of \sigOffsetShen\ dec) between the Shen et al. FWHM(\Hb) measurement and our FWHM(\Ha) measurement, derived based on common objects. The value of $\lbol$ in the T1 sample is derived from eq. 6, and in the Shen et al. sample $\lbol$ is derived 
from $\nln$(5100\AA). The plots shows clearly how the Eddington luminosity limit extends to quasar luminosities.
The Shen et al. FWHM measurements extend to higher luminosities using FWHM(\mgii) or FWHM(\civ) available for higher $z$ objects. We avoid utilizing these measurements since the relation of FWHM(\mgii) to FWHM(\Hb) is not linear, and the FWHM(\civ) shows only a very weak correlation with FWHM(\mgii) (e.g. Shen et al. 2008, Fine et. al. 2010). Note that with increasing $\lbol$ the range in the observed $\dva$ shrinks due to the upper limits on $l$ and $m$. A similar effect is seen for \mgii\ (Fine et. al. 2008), and \civ\  (Fine et. al. 2010). In fact, no broad line objects are expected to exist at a high enough $\lbol$ due to the Eddington limit. This merely reflects the fact that the minimal required $\mbh$ increases with $\lbol$, and AGN above some $\mbh$ just do not exist.

\subsubsection{The $\dva$ distribution}

How broad do the broad lines get? The lower panel of Fig. 6 shows there is a steep decline in the number of objects with $\dva>10\,000\ \kms$. The decline is independent of $\lbol$, over four orders of magnitude in $\lbol$. The decline does not result from our selection criteria, as shown in the upper panel. Figure 7 displays histograms of the FWHM distribution ${\rm d} N/{\rm d}\log \dva$ vs. $\dva$, at four luminosity bins. The distributions are remarkably similar, showing a roughly linear decline of $\log ({\rm d} N/{\rm d}\log \dva)$ vs. $\dva$, or equivalently ${\rm d} N/{\rm d}\log \dva \propto e^{-\dva/\dva_0}$, with $\dva_0 \approx 2700\ \kms$. 
The distributions peaks at $\dva \simeq 2500 - 4000\ \kms$, 
and drops steeply for smaller values. The sharp drop at the smallest $\dva$, and the increase in $\dva$ of the peak position with $\lbol$, is due to the Eddington limit ($l=0$, see Fig. 6), which gives a minimal $\dva\propto \lbha^{0.22}$ (eq. 3). The similar distributions for $\dva$ above the peak is surprising, as the $\dva$ distributions are set by the distribution of the number of objects vs. $l$ and vs. $m$. It is not clear why these two distributions combine to yield the same $\dva$ distributions at different $\lbol$. It may imply that the distributions of $\dva$ is directly set by the value of $\dva$, but it is not clear what may be the physical mechanism behind a dependence on $\dva$ only (see further discussion in \S 4.1).

\begin{figure}
\includegraphics{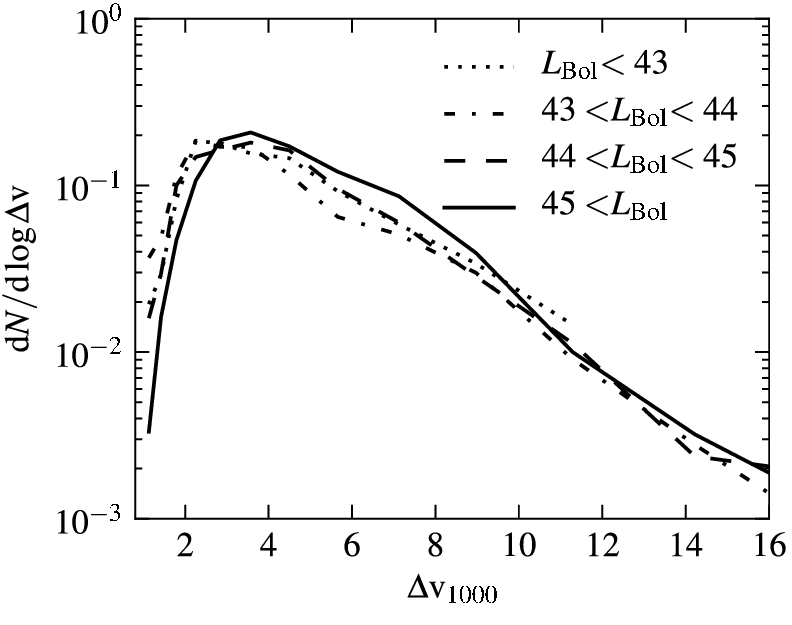}
\caption{
The FWHM distribution of the combined T1 + QCIV sample (Shen et al. 2008), presented in Fig. 6, at different $\lbol$ bins. The distributions are similar. The origin of this similarity is not clear. The $\dva$ distributions should be set by the distribution of $m$ and $l$ values. Either these distributions somehow lead to a $\dva$ distribution which is independent of $\lbol$, or it may imply that $\dva$, rather than $m$ and $l$, sets the observed $\dva$ distribution through some unknown mechanism. Note the effect of the Eddington limit at $\dv<3$, which increases the minimal $\dva$ and the peak position with increasing $\lbol$.}
\label{fig: FWHM hist}
\end{figure}

\subsubsection{The $\lledd$ distribution}
Figure 8 compares the Eddington ratio distributions of the T1 sample with the BQS (using the BG92 $z<0.5$ subsample, with values taken from Baskin \& Laor 2005) and the Kollmeier et al. (2006, hereafter K06) quasar samples. Both the BQS and K06 were selected as point sources, in contrast with the T1 sample, which includes both point sources and extended sources. The T1 sample is centred around $l \approx -1.3$, while the BQS and the K06 samples are centred around $l \approx -0.6$. To determine if the higher $l$ distributions are driven by the point source selection, we show the $l$ distribution of the point sources in our T1 sample. The T1 point sources are also shifted to higher $l$ values. This clearly demonstrates that the higher $l$ values of BQS and K06 samples are largely driven by the point-like selection effect. This trend is well understood by the $\mbh$ versus bulge luminosity relation
(Magorrian et al. 1998), which implies that $\lledd$ is equivalent to $L_{\rm AGN}/L_{\rm bulge}$. Thus quasars, i.e. objects which are selected by their quasi-stellar appearance, are by their selection high 
$\lledd$ AGN. 

Below we compare T1 hosts to other galaxy types (in \S 3.1.4 and \S3.5), where we use the subsample of \nResolved\ T1 AGN which are part of the SDSS main galaxy sample (Strauss et al. 2002). These objects are selected to have extended morphologies, and therefore by selection exclude high $\lledd$ objects (Fig. 8).

\begin{figure}
\includegraphics{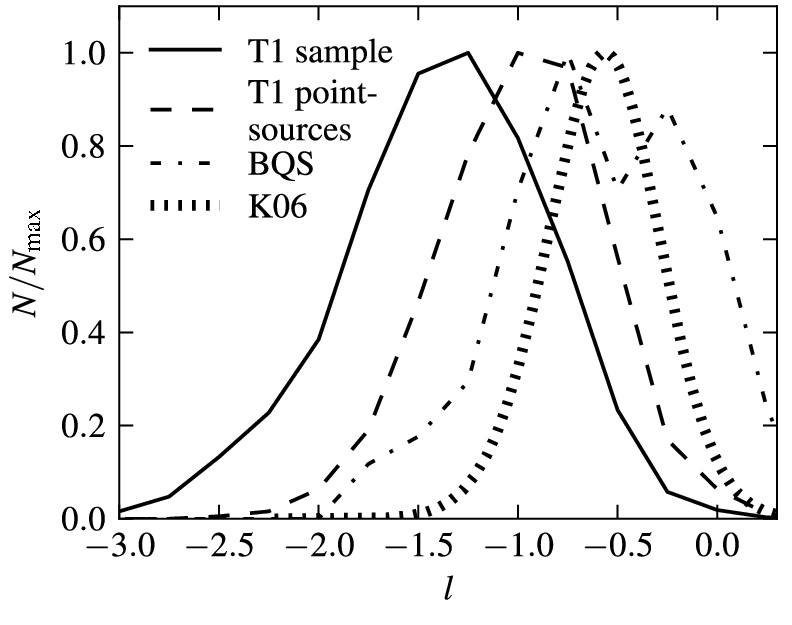}
\caption{A comparison of the Eddington ratio distribution in various samples.
The $l$ bin size is 0.25 decades. Both the BQS and the K06 samples were selected to be point sources. The T1 distribution is shifted to significantly lower $l$ values. To test the origin of this shift we plot the distribution of the T1 objects originally identified as point like sources, which is clearly shifted to higher $l$ values. Therefore, the notion that AGN tend to shine at $l\sim-0.5$ (K06) reflects the point like selection criterion. AGN termed quasi-stellar have a high $l$ by construction.}
\label{fig: lledd}
\end{figure}

\subsubsection{Other low luminosity type 1 samples}

Below we compare the completeness of the T1 sample to the other SDSS based samples of low luminosity type 1 AGN of Hao et al. (2005a), Vanden Berk et al. (2006) and GH07. 

The Vanden Berk et al. (2006) sample is based on the SDSS DR3 and includes 4\,666 objects. The AGN selection criterion is the detection of at least one emission line profile with FWHM $> 1000~\kms$. This criterion excludes AGN where the narrow component dominates the Balmer line profiles (e.g. panels `b' and `c' of Fig.  1). As a result, the Vanden Berk et al. (2006) sample misses $27\%$ of the T1 objects present in SDSS DR3.

The Hao et al. (2005a) sample is based on SDSS DR2 and includes 1\,317 objects, of which 52.5\% (692 objects) entered our parent sample (passing our S/N$>10$ and bad pixel criteria). Hao et al. fit the narrow and broad \Ha\ profile with a simpler procedure than described in \S2, using one or two Gaussians. Of their objects that entered our parent sample, 27\% were rejected due to a lack of broad line flux (\S 2.3), and further $20\%$ were rejected based on the excess flux shape (\S 2.5). A similar percentage of 47\% of the Hao et al. (2005a) sample were rejected by GH07. We therefore suspect that these broad \Ha\ detections by Hao et al. are not significant.

The GH07 sample is based on SDSS DR4 and includes 8\,495 objects. Our selection method of excess broad flux near \Ha\ is similar to the GH07 selection criterion. The broad \Ha\ detection rates in the T1 and GH07 samples are both 1.5\%. However, GH07 fit the broad \Ha\ feature in a smaller velocity range of $-7.4<\v_{1000}<6.3$ from the \Ha\ peak position, compared to $-14.3<\v_{1000}<14.5$ used in the T1 sample. Thus, we find that 1.6\% of our objects (58/3\,596) have $\dv > 10$, compared to 0.51\% (43/8\,495) in GH07. Also, 17\% of the 2\,273 GH07 which appear in our parent sample, were rejected from the T1 sample, potentially due to more strict selection criteria applied here. Comparing the measured $\dva$ values in individual objects which appear in both samples, we find that the T1 values are on average a few percent larger, with a mean offset of $\dva - \dva_{\rm{GH07}}=222\ \kms$. The RMS dispersion of the differences in the individual $\dva$ values is $875\ \kms$, and the RMS of the fractional difference in $\dva$ is 20\% (0.13 for $\log\ \lbha$). 
As noted in \S 2.6, 3\% of GH07 have $\dv < 1$, while such objects are excluded from the T1 sample.

\subsubsection{The fraction of SDSS objects which are T1}\label{subsec: detection fraction}
It is also interesting to compare the T1 sample to samples of type 2 AGN and samples of inactive galaxies. We use samples derived from the SDSS, to avoid biases due to different selection effects of other surveys. We use the classifications of Brinchmann et al. (2004, hereafter B04), who classified SDSS galaxy spectra according to the equivalent width and line ratios of narrow emission lines. Galaxies with narrow lines were divided according to their positions in the BPT diagrams (Baldwin, Phillips \& Terlevich 1981; Veilleux \& Osterbrock 1987). This classification was performed on all DR7 spectra in which the galaxy is dominant\footnote{Results available at http://www.mpa-garching.mpg.de/SDSS/DR7/}. Objects which are clearly type 1 AGN were excluded, to avoid AGN contamination of the host galaxy colours. So, the B04 classification includes only type 2 AGN. However, as mentioned in Kauffmann et al. (2003a), objects with weak broad lines were not excluded. Since our T1 sample reaches low $\lagn/\lhost$ values, it can include apparently type 2 AGN from B04. We find that out of 23\,279 objects classified as `AGN' by B04, 12\,340 appear in our parent sample (following the S/N $>10$ and bad pixel cuts, \S 2.1). Of these, 472 (4\%) are part of our T1 sample. This fraction is comparable with the Kauffmann et al. (2003a) estimate of 8\% weak type 1 objects in their type 2 SDSS DR1 sample, which were not specifically identified. 

It is interesting to explore the fraction of T1 objects classified by B04 as non AGN. In addition to the 472 T1 objects classified as type 2 AGN, 261 T1 objects were classified as composites, 42 as star forming galaxies (SFG), 20 as LINERs, and one as a non-emission line galaxies (NEG). So, a total of 324 objects out of 796, or 41\%, of low luminosity broad line AGN are not clearly identified as AGN based on their narrow lines. In a following paper we measure the narrow lines for all the T1 objects, and discuss their BPT based identifications.  

Figure 9 presents the classification of SDSS galaxies as a function of the spectroscopically measured $\nln(6166\AA)$, which corresponds to the SDSS $r$ band (but measured within a 3\arcsec -diameter fibre). We use all objects in the parent sample (\S 2.1) which are part of the SDSS main galaxy sample (Strauss et al. 2002). 
The T1 classification includes \nResolved\ objects (about 0.1\% of the objects were not classified neither here nor by B04, and are disregarded). 
For the T1 objects we derive $\nln(6166\AA)$ from the mean of 20-pixels around 6166\AA. For non-T1 objects we convert the synthetic\footnote{i.e., the spectrum convolved with the filter pass function.} $r$-magnitudes published by B04 to $\nln$. We correct the B04 measurement by $\Brinchmannoffset$ dec, which is the mean offset between the two methods in common objects. 

The classification fractions in Fig. 9 are somewhat different from B04 since some of the classifications depend on the emission lines S/N, which are higher on average in the S/N $>10$ parent sample presented here. Note that SFG heavily dominate ($>90$\%) at the lowest luminosities, and NEG dominate at the highest luminosities. The fraction of composite galaxies and type 2 AGN peaks at intermediate luminosities. In the T1 sample the luminosity can be dominated by the AGN. We therefore remove the AGN contribution to $\nln(r)$ using the result presented in \S 3.4 (eq. 5, 6\% of T1 objects with an implied negative host are disregarded). 
The fraction of AGN-subtracted T1 objects is marked as `T1 host'. The error bars denote the error in the fraction due to the dispersion of eq. 5. 

Remarkably, the T1 fraction follows well the NEG fraction, or equivalently, their galaxy luminosity distribution is similar. This suggests the two populations are related. If NEG are the host galaxies of type 1 AGN (with extended morphologies), 
then a fixed fraction of $\sim 3$\% NEG host broad line AGN, at the level detectable in this study. Alternatively, if the T1 occur in SFG, then the fraction of SFG which host broad line AGN increases sharply from $2\times 10^{-3}$ to about unity with luminosity. Further results on the T1 hosts are given in \S 3.5, and discussed in \S 4.3.

\begin{figure}
\includegraphics{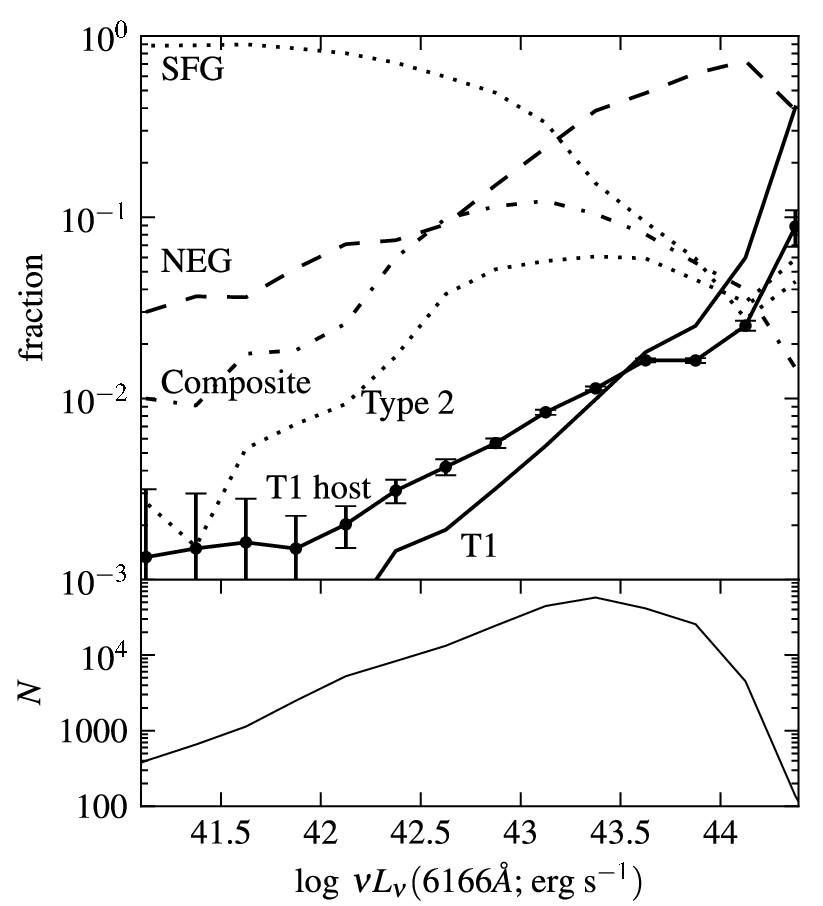}
\caption{The distribution of spectral types of \extendedParentSample\ $z<0.2$ galaxies in the parent sample (\S 2.1), as a function of the spectroscopy derived luminosity (roughly at the $r$-band). The type 2, SFG, NEG and composite classifications (based on the EW of the narrow lines and the BPT diagrams) are taken from B04. At low luminosities $>90$\% of the SDSS galaxies are SFG, while at high luminosities the NEG dominate. The T1 curve is based on the \nResolved\ extended objects from the T1 sample. The `T1 host' band is corrected for the AGN contribution to the luminosity. The error bars are derived from the dispersion in eq. 5. Note that the T1 host luminosity constitute a nearly fixed fraction of $\sim 3$\% of the NEG luminosity distribution. This may suggest that the T1 reside in NEG. The lower panel plots the total number of objects per $\nln$ bin of size 0.25 dec.
} 
\label{fig: type detection fraction}
\end{figure}

\subsection{The IR to X-ray luminosity versus $\lbha$} \label{subsec: lcont vs lbha}
Below we explore the correlation of the continuum luminosity in the X-ray, UV, optical, and near IR with $\lbha$, which provides a measure of the AGN luminosity without host contribution. We divide the T1 sample into $\lbha$ bins, each half a decade wide (the highest bin, $\lbha > 10^{43.8}\ \ergs$, includes only \highBLR\ objects and is ignored). Table 2 lists the number of objects per bin and the fraction of objects with detections in the various bands. The NIR detection is nearly complete, with a mean detection fraction of $96$\%. The mean UV detection is high ($93$\%), remaining $\ge 80\%$ even at the lowest $\lbha$ bins. The high detection rates ensures the derived mean continuum properties are not significantly biased in the UV and in the NIR. The mean X-ray detection fraction is \xraydetectionprcnt\%, and drops below 50\% for $\lbha<{42.3}\ \ergs$. 
The effects of the X-ray band detection rate are discussed in \S 3.6. A search for matched radio detections in the FIRST survey revealed an average detection rate of only 18\%. We therefore preferred not to explore further relations with the radio properties.

\begin{table*}
\begin{tabular}{c|c|c|c|c|c|c|c|c}
$\lbha$ range & mean $\lbha$ & $N$ & mean $z$ & resolved& \multicolumn{3}{|c|}{detection fractions} & R06 scaling  \\ 
 ($\log \ergs$)& ($\log \ergs$) & & & fraction  & $d_{\rm{NIR}}$ & $d_{\rm{UV}}$ & $d_{\rm{X}}$ & (dec) \\ 
\hline
43.3 - 43.8  &  43.48  &  136  &  0.237  &  0.03  &  0.99  &  1.00  &  0.69  &  -0.75  \\
42.8 - 43.3  &  43.01  &  406  &  0.205  &  0.33  &  0.94  &  0.98  &  0.63  &  -1.22  \\
42.3 - 42.8  &  42.54  &  662  &  0.163  &  0.65  &  0.95  &  0.96  &  0.56  &  -1.68  \\
41.8 - 42.3  &  42.04  &  859  &  0.119  &  0.86  &  0.97  &  0.94  &  0.47  &  -2.19  \\
41.3 - 41.8  &  41.56  &  853  &  0.094  &  0.96  &  0.97  &  0.92  &  0.33  &  -2.67  \\
40.8 - 41.3  &  41.09  &  488  &  0.070  &  0.99  &  0.99  &  0.90  &  0.21  &  -3.14  \\
40.3 - 40.8  &  40.60  &  138  &  0.041  &  1.00  &  0.96  &  0.89  &  0.20  &  -3.62  \\
39.8 - 40.3  &  40.10  &   23  &  0.021  &  0.96  &  0.91  &  0.80  &  0.22  &  -4.12  \\
\end{tabular}
\caption{Detection fractions in the eight $\lbha$ bins. The morphology is taken from the SDSS (Stoughton et al. 2002). The detection fractions are in the 2MASS (NIR), GALEX (UV) and ROSAT (X-ray) surveys, and require a detection in all bands of the survey. The R06 scaling is discussed in \S 3.4.
}
\label{table: lbha bins} 
\end{table*}

Figure 10 shows the correlation of $\lbha$ vs. $\nlnl$ in nine rest frame bands from the X-ray to the NIR. The 2 \kev\ luminosity is from ROSAT, 1528\AA\ and 2271\AA\ from GALEX, 3940\AA, 5100\AA, and 7000\AA\ from SDSS, and 1.2\mic, 1.7\mic, and 2.2\mic\ (J, H, \Ks\ bands) from 2MASS. The optical $\nlnl$ are measured on 20-pixel windows from the SDSS spectrum. The longest rest-frame wavelength is chosen so that it is available in all objects\footnote{Objects with $<6$ good pixels in the relevant window are excluded (0.6\%, 2\% and 4\% in 3940\AA, 5100\AA, and 7000\AA, respectively).}. The non-optical bands are also in rest wavelengths, calculated from the photometric values as described in \S 2.7. The average and the dispersion in $\log \nlnl$ in each $\lbha$ bin are marked in the plot. The ROSAT upper limits are shown in \S 3.6. The upper limits for the small fraction of objects with no GALEX detections are not plotted (it is not obvious how to derive them from the exposure time). The fraction of upper limits for 2MASS is tiny, so we do not plot the upper limits.

\begin{figure*}
\includegraphics{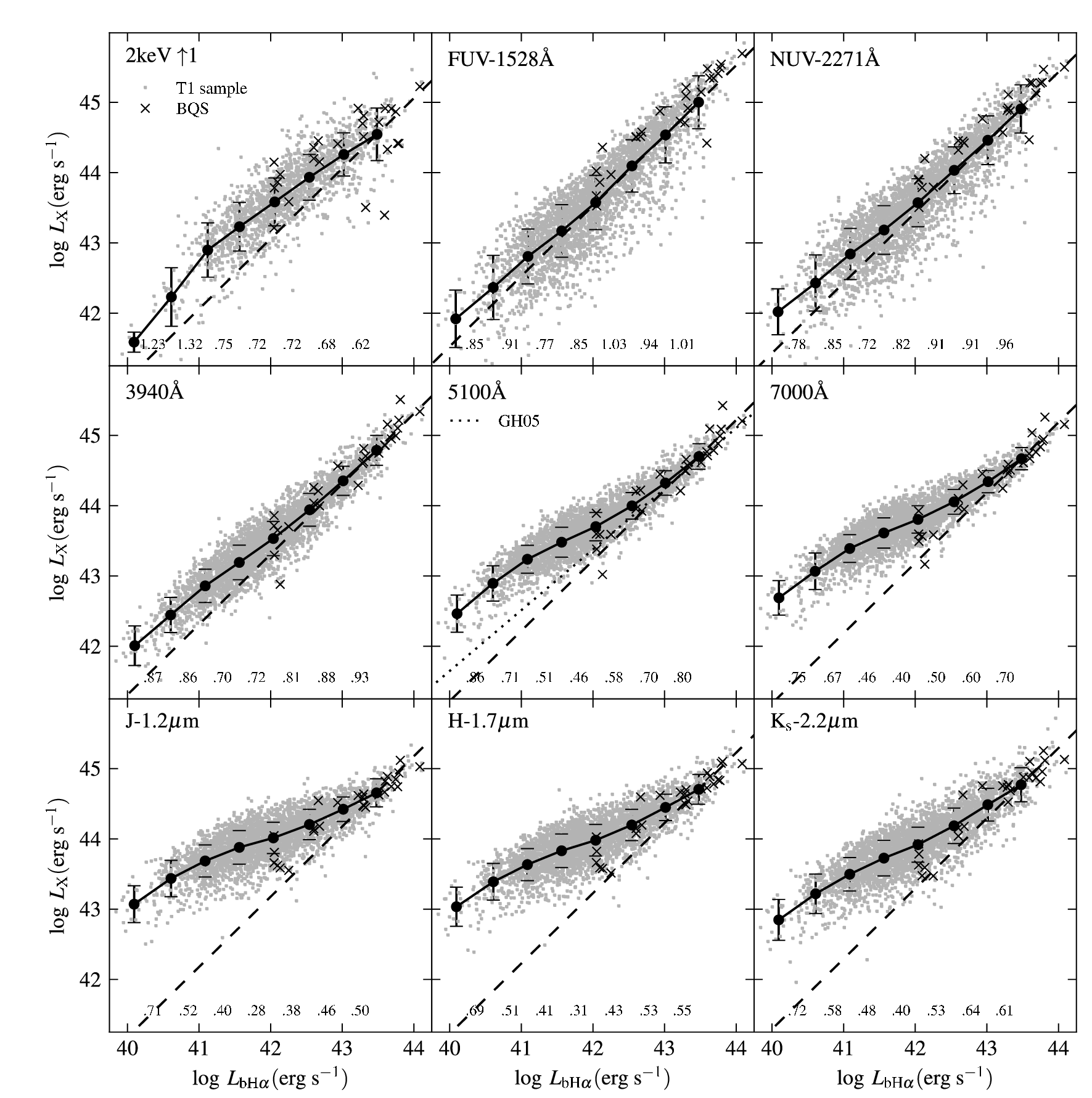}
\caption{The correlation of $\lbha$ vs. observed luminosity from the rest frame X-ray to the NIR
(the y-axis values in the X-ray panel are shifted upward by 1 decade for presentation purpose).
The black dots and the associated error bars mark the mean and the dispersion of $\nln$ in each 0.5 decade wide $\lbha$-bin. The dispersions in the UV and X-ray bands are larger than in the optical and NIR bands, probably reflecting the higher variability. The solid lines connect the averages of adjacent bins, with the local slopes written underneath. Dashed lines show linear relations, normalized by the highest $\lbha$-bin.
In the luminous bins in the FUV, NUV and 3940\AA\ panels, the slopes are consistent with a linear relation, in all other bands and luminosities the slopes are flatter (except the X-ray panel at the lowest $\lbha$ bins). There is a transition in all bands, excluding X-ray, from a steeper slope to a flatter slope with decreasing $\lbha$. The transition $\lbha$ increases with wavelength, up to a maximum of $\simeq 10^{43}\ \ergs$ at 1.2\mic. As shown below (Fig. 11), the transition occurs due to the host contribution, which peaks at 1.2\mic. 
The 5100\AA\ panel also presents the $\lbha - \lagn(5100\AA)$ relation from Greene \& Ho (2005), derived for a sample selected to minimize the host contribution. Objects that appear in the BQS sample are marked by a cross. Note that for $\lambda\ge 5100\AA$ the BQS objects are near the linear relation, consistent with their point like selection criterion which excludes significant host contribution. At shorter $\lambda$ the BQS are above the line, consistent with their blue optical colour selection. 
Note that the EW of \Ha, as implied by the 7000\AA\ panel, shows an apparent `inverse Baldwin effect' if the host light is not properly subtracted. } 
\label{fig: ridge line}
\end{figure*}
Fig. 10 shows there is a larger dispersion in the UV and X-ray luminosities for a given $\lbha$, compared to the dispersion in the optical and NIR bands. This likely reflects the larger variability in these bands. The relations of $\log\ \nlnl$ vs. $\log\ \lbha$ in the different bands clearly do not follow a simple linear (i.e. single power-law) relation. Instead of attempting to fit the distributions with a more complicated relation, say a double power-law, we just form a `ridge' line by connecting the averages of the adjacent $\lbha$ bins. The slope in each segment (noted in the bottom of each panel) is flatter at lower $\lbha$, excluding the X-ray panel (further discussed in \S 3.6). The change of slope in the FUV occurs at $\lbha \simeq 10^{41.5}\ \ergs$. The $\lbha$ where the slope change occurs increases at longer wavelengths, and reaches a maximum of $\simeq 10^{43}\ \ergs$ at 1.2\mic. It then decreases back to $\simeq 10^{42.5}\ \ergs$ at 2.2\mic. As we show in \S 3.3, this trend results from the relative host contribution at different bands, which peaks at 1.2\mic, decreases sharply in the FUV, and nearly disappears in the X-ray. 

The $\log \nlnl$(FUV) vs. $\log\ \lbha$ slope is close to unity in the four highest $\lbha$ bins, i.e. a linear relation between $\nlnl$(FUV) and $\lbha$.
Since the FUV occurs close to the position of the peak SED in AGN (e.g. Zheng et al. 1997), 
this linear relation suggests that $\lbha$ provides a good estimator of $\lbol$. 
The simple linear relation of $\nln$(FUV) and $\lbha$ suggests two independent linear relations. A linear relation of $\nln$(FUV) and the ionizing luminosity, as expected if the mean SED of AGN is luminosity independent. In addition, it requires a linear relation of $\lbha$ and the ionizing luminosity, as expected if the mean covering factor of the BLR is also independent of luminosity. 

In each panel of Figure 10 we also plot a linear relation, normalized to match the highest $\lbha$-bin. If the AGN SED is indeed luminosity independent, than the deviation from linear relation results from the host contribution, and thus one can use the deviation from the linear relation to estimate the host contribution as a function of wavelength. Note that at $\lambda\ge 5100\AA$ the ridge line slope does not approach unity even at the highest $\lbha$ -bin, which may suggest the sample does not quite reach the `pure' AGN emission regime (see \S 3.3).

Greene \& Ho (2005) selected a subset of low $z$ SDSS AGN where the host contribution to the optical emission is minimal, based on the weakness of the \caii\ K stellar absorption feature. Their measured $\nln$(5100\AA) vs. $\lbha$ relation is plotted in the 5100\AA\ panel of Fig. 10, and indeed passes close to the assumed linear relation, normalized at the highest $\lbha$-bin. This strongly supports the host contribution interpretation for the slope changes with decreasing luminosity seen in Fig. 10.

\subsubsection{The inverse Baldwin relation for \Ha}

We note in passing that the EW of \Ha, given to a good approximation by $\lbha/\nln(7000\AA)$, decreases with decreasing luminosity due to the increasing relative contribution of the host (Fig. 10, 7000\AA\ panel). This corresponds to the `inverse Baldwin effect' of the Balmer lines, noted by Croom et al. (2002) for \Hb\ and \Hg\ in low luminosity AGN found in the 2dF quasar survey. Unlike the suggestion that the inverse Baldwin effect of the Balmer lines is an intrinsic property of AGN (Croom et al. 2002), our analysis suggests the intrinsic \Ha\ EW is luminosity independent, i.e. \Ha\ shows no Baldwin effect, as was found for \Hb\ in more luminous AGN (Dietrich et al. 2002; cf. Netzer et al. 2004).

\subsubsection{The $\nln$ vs. $\lbha$ for the BQS sample}
There are \nInPG\ objects in the T1 sample that also appear in the BQS sample (Neugebauer et al. 1987). These are marked in Fig. 10. The BQS selection criteria are of point sources with a relatively blue optical colour, in contrast with the T1 selection criteria which are independent of morphology and colour. It is therefore interesting to see how the BQS selection criteria affect their $\lbha$ vs. $\nlnl$ correlations. In the $\lambda\ge 5100\AA$ bands the BQS objects have a lower $\nlnl$ at a given $\lbha$, and lie closer to the linear relation, in particular for the lowest $\lbha$ objects. This offset from the T1 sample is consistent with their selection of a weaker host contribution. At the $\lambda< 5100\AA$ bands the BQS objects are offset to higher $\nlnl$ values. The offset increases into the UV, consistent with the optical blue colour selection. In the X-ray band the BQS appear to be a random subset of the T1 sample. Note that the BQS sample objects do not fall outside the observed T1 objects distributions, but are offset towards weaker hosts and stronger UV emission, at a given $\lbha$.

\subsection{Average SED versus $\lbha$}\label{subsec: average SED} 
Above we have found indirect indications that 1. the mean SED of AGN and the mean covering factor of the BLR are independent of luminosity, 2. the host galaxy contribution flattens the $\nlnl$ vs. $\lbha$ relations with decreasing $\lbha$. We explore these points qualitatively in this section. A quantitative analysis is performed in \S 3.4 -- \S 3.5. Below we form the observed mean AGN SED for the eight $\lbha$ bins (Table 2), 
subtract a mean net AGN SED scaled linearly by $\lbha$, and compare the residual luminosity to the mean SED of a matched sample of inactive galaxies.

Figure 11, left panel, shows the mean SED in the 2.2\mic\ to 1528\AA\ region for the eight $\lbha$ bins. The integration apertures of 2MASS, SDSS, and GALEX are different, thus the fraction of the host light included is different in different bands. Furthermore, these differences change with $z$ and depend on the host type. Although the plotted SEDs likely differs from the intrinsic SEDs due to these aperture effects, the difference is irrelevant to our purpose of understanding the SEDs, as long as all objects used in the analysis presented below are similarly affected.

\begin{figure*}
\includegraphics{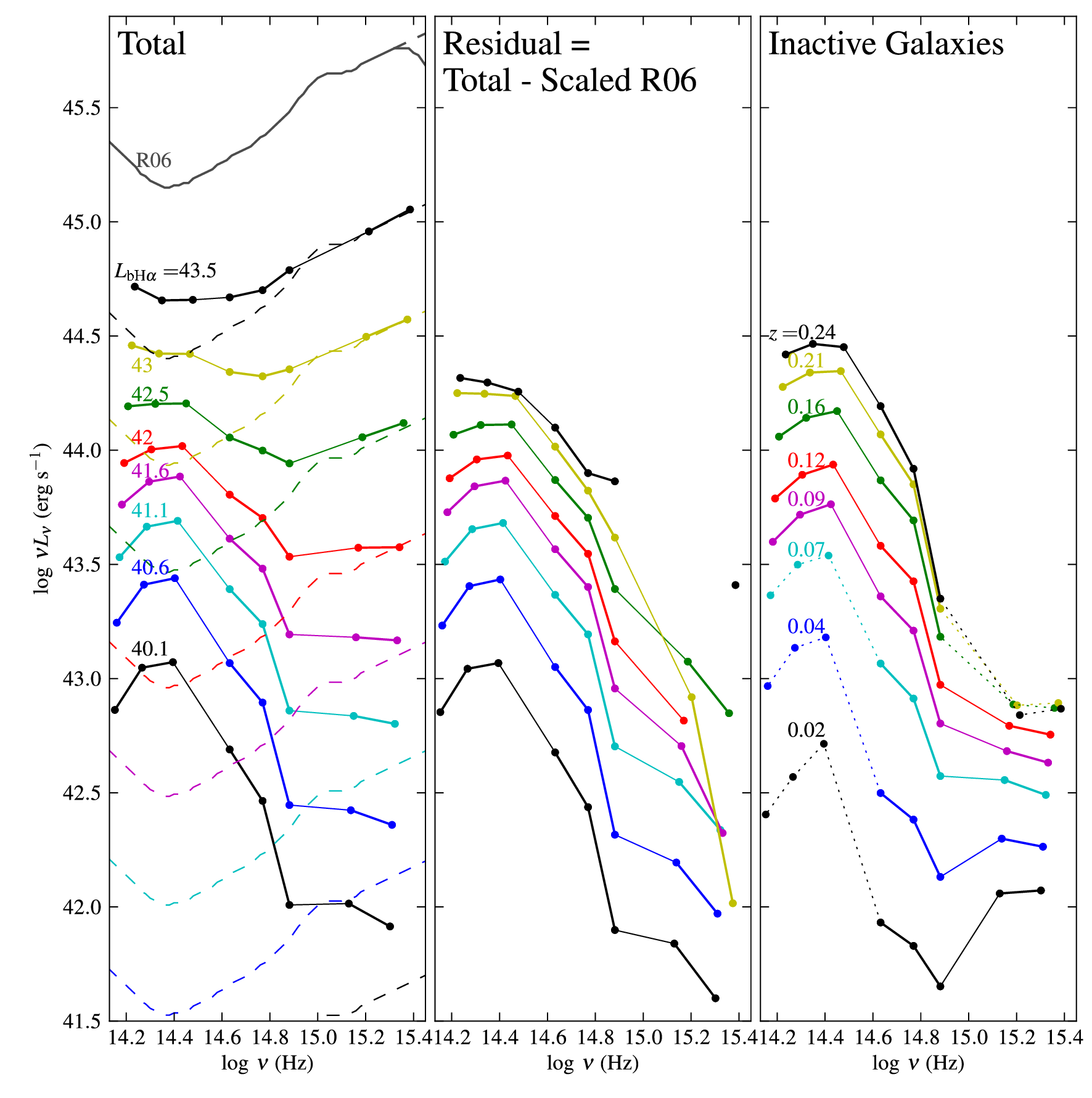}
\caption{
The luminosity dependence of the mean SEDs at 2.2\mic --1528\AA. {\bf (Left panel)} The mean T1 objects observed SEDs for the eight $\lbha$ bins (solid lines, see \S 3.3). Bands from different surveys (different apertures) are connected by a thinner line. Bands with $<70\%$ detection are connected by dotted lines. Note the transition from an AGN-dominated UV peaked SED to a galaxy-dominated NIR peaked SED with decreasing $\lbha$. The R06 SED of luminous quasars (top solid gray line), is assumed as the net AGN SED. The dashed lines show the expected net AGN contribution, derived by matching the R06 SED to the UV in the highest $\lbha$ bin, and rescaling by the relative $\lbha$ of the lower seven bins. The scaled R06 SEDs match the mean SEDs in the UV for the 2\nd --4\th\ $\lbha$ bins, as expected from their linear relation (Fig. 10). {\bf (Middle panel)} Residual SEDs derived by subtracting the scaled R06 SED from the total SED (colours match the left panel).
For clarity, errors are not shown, though note they are high in the luminous bins, as the residual is a small fraction of the total light. The residuals all appear to have a typical galaxy SED. {\bf (Right panel)} Average SEDs of SDSS inactive galaxies, with $z$-distributions matched to the $z$-distributions of the T1 objects in each $\lbha$ bin (mean $z$ noted near each curve). The increase of $\lhost$ with $\lbha$, seen in the middle panel, is just a selection effect, as inactive galaxies selected at matched $z$ show the same trend, inherent to a flux limited sample. 
Note the significant excess luminosity in the \Ks\ and H-bands in the most luminous residuals (middle panel), implying that a linear scaling of the net AGN SED is applicable only to the optical-UV region.}
\label{fig: SED}
\end{figure*}

We cannot use our highest $\lbha$ SED bin as a template net AGN SED, as the $\nlnl$ vs. $\lbha$ relation remains flatter than unity at $\lambda>5100\AA$ even in that bin, indicating possible significant host contribution in those bands. We need a template of more luminous AGN, where the host contribution is likely negligible. We therefore use Richards et al. (2006, hereafter R06), which provide a mean AGN SED constructed using SDSS quasars (with Spitzer detections). The R06 template SED is derived from the SDSS photometry, supplemented with photometry from various other surveys, including 2MASS and GALEX. The R06 SED is a factor of $\sim 5$ more luminous than our highest $\lbha$-bin SED (see Fig. 11), which ensures that host contribution should be lower. In addition, R06 removed the host contribution, which they find is generally $<0.1$ dec, even in the J-band. Since the FUV of R06 is mostly derived from SDSS photometry of high $z$ quasars, it shows a turnover at $\log \nu \gtrsim 15.32$, induced by the \La -forest, which is not present in the T1 sample SEDs. We therefore extrapolate the R06 SED up to $\log \nu = 15.4$ (the highest relevant rest frame $\nu$) with $L_{\nu} \propto \nu^{-0.5}$, as seen at lower $\nu$ (see also figure 19 of Trammell et al. 2007).

The R06 SED was scaled to match the NUV data points of the highest $\lbha$ bin ($\lbha = 10^{43.5}\ \ergs$, left panel of Fig. 11). For each of the seven lower luminosity bins, we further rescaled the R06 SED by the mean $\lbha$ in each bin, relative to $\lbha$ in the highest bin. The scale applied to each bin is listed in Table 2. Note that the $\lbha$-rescaled R06 SEDs match well the FUV data points of the mean SEDs also in the second to fourth $\lbha$ bins, as expected from their linear relation with $\lbha$ (FUV panel in Fig. 10).

We now subtracted the scaled R06 SED, i.e. the matched `pure' AGN SED, from the mean SED in each bin, and the residuals are plotted in the middle panel of Fig. 11. This residual includes both the host emission, and any change in the mean AGN SED with decreasing $\lbha$. The errors on the residual are high in the luminous bins and in the UV, since the residual is a small fraction of the total light. For clarity sake, error bars are not shown. The possible systematic errors in this method are discussed in \S3.4. The residuals have similar overall SEDs, with some small differences at the NIR. They all appear to be similar to a typical galaxy SED, which peaks at $\log \nu = 14.4$. The residuals imply that the mean host galaxy luminosity increases with the mean $\lbha$ luminosity, which suggests a tight relation between the AGN luminosity and the host luminosity. However, as we show below, this apparently tight relation is just a selection effect.
 
To compare the residuals to the mean SED of typical galaxies, we formed a matched-$z$ sample of 298 SDSS galaxies classified by B04 as inactive (excluding `type 2 AGN' or `composites', see \S 3.1.4). The galaxies were drawn from the GALEX Deep Imaging Survey region, to maximize UV detections, and thus minimize bias in terms of their UV emission. Their SEDs were formed in a manner identical to the T1 sample. For each $\lbha$-bin of the T1 sample we formed a $z$-matched sample of inactive galaxies in the following manner. First, the inactive galaxies are divided into eight $z$ bins, evenly spaced in $z^{1/2}$, and the mean-SED of each $z$-bin is computed, forming `$z$-specific' SEDs. Then, the $z$-distribution of each AGN $\lbha$-bin is measured. Finally, the `$z$-specific' SEDs are convolved with the different $z$-distribution functions. 
Note that only a fraction of the high luminosity (and high $z$) AGN are resolved (Table 2), while all inactive galaxies are resolved by definition (\S 2.1). Therefore, the high $z$ AGN are subject to somewhat different selection effects from inactive galaxies with the same $z$.  

The mean-SEDs of the $z$-matched inactive galaxies, are shown in the right panel of Figure 11. The convolved detection fraction is $<70\%$ at high $z$ in the FUV, and at low $z$ in the NIR, so the SEDs may be somewhat biased in these regions. Except in the lowest $z$ bins, the residuals largely resemble the inactive galaxies, both in shape and in luminosity. The trend of increasing host luminosity with the AGN luminosity, seen in the middle panel, just reflects the fact that higher luminosity AGN reside at higher $z$. Since most of the T1 objects are selected from the SDSS sample of extended objects (galaxies), and the SDSS sample is flux limited, the higher $z$ AGN must reside in higher luminosity host galaxies. Now, for the unresolved AGN, which dominate at the highest $\lbha$ (see Table 2), the hosts are not forced to be more luminous to be detectable at higher $z$, and indeed the inactive galaxies are somewhat more luminous than the AGN hosts at the highest $\lbha$ bins. Unresolved AGN flag the position of their host galaxies, galaxies which would have not necessarily passed the flux limit of the SDSS survey.

The subtraction of the scaled R06 AGN SED leads to a residual SED which is generally consistent with the mean SED of inactive galaxies in the middle and high $\lbha$ bins, indicating the linear scaling of the R06 SED with $\lbha$ is roughly correct. At the lowest $\lbha$ bin the inactive galaxies show excess UV emission. This is unlikely to be an AGN SED effect, as the host dominates the emission at all bands (Fig.11, left panel), and likely reflects the difference in the mean AGN host galaxy (NEG) and the mean SDSS inactive galaxy at low $z$ (SFG, see Fig. 9).
There is also significant excess luminosity in the \Ks\ and H-bands in the most luminous residuals. This could be due to a lower dust 
covering factor with increasing luminosity, as observed by Maiolino et al. (2007) and Treister et al. (2008), and thus weaker dust emission in the R06 template compared to the T1 AGN. We do not attempt to quantify this relation or pursue it further, yet we note that the simple scaling law of a fixed AGN SED does not apply in the near IR. Note that the X-ray emission also does not scale linearly with $\lbha$ either (Fig. 10 and \S 3.6). Therefore, we emphasize the applicability of this simple scaling law is limited only to AGN optical-UV emission, the region most likely dominated by the accretion disc emission.

\subsection{The AGN SED scaling with $\lbha$}

We express the net AGN $\nln$ ($\equiv \lagn$) by scaling the R06 SED by $\lbha$: 
\begin{equation}
\lagn(\lambda) \equiv (L_{\rm{R06}}(\lambda) \pm 10\%) \times  (\frac{\lbha}{10^{\pureAGNbHa}\ \ergs})^{\beta}
\end{equation}
The normalization of $\lbha$, $10^{44.2}\ \ergs$, corresponds to $\lbha=10^{43.5}\ \ergs$ of the highest luminosity bin, times the difference of $10^{0.7}$ in $\nln$(NUV) between this bin and the R06 SED, as described in the previous section. 
The scaling index is denoted by $\beta$, where a $\beta$ independent of $\lambda$ indicates that the mean SED shape remains fixed with $\lbha$. 
We use a 10\% systematic error on the scaled R06 luminosity to get an estimate of the implied errors in the residual fluxes by systematic errors of a given level.

\subsubsection{Constraining $\beta$}

In \S3.3, we found indications that $\beta \approx 1$. The left panels in Figure 12 provide additional support for this linear scaling law. It compares the implied host luminosity ($\lhost \equiv {\rm mean}\ \nln - {\rm mean}\ \lagn$) 
at different $\lbha$ bins, for different $\beta$, to the mean luminosities of SDSS type 2 AGN with the same $z$ distribution. The luminosity is presented in two optical bands: the SDSS $z$ band (8932\AA) and the SDSS $u$ band (3551\AA). The $z$ band is selected as it follows the stellar mass. The SDSS $u$ band is a measure of the hotter, more massive stars, and thus a measure of the more recent star formation rate (SFR). Going further to the UV bands probes hotter and more massive stars, and provides a measure of the yet more recent SFR. However, given the little residual emission left in the UV, the uncertainty can be large. 
For all objects, we use the $u$ band and the $z$ band photometric luminosities in the 3\arcsec\ SDSS fibre aperture, fixed by 0.35 mag, the mean offset for a PSF (Adelman-McCarthy et al. 2008, see also \S 2.7). Magnitudes are converted to physical units using the SDSS-AB offsets in Abazajian et al. (2004). 
We use only the \nResolved\ objects from the T1 sample drawn from the SDSS sample of galaxies, which have the same selection criteria for spectroscopy as the type 2 AGN (\S 2.1). We use only type 2 AGN (classified by B04, see \S 3.1.4) that pass the S/N and bad pixel criteria applied to the T1 sample (\S 2.1). 

Fig. 12 presents $\lhost$ derived for $\beta=0.9, 1.0$ and $1.2$. The three left panels present the mean $\lhost(u\ \rm{band})$, $\lhost(z\ \rm{band})$ and $u-z({\rm host})$ colour for the T1 and for $z$ matched type 2 AGN (the mean $z$ for each $\lbha$ bin is marked at the top of the figure). We also plot the relations for inactive galaxies here selected from the complete SDSS survey (not limited to the deep survey field of GALEX).
The error bars on the $\beta=1$ line depict the effect of a 10\% systematic error in the scaling of the R06 SED (eq. 4), added to the error in the mean. The possible systematic error has a significant effect at the highest $\lbha$ bins, where $\lhost$ becomes a small fraction of the total luminosity, in particular in the $u$ band. At the highest $\log\ \lbha = 43$ bin, the scaled R06 becomes higher than the mean $u$ band luminosity, leading to an implied negative $\lhost$, and only the upper limit from the systematic error is shown. At lower $\lbha$ bins the scaled R06 SED becomes much weaker than the observed mean SED (left column of Fig. 11), and the host SED nearly equals the observed mean SED.

\begin{figure*}
\includegraphics{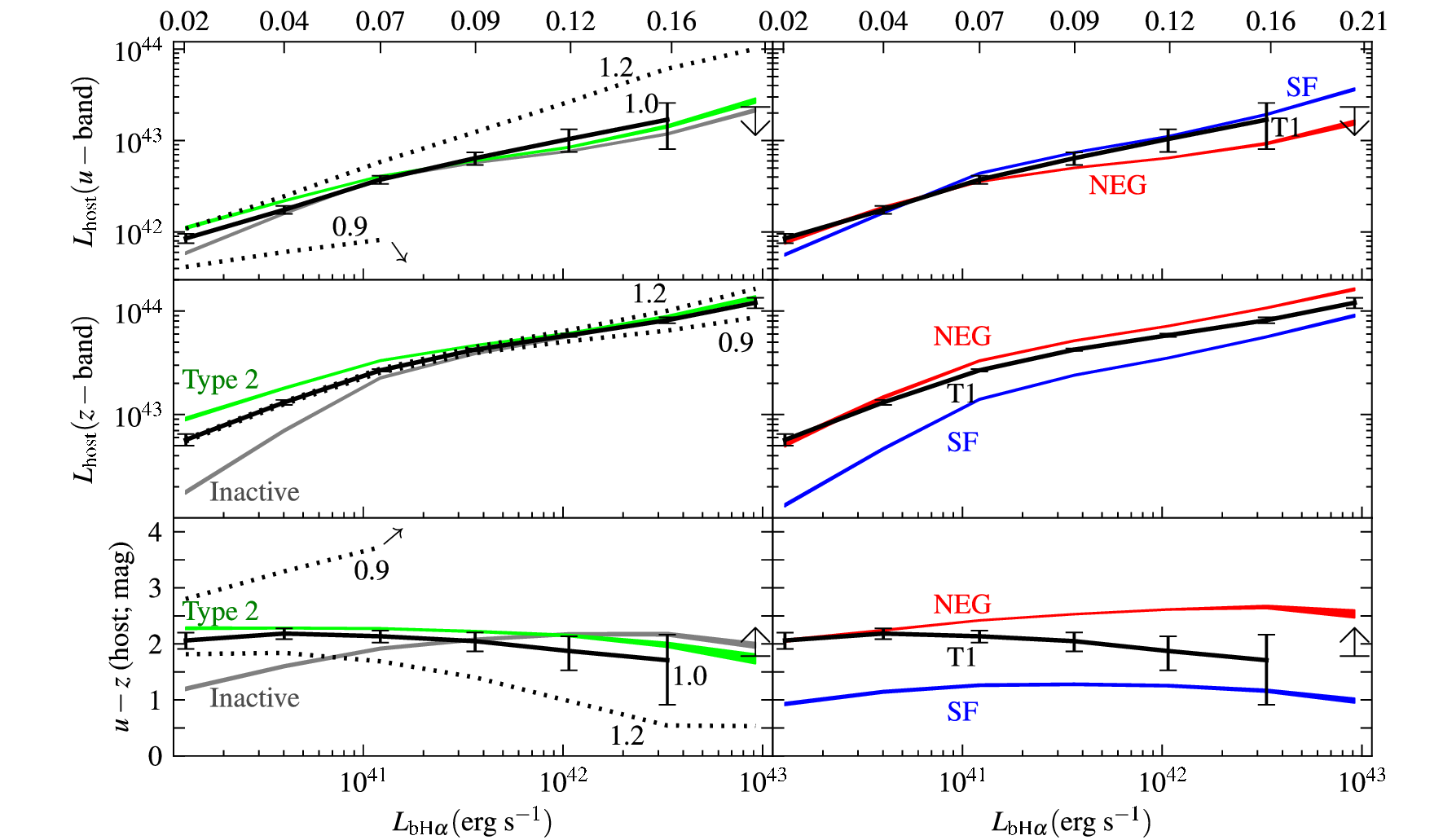}
\caption{
{\bf (Left panels)} A comparison of the implied host luminosity (total - scaled R06), for different scaling laws, with $z$-matched type 2 AGN, in the $u$ and $z$ bands. The residuals vs. $\lbha$ are shown as solid/dotted lines for linear/non-linear scaling laws. The scaling index, $\beta$ (eq. 4), is noted near each line. Error bars are the sum of a 10\% systematic error (eq. 4) and the error in the mean.
The mean $z$ is noted on top. Also shown are $z$-matched inactive galaxies. In the $u$ band, $\beta < 0.9$ is ruled out, as the implied residual is negative for $\lbha > 10^{41}\ \ergs$. It is remarkable that the simplest scaling, $\beta=1$ in both bands, leads to the closest match of the mean type 1 host to the mean type 2 host. Together with the linear $\nln$(UV)-$\lbha$ relationship at high luminosities (Fig. 10), this implies the mean optical-UV net AGN SED shape is fixed and scales linearly with $\lbha$.  The small offset of T1 and type 2 at low $\lbha$ cannot be explained by $\beta \neq 1$ as the AGN emission is a small fraction of the total light. 
{\bf (Right panels)} A comparison of the residual implied from the $\beta=1$ scaling law, with inactive galaxies (left panel) now broken to NEG and SFG. At low $\lbha$, hosts of T1 AGN resemble NEG, both in mass ($z$ band) and in specific SFR (i.e. colour), 
consistent with the tendency of low luminosity AGN to reside in elliptical hosts. With increasing $\lbha$,
the T1 hosts become less luminous in the $z$ band and bluer than the average NEG. The implied SFR is noted in Table 3 and \S 3.5.
}
\label{fig: Residual comparison}
\end{figure*}

As Fig. 12 clearly shows, a $\beta=1$ leads to a good match of the T1 host galaxies to the host galaxies of type 2 AGN, as expected based on AGN unification. Values of $\beta<0.9$ and $\beta>1.2$ in the $u$ band lead to highly discrepant hosts. The $z$ band is more host dominated, and less sensitive to the scaled AGN subtraction, so it does not provide a strong constraint on $\beta$. 
It is remarkable that the simplest scaling, $\beta=1$, i.e. that the net AGN SED maintains a fixed shape, which scales linearly with $\lbha$, leads to the closest match of the implied type 1 host to the observed type 2 host. The small differences in overall luminosity between type 1 hosts and type 2 AGN at low $\lbha$ (low $z$) cannot be due to a different $\beta$, as the AGN light is too weak. They could either be a result of the different selections used to create the type 1 and type 2 samples, or some modification of the unified model at low luminosities.

\subsubsection{The mean net AGN SED}

Combining the $\beta=1$ deduced from the $u$ and $z$ bands (Fig. 12) with the linear relation of $\nln$(FUV) and $\lbha$ (Fig. 10), we get that three bands in the 1500\AA\ -- 9000\AA\ range scale linearly with $\lbha$, implying that the optical-UV net AGN SED shape is independent of luminosity. This result is consistent with the finding of Bentz et al. (2009) for a sample of 35 reverberation mapped AGN, where they found that the dependence of the BLR radius on AGN optical luminosities, corrected for the host emission, is the same as the dependence on UV luminosities (Kaspi et al. 2005), which implies that the net AGN optical luminosity scales linearly with their UV luminosity. 

Below we list the ratios of AGN continuum emission in different rest-frame wavelengths to $\lbha$, as implied by the linear relation and by the R06 SED:
\begin{equation}
\lagn \left( \begin{array}{c}
  1450\AA  \\   2500\AA  \\   3940\AA  \\  5100\AA  \\  6166\AA \\ 7000\AA  \\  1.2\mic  \\
\end{array} \right) ~~=~~ \left( \begin{array}{c}
33^{\times \luvscatter}_{\div \luvscatter} \\ 26 \\ 17 \\ 14 \\ 12 \\ 11 \\ 8.5

\end{array} \right) \times \lbha
\label{eq: lagn specific bands}
\end{equation}
The implied mean EW of the broad \Ha\ is 570\AA. The dispersion is derived from the scatter around the linear relation between $\log \nln$(FUV) and $\log \lbha$, in the four luminous bins (Fig. 10). We do not attempt to measure the dispersion in the $\lbha$ vs. $\lagn(\lambda)$ relations at optical $\lambda$, due to the unknown dispersion in the host contribution. However, the dispersion at optical $\lambda$ is smaller than in the UV (Fig. 10), despite the additional dispersion due to the host contribution. Therefore, for $\lambda>$ FUV, we treat the noted scatter of 2.4 as an upper limit on the scatter of $\lagn(\lambda)$ vs. $\lbha$.

Using the conversion of R06 from $\nlnl$ at various bands to the bolometric luminosity (see their figure 12), $\lbol$, we get 
\begin{equation}
\lbol = \lbolcoeff^{\times \luvscatter}_{\div \luvscatter} \times \lbha
\label{eq: lbol}
\end{equation}

\subsection{The AGN host}

\subsubsection{Host mass}

As discussed above (\S 3.4, and Fig. 11), the apparent correlation of $\lhost$ and $\lbha$, is most likely just a selection effect, as the mean luminosity of inactive galaxies, of similar $z$ distribution, rises similarly with $z$. This similarity strongly suggests that the apparent correlation between $\lbha$ and the AGN $\lhost$ is induced by the correlation of both quantities with $z$. To test if the $\lhost$ versus $\lbha$ correlation is significant, we use the T1 sample presented in Fig. 12 (selected from the galaxy catalogue), to perform a partial correlation analysis. The T1 $\lhost(z~{\rm band})$ is indeed strongly correlated with $\lbha$, with a Pearson correlation coefficient $P_r(\lhost,\lbha)=0.66$. However, both quantities are indeed strongly correlated with $z$, with $P_r(\lhost,z)=0.82$, and $P_r(\lbha,z)=0.74$\footnote{The values of $P_r$ are calculated using $\log \lbha, \log\ \lhost$ and $\log\ z$.}, 
and the partial correlation of $\lhost(z~{\rm band})$ versus $\lbha$, at a fixed $z$, is only $P_r(\lhost,\lbha)|_z=0.14$. Furthermore, a fit of the residuals in the $\lhost(z~{\rm band})$ - $z$ relation versus the residuals in the $\lbha$  - $z$ relation gives $\lhost(z~{\rm band}) \propto \lbha^{0.07}$, which again demonstrates there is no significant dependence of $\lhost(z~{\rm band})$ on $\lbha$.

Note that $\lhost$ and $L_{\rm AGN}$ are not entirely unrelated, as one expects a lower limit to $\lhost$ at a given $\lagn$, in order not to exceed $L_{\rm Edd}$. But the T1 subsample of extended objects excludes AGN close to $L_{\rm Edd}$ (\S3.1.2, Fig. 8). 

As discussed above (\S 3.1.4), the T1 $\lhost$ distribution follows the NEG luminosity distribution (Fig. 9). We therefore break the group of inactive galaxies (Fig. 12, left panels) to its two subgroups of NEG and SFG (Fig. 12, right panels). At low $\lbha$, $\lhost$($z$ band) follow well NEG. This similarity is consistent with the tendency of low luminosity AGN to reside in early type galaxies (Ho et al. 1997b). With increasing $\lhost$, the T1 hosts become less luminous in the $z$ band than NEG, and at the highest luminosities the T1 hosts are intermediate between NEG and SFG.

\subsubsection{Host color}

The $u-z$ colour of the T1 host follows well NEG at the lowest $\lbha$. With increasing $\lbha$ the T1 hosts become bluer, and are intermediate between SFG and NEG at the highest $\lbha$. 
Kauffmann et al. (2003a) find that low luminosity type 2 AGN have NEG colours, while high luminosity type 2 AGN tend to have colours intermediate between NEG and SF galaxies. The results above indicate that type 1 AGN have the same hosts as type 2 AGN. 
Note that our selection of type 1 AGN is independent of the distinction between SFG and NEG, which is based on narrow lines. In contrast, a type 2 AGN sample, selected based on narrow line ratios, excludes a SF-dominated host galaxy. Although, as shown by Kauffmann et al. (2003a, fig. 6 there), only the lowest luminosity AGN can be dominated by the host SF line emission.

In Table 3, we derive the SFR and specific SFR for each $\lbha$ bin, using the rest-frame $u$ band luminosity and $u-r$ colour of the mean hosts. We calculate the rest frame $u$ band and $r$ band luminosities using a power law interpolation between adjacent observed bands.
The third column lists $b_{300}(u-r)$ (Blanton \& Roweis 2007), the ratio of stars formed in the last 300 million years to the total star formation history of the galaxy, based on SED modeling of SDSS galaxies using the stellar models of Bruzual \& Charlot (2003). 
The fourth column lists the current SFR, based on the correlation of the $u$ band fibre luminosity with $L_{\Ha}$ in SDSS SFG (Moustakas et al. 2006). These estimates are subject to a host of systematic effects, such as reddening, the time dependence of the star formation, and obviously the estimated host luminosity.

\begin{table}
\begin{tabular}{c|c|c|c}
$\log\ \lbha$ & $z$ & $\log b_{300}$ & SFR ($\msun$ yr$^{-1}$) \\
\hline
40.1 & 0.02 & -2.89$^{+0.16}_{-0.17}$ & 0.11$^{+0.01}_{-0.01}$ \\
40.6 & 0.04 & -2.92$^{+0.13}_{-0.14}$ & 0.24$^{+0.02}_{-0.02}$ \\
41.1 & 0.07 & -2.73$^{+0.12}_{-0.13}$ & 0.60$^{+0.05}_{-0.05}$ \\
41.6 & 0.09 & -2.56$^{+0.16}_{-0.18}$ & 1.1$^{+0.1}_{-0.1}$ \\
42.0 & 0.12 & -2.38$^{+0.25}_{-0.32}$ & 1.9$^{+0.4}_{-0.4}$ \\
42.5 & 0.15 & -2.28$^{+0.40}_{-0.65}$ & 3.5$^{+1.1}_{-1.3}$ \\
\end{tabular}
\caption{The dependence of mean aperture SFR (within the SDSS fibre) on AGN luminosity. Col. 3 lists the mean ratio of stars formed in the last $3 \times 10^8$ years compared to the total star formation of the galaxy, derived from rest-frame $u-r$ colour (Blanton \& Roweis 2007). Col. 4 lists the SFR derived from $u$ band luminosity (Moustakas et al. 2006). Both measures are based on the mean T1 AGN, after subtracting the scaled R06 SED (Fig. 11). Errors are the sum of a 10\% systematic error (eq. 4) and the error in the mean.}
\label{table: SFR}
\end{table}

\subsubsection{The mean host/AGN luminosity ratio}

The results described above provide a practical measure of the average host/AGN luminosity ratio in various bands, as a function of the AGN luminosity for SDSS, GALEX, and 2MASS. Figure 13 plots the host luminosity (i.e. residual luminosity) to the net AGN luminosity (i.e. the scaled R06 SED emission), as a function of $\lbol$ (eq. 6) for different rest-frame bands. In the J band the host dominates, on average, up to $\log \lbol=45.5$, well into the quasar-scale emission, including the highest $\lbha$-bin in our analysis. At 5100\AA\ the host dominates at $\log \lbol\le 44.7$, consistent with the relation obtained by Shen et al. (2011) for the SDSS quasar sample.  As expected, in the FUV the AGN becomes more dominant already at $\log \lbol > 42.5$.  The plotted ratios are strongly dependent on the apertures used in each band.
Also, the dispersion of $\lhost/\lagn$ in each $\lbha$ bin is significant, so individual objects may differ significantly. The plotted ratio only serves as a practical guide for the AGN catalogues of SDSS, 2MASS and GALEX samples used here. 

\begin{figure}
\includegraphics{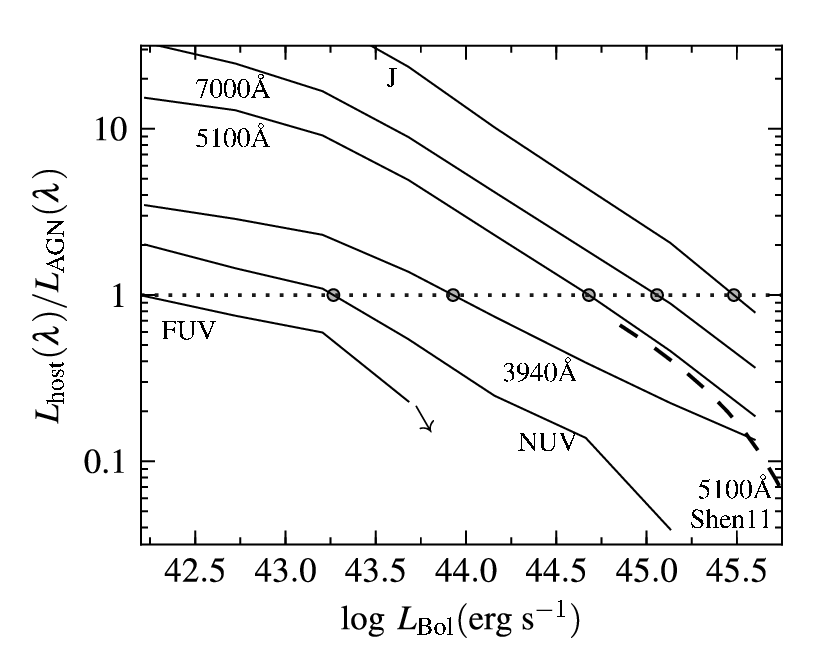}
\caption{The mean host to AGN luminosity ratio as a function of AGN bolometric luminosity at different rest-frame bands, based on the decomposition in Fig. 11. The small gray circles mark the transition luminosity from host dominance to AGN dominance. In the J band the host dominates up to $\lbol = 10^{45.5}\ \ergs$, well into the quasar regime, while in the FUV the AGN dominates already at $\lbol > 10^{42.5}\ \ergs$. The ratio at $\lambda=5100\AA$ is consistent with the Shen et al. (2011) result (dashed line) which is based on the QCV sample. Note that the results are relevant for the specific apertures used by GALEX, SDSS and 2MASS.}
\label{fig: hostAGNratio}
\end{figure}

\subsection{Unobscured type 2 AGN based on the $L_{\rm X}$ versus $\lbha$ relation}\label{subsec: x-ray vs lbha}

Unobscured type 2 AGN, or the so-called `true type 2' AGN, are AGN in which the BLR is not observed because it does not exist, as opposed to standard type 2 AGN in which the BLR is obscured. Candidates are often found by looking for low luminosity AGN, detected through a nuclear X-ray point source, that do not show a sign of a BLR in their optical continuum, but their X-ray spectrum does not indicate obscuration (Tran et al. 2011, Shi et al. 2010, Rigby et al. 2006, Trump et al. 2009).
Below we describe how the $L_{\rm X } (\equiv \nln(2\ \kev))$ versus $\lbha$ relation can be used to test whether the lack of detection of broad lines is indeed significant, and justifies the true type 2 identification.

Our X-ray band detection rate for the complete T1 sample is $\xraydetectionprcnt\%$. The detection rate depends on $\lbha$, and becomes as low as \xraylowbindetectionprcnt\% in the lowest $\lbha$ bins (Table 2). Therefore, in the following analysis of the $L_{\rm{X}}$ versus $\lbha$ relation we utilize only objects with $\fbha > 10^{-13.5}\ \flux$, which have an X-ray detection rate of \xraydetectionprcnthighf\% (see Fig. 3). Figure 14 presents the $L_{\rm X }$ versus $\lbha$ relation. The $L_{\rm X}$ of inactive galaxies is generally 
$<10^{42}\ \ergs$ (Bauer et al. 2004), so host contribution should not significantly affect our analysis.

A least squares minimization yields 
\begin{equation}
\label{eq: X-ray}
\log L_{\rm{X},42} = (\xtolbhaIndex \pm \xtolbhaIndexError)\log L_{\rm{b\Ha}, 42}\ + (\xtolbhaOffset \pm \xtolbhaOffsetError)  \\
\end{equation}
\begin{equation} 
\label{eq: X-ray2}
\alpha_{\rm{ox}} = (-0.09 \pm 0.01) \times \log l_{2500\AA, 29} - 1.42
\end{equation}
where $\aox \equiv 0.38 \times \log ( l_{2\ \kev} / l_{2500\AA} )$, $\lbha$ and $L_{\rm X}$ are in units of $10^{42}\ \ergs$ and $l_{2500\AA}$ in $10^{29}\ \ergs\ \Hz^{-1}$. 
Equation 8 is similar to eq. 2 in Steffen et al. (2006), where $\aox \propto \log l_{2500\AA}^{(-0.137 \pm 0.008)}$. The fit relation (eq. 7) is depicted in Fig. 14. For each $\lbha$-bin, a triangle marks the mean upper limit on $L_{\rm X}$ of the T1 objects without ROSAT detections. 

\begin{figure}
\includegraphics{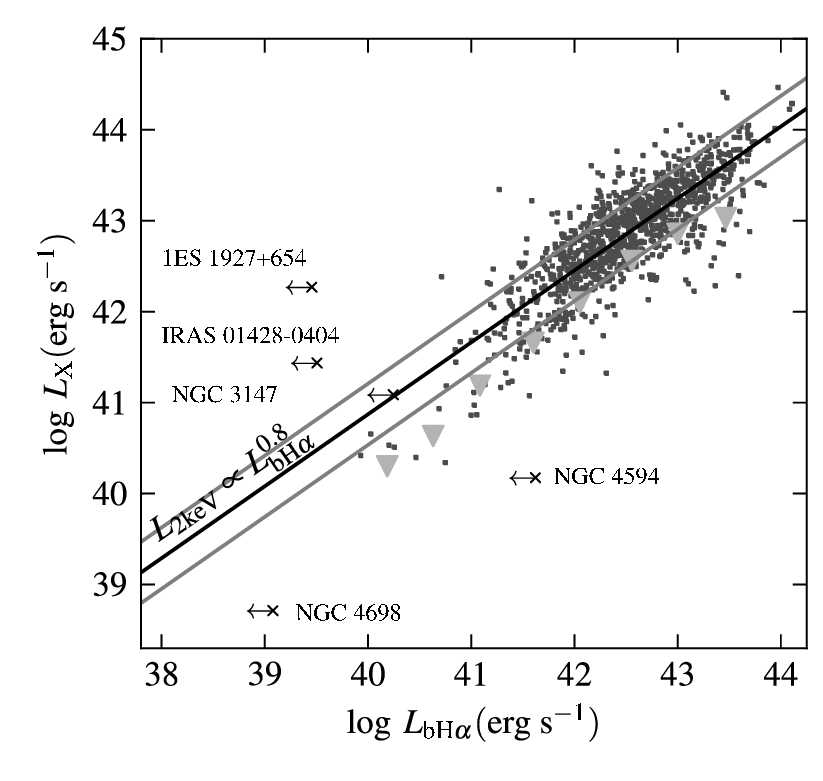}
\caption{The $L_{\rm X}$ vs. $\lbha$ relation as a probe for true type 2 AGN. The plot includes only T1 objects with $\fbha>10^{-13.5}\ \flux$, where $>50$\% have ROSAT detections (Fig. 3). The non detections are marked by triangles, which show the mean upper limits, averaged in half decade wide $\lbha$ bins. The best power law fit for the X-ray detections is shown as a black line, with gray lines indicating the dispersion. The left pointing arrows mark five true type 2 candidates from Tran et al. (2011) and Shi et al. (2010). These objects have an apparently unobscured X-ray emission, but no broad \Ha\ detection. 
Only 1ES 1927+654 and IRAS 01428-0404 have upper limits on $\lbha$ which are significantly lower than expected based on their $L_{\rm X}$, and are therefore potentially `true type 2' AGN. In NGC 4594 and NGC 4698 the expected $\lbha$, based on their $L_{\rm X}$, is a factor of $\sim 100$ below the upper limit on $\lbha$, and in NGC 3147 50\% of the objects are expected to be below the $\lbha$ upper limit.
The upper limit on $\lbha$ depends on the expected $\dva$, which can be derived from the $\mbh$ and $\lledd$ estimates for each object.
} 
\label{fig: X ridge line}
\end{figure}

Eq. 7 provides a useful relation for the search for the putative unobscured type 2 AGN. 
Below we compare the upper limit on $\lbha$ for a number of potentially `true type 2' AGN presented by Tran et al. (2011) and Shi et al. (2010) with their predicted $\lbha$ based on their observed $L_{\rm X}$.

For each object we set an upper limit on the broad \Ha\ flux as 10\% of the continuum flux density near \Ha, multiplied by the expected $\dva$. The expected $\dva$ is estimated from $\mbh$ and $\lledd$, by applying eqs. 2 and 3. We use the three objects in Tran et al. (2011) where estimates of $\mbh$ and $\lledd$ are provided, and also two objects from Shi et al. (2010), NGC 4594 and IRAS 01428-0404, where optical spectra and $\mbh$ estimates are provided. These five objects have a measured $L_{\rm X}$, and we use eq. 7 to derive the expected $\lbha$ based on their observed $L_{\rm X}$, and compare it to the upper limit on $\lbha$ derived above. The five objects are plotted in Fig. 14 (left-pointing arrows). Three of the potentially true type 2 candidates, NGC~4594, NGC~4698, and NGC~3147 have upper limits on $\lbha$ that can accommodate the expected values of $\lbha$ based on the extension of eq. 7 to lower luminosities. Thus, the absence of a broad \Ha\ in these objects is not significant. A high angular resolution spectrum is required to exclude the strong host contribution near 7000\AA, and determine if the expected weak \Ha\ is indeed missing. In the other two objects, IRAS 01428$-$0404 and 1ES 1927+654, the expected $\lbha$ is well above the upper limits.
The deficiency in $\lbha$ is significant at the levels of $4\sigma$ and $7\sigma$, respectively\footnote{Our calculation of the dispersion assumes $\lbha$ is the independent variable, since the T1 sample is selected by the broad \Ha. When estimating $\lbha$ from $L_{\rm X}$ one should use an X-ray selected sample and treat $L_{\rm X}$ as the independent variable.}, and these two objects appear to be true type 2 AGN.

An estimate of the expected $\lbha$ can also be derived directly from $\lbol$, using eq. 6. However, the estimated $\lbol$ in Tran et al. and Shi et al. is based on $L_\oiii$, and its relation with AGN luminosity has significant dispersion (Baskin \& Laor 2005, Caccianiga \& Severgnini 2011). This large dispersion is less inhibiting in the estimation of $\dva$, used to derive the upper limit on $\lbha$, since $\dva \propto \lbol^{\sim -1/4}$ when $m$ is known (eq. 2). An analysis of the relation between $L_{\oiii}$ and $\lbha$ in the T1 sample, and its implication on the identification of true type 2 AGN, will be performed in a following paper.

\subsection{The SED dispersion}\label{subsec: sed by lbha and slope} 

Above we have shown that the mean SED of broad line AGN is well reproduced by the sum of the mean SED of luminous quasars, scaled down by $\lbha$, plus the mean SED of inactive galaxies (NEG-like at low $\lbha$, and bluer with increasing $\lbha$). However, what is the dispersion in the shapes of individual SEDs? Below we provide some measures of the distribution of the SEDs, and discuss its possible origin.

\subsubsection{The distribution of $\nln(\rm FUV)/\lbha$ and $\an$}

Figure 15, left panel, presents the distribution of $\nln$(FUV)/$\lbha$ in four decade-wide bins of $\lbha$ (coloured solid lines, each bin in Fig. 15 is composed of two consecutive $\lbha$ bins from Table 2). 
The distributions for $\lbha \ge 10^{41.4}\ \ergs$ show a narrow peak at $\nln$(FUV)/$\lbha=45-75$, compared to the mean value of 33 (eq. 5). There is an extended tail towards lower values, which is similar in all three bins. The amplitude of the narrow peak drops with decreasing mean $\lbha$. In the lowest bin, $\lbha=10^{40.5}\ \ergs$, the peak becomes broader and is shifted to higher $\nln$(FUV)/$\lbha$ values. As found above (Fig. 10 FUV panel), this bin is affected by significant host contribution to the FUV band, which produces the offset to higher $\nln$(FUV)/$\lbha$ values, and a larger dispersion. The higher $\lbha$ bins are not affected by the host contributions, and the observed distribution of $\nln$(FUV)/$\lbha$ values is an intrinsic AGN property. The distribution of these three bins is similar, despite being selected differently by the SDSS, where the $\log \lbha=43$ objects are mainly point sources, selected by their non-stellar colours, while lower $\lbha$ objects are selected to have extended morphologies (\S 2.1 and Table 2). Since \Ha\ is generally powered by recombination, it provides a measure of the ionizing photons intercepted by the BLR. Thus, the narrow peak suggests there is a small dispersion in $L_{\rm ionizing}/\nln$(FUV), and a small dispersion in the covering factor of the BLR. The tail towards lower $\nln$(FUV)/$\lbha$ values may be due to reddening, as further discussed below based on additional indicators.

\begin{figure}
\includegraphics{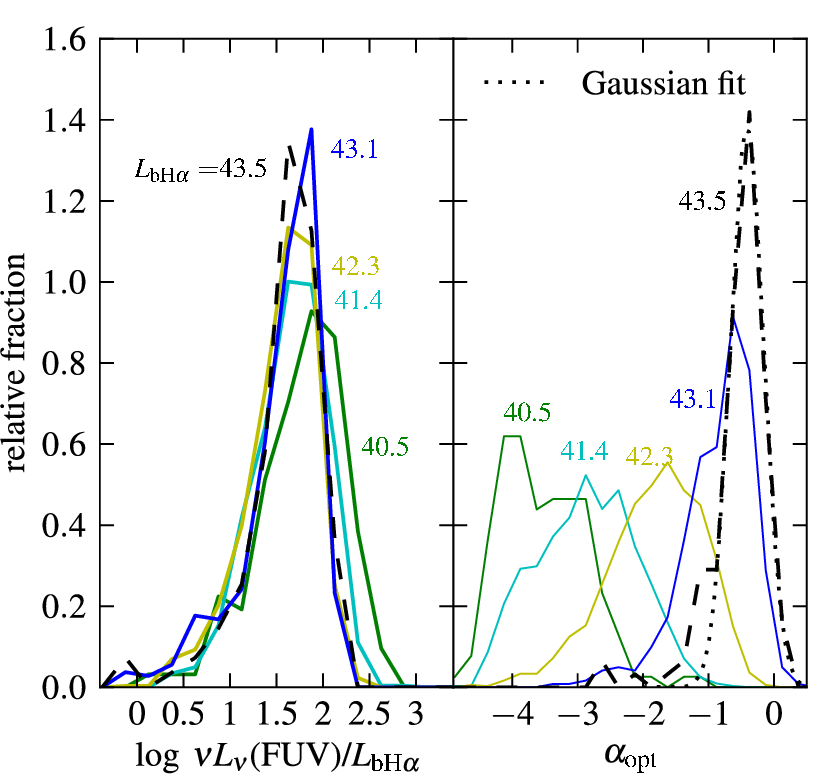}
\caption{{\bf (Left panel)} The distribution of the $\nln$(FUV)/$\lbha$ values as a function of $\lbha$. Four $\lbha$ bins, a decade wide, are plotted (solid lines). The mean $\lbha$ is noted near each distribution. The upper half of the upper bin, with $43.3<\log \lbha<43.8$, is also shown (dashed line). The distributions for
$\log \lbha>40.5$ have a very similar shape. All distributions
are asymmetric, with an extended tail towards low $\nln$(FUV)/$\lbha$ values, which may be produced by reddening. In the $\log \lbha=40.5$ bin, host contribution of the UV broadens the distribution and increases the mean value. {\bf (Right panel)} The distribution of optical slope, also binned by $\lbha$. The host emission makes $\an$ redder with decreasing AGN luminosity. The upper half of the highest luminosity bin is least affected by the host contribution, and is well fit by a Gaussian (dotted line) and an extended red tail, which may also be produced by reddening. There is therefore a small dispersion in $\nln$(FUV)/$\lbha$ and $\an$ in the net AGN emission, based on the objects least affected by the host contribution.}
\label{fig: slope and UV hist}
\end{figure}

Figure 15, right panel, shows the distribution of the SDSS based optical slope $\an \equiv 4 \times \log ( l_{\nu}(3940\AA)/l_\nu(7000\AA) )$ as a function of $\lbha$. The $\an$ distribution moves to redder values with decreasing $\lbha$, reflecting the increasing amount of host contribution. The highest $\lbha$ bin (43.1) is least affected by the host contribution, and to further decrease the effect of the host contribution we show the upper half of the $\log\ \lbha = 43.1$ bin, where $\log\ \lbha = 43.5$ ($43.3<\log\ \lbha<43.8$). The distribution is well characterized as a Gaussian (see plot) with a red tail excess. This distribution is similar to the slope distribution found in luminous quasars by Richards et al. (2003), measured between 1450\AA\ and 4040\AA\ of composite spectra. Their slope distribution peaks between --0.54 and --0.41 (table 1 there), while $6\%$ of their objects are in a red tail, attributed to a reddening effect. Here, the peak of the distribution of the $\lbha \sim 10^{43.5}\ \ergs$ objects is at $\an = -0.39$, and 12\% of the objects are in the red tail with $\an < -1$. This similarity with Richards et al. (2003) indicates that the AGN slope distribution does not change significantly with luminosity, and that the host effect on the scatter in $\an$ is negligible at $\log \lbha \ge 43.3$. 
Both the weaker FUV emission and the steeper optical slope may indicate dust reddening, and their relation is explored below. Thus, we conclude that both the mean net AGN SED shape is independent of luminosity, and
it also shows a small dispersion, as indicated by the distribution of $\nln$(FUV)/$\lbha$ and $\an$ in the AGN dominated luminosity range.

\subsubsection{The dependence of the SED on $\an$}
 
How does the overall SED change with $\an$? Figure 16 shows the mean SED at the eight different $\lbha$ bins presented above (Fig. 11), further divided into 0.5-wide $\an$ bins. To minimize statistical noise, we present only bins with $\geq \slopebinmingroupsize$ objects. The inset in each panel shows a zoom of the spectrum around 4000\AA, where the AGN emission lines of \Hd, \He, \neiii, and the stellar absorption features of the \caii\ H and K lines, can be seen. The stellar absorption features provide information on the host stellar population (e.g. Wild et al. 2007). Table 4 lists for each bin, the mean-$\lbha$, mean-$\an$, number of objects, and detection fractions in the X-ray ($d_{\rm{X}}$) and UV ($d_{\rm{UV}}$).
Note that at lower luminosities the mean X-ray detection fraction becomes well below 50\% (Table 4), so the mean $L_{\rm X}$ becomes significantly biased. Furthermore, at $\log L_{\rm X}<42$ the host contribution may become significant (e.g. Bauer et al. 2004).

\begin{figure*}
\includegraphics{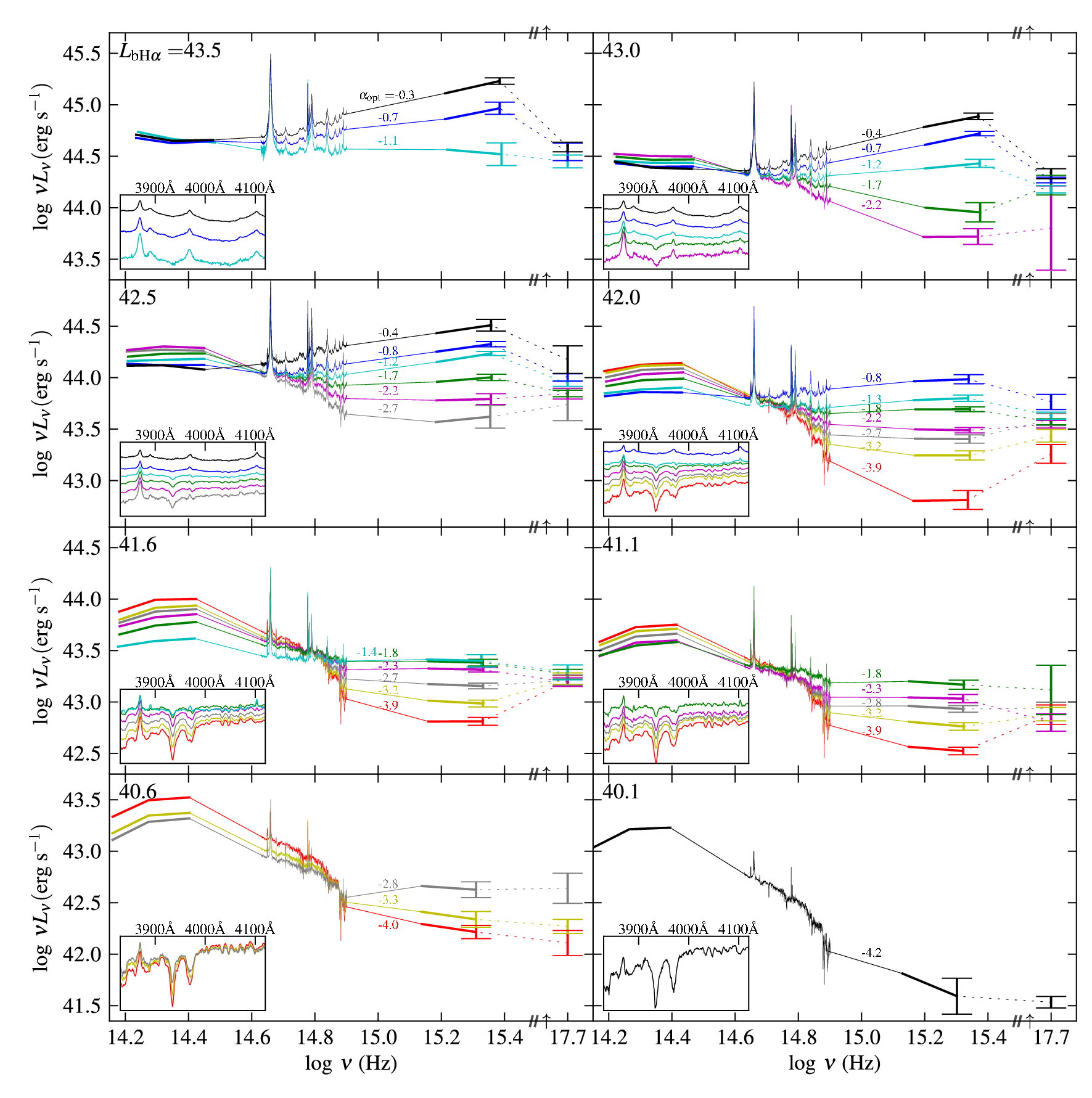}
\caption{The mean IR to X-ray observed SEDs and optical spectra near 4000\AA\ as a function of $\lbha$ and $\an$. 
The mean $\lbha$ is noted in each panel, and the mean $\an$ is noted near each curve. Only bins with $\geq \slopebinmingroupsize$ objects are shown. The value of $L_{\rm X}$ is shifted in the y-axis (+1 dec) for presentation. 
The error bars denote the error in the mean $L_{\rm X}$ and $\nln$(FUV). Data from different surveys are connected by thinner lines.
The insets show a zoom of the optical spectrum around 4000\AA, showing the AGN emission of \Hd~$\lambda4102$, \He~$\lambda3970$, and \neiii~$\lambda3869$, and \caii~H,K $\lambda\lambda3934, 3968$ stellar absorption features. In the highest $\lbha$ panel the difference between the SEDs increases from the NIR to the UV and disappears in the X-ray, consistent with a reddening effect. At $42 \leq \log \lbha \leq 43$, lower $\an$ SEDs also show a stronger bump at $\log \nu =14.4$, and stronger stellar absorption features, indicating a stronger host contribution, in addition to reddening.  
At $\lbha \leq 41$, the host dominates the emission at all bands (see Figs. 10, 13), and the increase of the equivalent widths of the stellar absorption features with decreasing $\an$ reflects the change in the host stellar population.
Note that dust extinction appears as missing UV emission for a steeper $\an$, while host contribution appears as excess NIR emission for a steeper $\an$. 
The host contribution progresses to shorter wavelengths with decreasing $\lbha$. 
}
\label{fig: slope binned}
\end{figure*}

\begin{table*}
\begin{tabular}{c|c|c|c|c|c|c|c|c|c}
$\lbha$ range   &  $\lbha$  &  $\an$  & $N$  & \multicolumn{2}{|c}{detection fractions} & $\ewha$ & $\frac{\lhost}{\lagn}$(J) & $\dbreak$ & $\dva$\\
 ($\log \ergs$) & ($\log \ergs$)&  &  & $d_{\rm{UV}}$ & $d_{\rm{X}}$ & (\AA) &  & & ($\kms$) \\ 
\hline
\hline
43.3 - 43.8  &  43.50  &  -0.30  &  66  &  1.00  &  0.73  &  453 $({\pm 13})$  &  0.8  &  0.92  &  4000  \\
             &  43.45  &  -0.66  &  38  &  1.00  &  0.71  &  456 $({\pm 20})$  &  1.0  &  0.93  &  4000  \\
             &  43.42  &  -1.13  &  11  &  0.89  &  0.55  &  503 $({\pm 34})$  &  1.0  &  0.94  &  4500  \\
\hline
42.8 - 43.3  &  43.07  &  -0.35  &   64  &  1.00  &  0.89  &  353 $({\pm 13})$  &  1.6  &  0.93  &  2900  \\
             &  43.03  &  -0.74  &  145  &  0.97  &  0.66  &   349 $({\pm 9})$  &  2.1  &  0.94  &  3500  \\
             &  42.98  &  -1.20  &  102  &  1.00  &  0.59  &  318 $({\pm 10})$  &  2.7  &  0.97  &  4100  \\
             &  42.98  &  -1.70  &   32  &  0.96  &  0.41  &  296 $({\pm 17})$  &  3.0  &  1.02  &  5000  \\
             &  42.97  &  -2.23  &   10  &  0.90  &  0.20  &  303 $({\pm 27})$  &  3.2  &  1.07  &  7000  \\
\hline
42.3 - 42.8  &  42.62  &  -0.39  &   14  &  1.00  &  0.71  &  217 $({\pm 24})$  &  3.2  &  0.95  &  2600  \\
             &  42.60  &  -0.80  &  127  &  0.97  &  0.68  &   252 $({\pm 8})$  &  3.4  &  0.96  &  2700  \\
             &  42.55  &  -1.24  &  211  &  0.97  &  0.58  &   220 $({\pm 6})$  &  4.7  &  1.00  &  3300  \\
             &  42.52  &  -1.71  &  153  &  1.00  &  0.49  &   198 $({\pm 6})$  &  5.9  &  1.04  &  4000  \\
             &  42.49  &  -2.22  &   75  &  0.91  &  0.56  &   189 $({\pm 9})$  &  7.3  &  1.09  &  4300  \\
             &  42.48  &  -2.72  &   25  &  0.64  &  0.36  &  200 $({\pm 15})$  &  6.7  &  1.14  &  4900  \\
\hline
41.8 - 42.3  &  42.13  &  -0.81  &   40  &  0.97  &  0.80  &  145 $({\pm 10})$  &   6.5  &  0.98  &  2300  \\
             &  42.08  &  -1.29  &  126  &  0.95  &  0.60  &   153 $({\pm 6})$  &   7.8  &  1.02  &  2300  \\
             &  42.06  &  -1.76  &  226  &  0.99  &  0.50  &   125 $({\pm 3})$  &  10.3  &  1.08  &  2700  \\
             &  42.02  &  -2.24  &  217  &  0.95  &  0.40  &   112 $({\pm 3})$  &  13.3  &  1.13  &  3400  \\
             &  42.01  &  -2.70  &  122  &  0.93  &  0.38  &   107 $({\pm 4})$  &  14.7  &  1.20  &  3800  \\
             &  41.98  &  -3.21  &   62  &  0.85  &  0.29  &    93 $({\pm 6})$  &  18.0  &  1.31  &  4200  \\
             &  41.97  &  -3.86  &   29  &  0.70  &  0.41  &    94 $({\pm 9})$  &  19.1  &  1.47  &  5400  \\
\hline
41.3 - 41.8  &  41.64  &  -1.35  &   28  &  0.96  &  0.68  &  108 $({\pm 9})$  &  12.1  &  1.04  &  1900  \\
             &  41.61  &  -1.79  &  115  &  0.97  &  0.36  &   78 $({\pm 4})$  &  20.1  &  1.10  &  2200  \\
             &  41.56  &  -2.27  &  212  &  0.96  &  0.35  &   64 $({\pm 2})$  &  25.2  &  1.18  &  2600  \\
             &  41.56  &  -2.74  &  186  &  0.93  &  0.31  &   61 $({\pm 2})$  &  28.5  &  1.25  &  3100  \\
             &  41.54  &  -3.25  &  146  &  0.88  &  0.31  &   56 $({\pm 2})$  &  31.5  &  1.36  &  3900  \\
             &  41.52  &  -3.86  &  129  &  0.78  &  0.22  &   47 $({\pm 2})$  &  38.5  &  1.51  &  4500  \\
\hline
40.8 - 41.3  &  41.17  &  -1.76  &   20  &  0.95  &  0.30  &  45 $({\pm 5})$  &  35.0  &  1.15  &  1800  \\
             &  41.11  &  -2.31  &   59  &  0.96  &  0.32  &  41 $({\pm 3})$  &  41.3  &  1.21  &  2400  \\
             &  41.10  &  -2.77  &  127  &  0.96  &  0.30  &  38 $({\pm 2})$  &  49.2  &  1.29  &  2700  \\
             &  41.08  &  -3.24  &  111  &  0.91  &  0.18  &  33 $({\pm 1})$  &  55.4  &  1.41  &  3000  \\
             &  41.06  &  -3.91  &  160  &  0.81  &  0.12  &  29 $({\pm 1})$  &  63.2  &  1.60  &  4100  \\
\hline
40.3 - 40.8  &  40.62  &  -2.80  &  23  &  0.95  &  0.09  &  32 $({\pm 4})$  &   64.9  &  1.29  &  2500  \\
             &  40.62  &  -3.26  &  30  &  0.90  &  0.27  &  28 $({\pm 3})$  &   76.3  &  1.42  &  3100  \\
             &  40.59  &  -3.98  &  74  &  0.85  &  0.20  &  20 $({\pm 1})$  &  110.8  &  1.63  &  3900  \\
\hline
39.8 - 40.3  &  40.10  &  -4.24  &  10  &  0.75  &  0.30  &  15 $({\pm 2})$  &  182.0  &  1.66  &  4100  \\
\end{tabular}
\caption{Mean quantities of the bins of $\lbha$ and $\an$ pictured in Fig. 16. The $\lbha$ ranges are as in Fig. 11 and Table 2. The $\an$ bins have a width of 0.5. Only bins with $N \geq \slopebinmingroupsize$ are listed. Col. 6 shows the mean and the error in the mean $\ewha$. The $\lhost / \lagn$ ratio (Col. 7), is calculated as in Fig. 13. The formulation of the 4000\AA-break (Col. 8) is given in Balogh et al. (1999). 
}
\label{table: slope bins}
\end{table*}

In the highest $\log \lbha = 43.5$ bin, the $\an$ distribution is narrow, as seen in Fig. 15, and there are only three $\an$ bins of SEDs. The difference between the SEDs is small in the IR, increases to the UV, and disappears again in the X-ray. This pattern is highly suggestive of a reddening effect, as the peak extinction of dust is in the UV, and it drops both towards the NIR and towards the X-ray. Also, as shown below, the \Ha\ EW is independent of $\an$. If the observed SEDs were intrinsic, the large differences in the FUV emission, of more than a factor of $>5$, should have produced similar amplitude in the mean \Ha\ EW, while the observed difference is $\lesssim 10$\%.
These arguments support the suggestion that the red tails discussed above (Fig. 15) are due to dust extinction. It also hints that some of the dispersion in the narrow Gaussian peak of the $\an$ distribution 
($\an=-0.3$ to $-0.7$) is due to reddening. The reddening scenario is further explored in \S3.7.3.

At $\log \lbha \sim 43.0$, there is a larger range of $\an$, which is also consistent with a larger range of dust extinctions. However, there is a small reversal in the NIR, where the steeper $\an$ objects, which are weaker in the UV, become stronger in the NIR. The different $\an$ SEDs cross at $\lambda \sim 1 \mic$. One may suspect that the reverse NIR effect results from a larger covering factor of dust, and thus stronger dust emission, in more reddened objects. However, the difference is most prominent at $\log \nu\simeq 14.5$, where dust emission is negligible. The emission of galaxies generally peaks at $\log \nu\simeq 14.5$, indicating increasing host contribution as $\an$ get steeper. The lower $\lbha$ bins show a stronger effect, as the NIR becomes significantly more prominent, and the SEDs crossing point shifts to shorter $\lambda$, going down to $\lambda\sim 6000$\AA\ at $\log \lbha \sim 42$, and $\lambda\sim 5000\AA$ at $\log \lbha \sim 41$. 
 
As the inset shows, with decreasing $\an$ at a given $\lbha$, the stellar \caii\ H and K absorption lines become stronger. The effect becomes more prominent at the lower $\lbha$ bins. Thus, apart from the highest $\lbha$ bin, the change in the SED with $\an$ also reflects an increasing host/AGN luminosity ratio, as also indicated by the increasing excess emission at $\log \nu\simeq 14.5$. For $\log \lbha \lesssim 42$, a steeper $\an$ at a given $\lbha$, is mostly produced by an increased host contribution, rather than by increasing dust extinction. This is indicated by the shift of the SED crossing point, i.e. spectra have a redder slope due to increasing NIR emission, rather than decreasing UV emission. 

What makes some AGN be more host dominated than others, within the same $\lbha$ (i.e. AGN luminosity) bin?  A higher host luminosity near 1~$\mic$ implies a higher host mass, and if the host is bulge dominated, also a higher bulge mass. A higher bulge mass implies a higher $\mbh$, and thus a lower $\lledd$. Thus, the range of host contributions reflects the range of $\lledd$ at a given AGN luminosity. This is further explored below using the relation between $\dva$ and $\an$, and in \S 3.9 by exploring directly the SED as a function of $\lledd$ and $\mbh$. 

An association of increasing reddening with increasing host/AGN ratio is clearly seen for the $\log \lbha = 42, 42.5$ AGN (Fig. 16). This suggests that the increasing reddening is associated with dust in the host galaxy. An alternative explanation for the range of reddening is dust local to the AGN, where an excess of dust along the line of sight implies a close to `edge on' view, i.e. an `almost' type 2 AGN. Such a local dust effect is not expected to affect host/AGN luminosity ratio near 1~$\mic$, in contrast with the observations. We note that
we cannot exclude an alternative option where the steeper $\an$ bins are composed of two separate populations, 
host dominated AGN, and reddened AGN, in particular if the mean broad \Ha\ equivalent width ($\ewha$)
tends to be higher for host dominated AGN.

The various notions raised above are shown quantitatively in Table 4. The mean $\ewha$ of each bin is tabulated in the seventh column. In the most luminous bin, the $\ewha$ are similar to within $\sim 10$\%. This is consistent with the reddening scenario, as dust, which must reside outside the BLR, is unlikely to change the $\ewha$. From the $\log \lbha \sim 43$ bin and downwards, there is a general decrease of $\ewha$ with $\an$, most likely due to the increase in host contribution. In the eighth column we list $\lhost/\lagn$(J), where $\lagn$(J)$=8.5\ \lbha$ (eq. 5), and $\lhost$(J)$=\nln$(J)$-\lagn$(J). This column gives quantitatively the dependence of $\an$ on the relative amount of host contribution. 
As can be seen, the relative host contribution is a significant factor in the determination of the slope already at the second luminous bin, as concluded above based on the SED. However, the small reduction of $\lesssim 10$\% in the $\ewha$ with steepening $\an$, for $\log \lbha\ge 42.5$, indicates that the steepening is not due to increasing host contribution, but due the increased reddening. 
At $\log \lbha\le 42$ the significant drop in $\ewha$ with steepening $\an$ indicates the steepening is due to increasing host contribution, and not due to reddening.
In column 9, the 4000\AA-break is tabulated, derived from the ratio of flux at $4000\AA<\lambda<4100\AA$ to the flux at $3850\AA<\lambda<3950\AA$ ($\dbreak$; Balogh et al. 1999). The net AGN emission, and the associated `small blue bump', produce $\dbreak<1$, while stars produce $\dbreak > 1$ which increases with decreasing surface temperature.
The value of $\dbreak$ increases with decreasing $\lbha$, and with decreasing $\an$ at a given $\lbha$, reflecting the diminishing AGN contribution compared to old stars in the host. However, in the three lowest luminosity bins, this trend saturates for $\an < -2.5$, and $\dbreak$ depends only on $\an$, and is independent of $\lbha$ (see also Fig. 16). The dependence of $\dbreak$ on $\an$ at these luminosities probably reflects the host type, where a bluer host galaxy (flatter $\an$) is associated with hotter stars, and thus lower $\dbreak$ values.
Finally, the last column shows the average $\dva$ of each bin. The $\dva$ follows $\an$ at all $\lbha$, similar to $\dbreak$ and $\lhost/\lagn$(J). This trend reflects the $\mbh$ versus bulge mass relation. As discussed above, a steeper $\an$ is associated with a more massive bulge, at a given AGN luminosity (given $\lbha$ bin). A higher bulge mass implies a higher $\mbh$, and therefore necessarily requires a higher $\dva$. Since $\mbh \propto \dva^2$ (eq. 2), 
at a fixed AGN luminosity, we expect a steeper than linear rise in $\lhost/\lagn$(J) with $\dva$, as Table 4 indicates. Also, at the three lowest $\lbha$ bins, which are dominated by the host emission, a bluer host colour is associated with a lower $\dva$, indicating a lower bulge mass, and thus a later type host galaxy. This result is consistent with the tendency of bluer galaxies to be of later type.

\subsubsection{The effect of dust extinction on the SED}\label{subsec: extinction law} 

Above we have shown that the effect of the host on the optical slope is small at $\lbha \sim 10^{43.5}\ \ergs$, and that the dispersion in optical slope at this luminosity seems to originate from a reddening effect. In this subsection we evaluate the implied general properties of the extinction law. 

The upper panel of Figure 17 compares $\nln({\rm UV})/ \lbha$ with $\an$ in the T1 objects from the $\lbha = 43.5$ bin. The $\nln(\rm{UV})$ value is the geometrical mean of the NUV and FUV luminosities. There is a strong correlation of $\an$ versus $\nln(\rm{UV})/\lbha$ ($P_r = 0.78$, significance of $11\sigma$), with a best-fitting least squares slope of $0.93 \pm 0.06$. 
This correlation strongly suggests reddening is a significant cause of dispersion in $\an$, already in the $-1 < \an < 0$ range ($P_r = 0.5$, significance of $6\sigma$ for objects with $\an>-1$). 

Over plotted on the data are the expected relations for different extinction curves, calculated using the formulations in Pei (1992) for MW, SMC and LMC dust, and also the `grey dust' formulation in Gaskell \& Benker (2007). We assume the intrinsic unabsorbed mean values of $\an=-0.3$ which extends to the UV, and $\nln(\rm{UV})/\lbha=47$. We apply the appropriate extinction law , and then convolve the extincted power law
with the response functions of the GALEX filters at $z=0.24$ (the mean $z$ of the $\log\ \lbha=43.5$ bin, Table 2). The SMC or LMC extinction laws are preferred over MW and grey dust. 

An SMC extinction correction of $\Delta \an=0.4$ translates to $\ebv = 0.1$. Therefore, the $\an=-0.7$ bin (30\% of the objects, Table 4) and $\an=-1.1$ bin (9\% of the objects) have an $\ebv$ of 0.1 and 0.2 larger than the $\an=-0.3$ bin. The mean $\ebv$ of the $\an=-0.3$ bin, which includes the majority of the objects, is hard to assess, as there may be some intrinsic dispersion in $\an$ (the best fitting slope in Fig. 17 yields a residual scatter of 0.29 in $\an$). If the 10 percentile of flattest $\an$ of all $\log \lbha =43.5$ T1 objects, $-0.13$, is the intrinsic unreddened slope, then the mean $\ebv$ of the $\an=-0.3$ bin is $\sim 0.04$.

In the lower panel we compare the UV slope $\auv\ (\equiv 5.8 \times \log (l_{\nu; \rm{FUV}}/ l_{\nu; \rm{NUV}}))$ with $\nln(\rm{UV})/\lbha$ for the T1 objects with $\lbha > 10^{42.3}\ \ergs$, where the host emission in the NUV is negligible (note $\auv$ and $\nln(\rm{UV})$ are independent quantities). The general trend is a decreasing $\auv$ with decreasing $\nln(\rm{UV})/\lbha$, as expected from reddening, but the scatter is significantly larger than the relation above with $\an$. The mean trend seems to favor the MW or the grey dust curve. Note that unlike $\an$, which is based on the continuum derived from the SDSS spectroscopy, $\auv$
and $\nln(\rm{UV})$ are derived from broad band photometry, which includes the UV continuum and emission features. The dispersion in emission line properties likely induces the large scatter observed. We also note that the extinction law lines are derived based on the mean $z$ of 0.18, while the apparent amount of reddening changes with $z$ as different rest-frame frequencies enter the filters. This is especially pronounced in the MW extinction curve, as the 2175\AA\ feature enters and leaves the two UV filters. UV spectroscopy is required to derive more accurate continuum slopes in the UV, and to test whether the extinctions laws implied by the UV and by the optical are indeed inconsistent.

\begin{figure}
\includegraphics{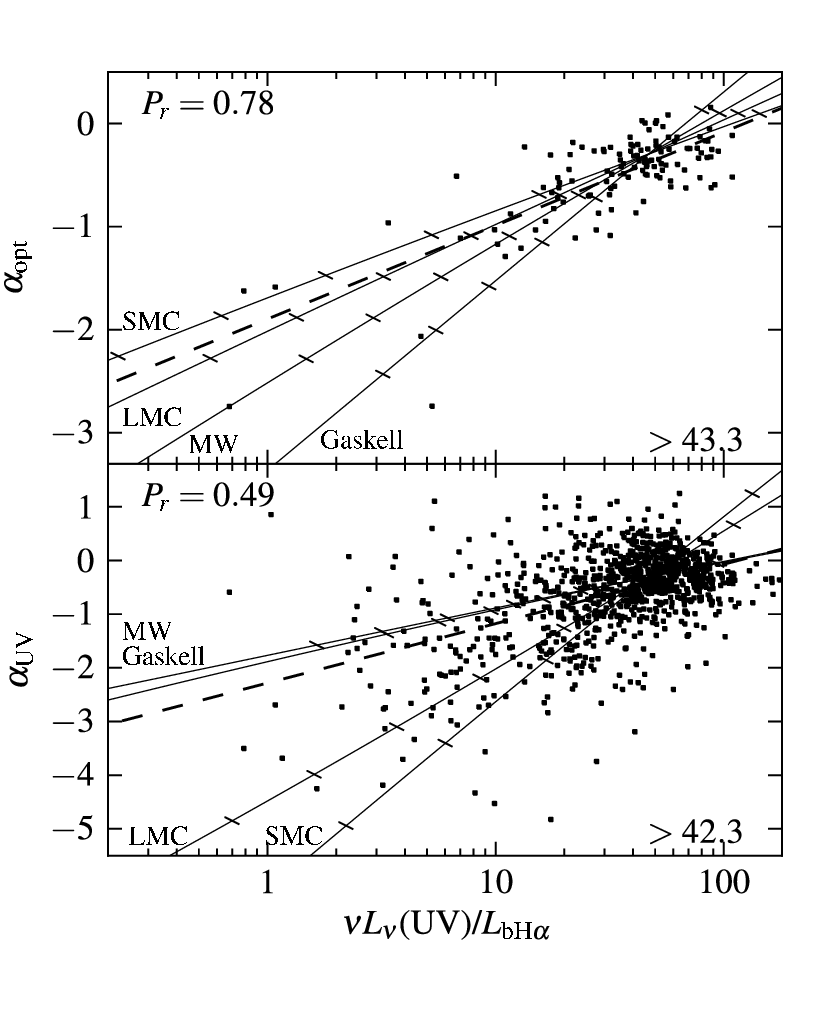}
\caption{The observed relation of $\an$ and $\auv$ with $\nln$(UV)/$\lbha$. Over plotted are the expected relations for different extinction laws of an intrinsic power law emission with $\an=\auv=-0.3$ and
a $\nln$(UV)/$\lbha=47$. Small dashes `-' mark jumps of 0.1 in $\ebv$. The minimum $\log\ \lbha$ of the T1 objects used, to avoid host contribution, are noted in each panel. The data points are fit by linear relations (dashed lines). 
{\bf Upper panel} 
The two independent ratios are highly correlated, and follow an SMC/LMC extinction law. This correlation suggests that reddening affects the dispersion in $\an$ also for $\an>-1$ (the narrow Gaussian peak of the $\an$ and $\nln{\rm (UV)}/\lbha$ distributions, Fig. 15). 
{\bf Lower panel} 
There is a general decrease of $\auv$ with $\nln$(UV)/$\lbha$, consistent with reddening, though with substantial scatter. The mean trend prefers the MW / Gaskell \& Benker (2007) extinction curves, although photometry-based slopes are significantly uncertain.
 }
\label{fig: extinction law}
\end{figure}

\subsection{The dependence of the SED on $\lbha$ and $\dva$}\label{subsec: sed by lbha and fwhm} 
In the following section we briefly explore the dependence of the mean SED on $\lledd$ and $\mbh$. Since both parameters are derived from the two directly observed quantities, $\lbha$ and $\dva$, we first explore the dependence of the mean SED on these two quantities. 

Figure 18 presents the mean optical spectra as a function of $\lbha$ and $\dva$. Each panel is for one decade wide $\lbha$ bins, and the plots in each panel are for $\log\ \dva$ bins a quarter decade wide.  Various prominent emission lines and \feii\ multiplets are noted in the plot (Vanden Berk et al. 2001; Phillips 1978). The inset in the lower two panels zooms on the spectrum near 4000\AA\ (see Fig. 16). The highest luminosity AGN at $\log\ \lbha =43.1$ (upper panel) show clearly the strong dependence of the \feii\ emission on $\dva$, which is part of the Boroson \& Green (1992) eigenvector 1 (EV1) trends. The $\log \dva=3.2$ spectrum shows that \feii\ multiplets are present from 7000\AA\ to below 4200\AA. The Balmer lines become asymmetric, with increasing $\dva$, as expected from the EV1 correlations. The EW of the \oi\ $\lambda$6300, \oii\ $\lambda$3727, and \neiii\ $\lambda 3869$ also increase with increasing $\dva$. 
Stellar absorption features such as \caii\ H,K~$\lambda\lambda 3934, 3968$, G4300, Mgb~$\lambda 5175$, and NaD~$\lambda 5893$ become clear at the highest $\log\ \dva=4.1$ plot, and some can be traced back to lower $\dva$ values. 

The $\log \lbha = 42.3$ bin shows a more prominent host contribution with increasing $\dva$, as indicated by the increasing stellar absorption features and by the steepening $\an$. These effects are more prominent than in the upper panel. However the increase in the narrow line EW with increasing $\dva$ is not as strong as in the $\log \lbha = 43.1$ bin. 
The lowest two panels show spectra which are dominated by the host continuum. This is demonstrated by the fact that the mean continuum luminosity density in the $\log \lbha = 41.4$ bin is typically only a factor of two lower than in the $\log \lbha =42.3$ panel, although the expected net AGN contribution, based on $\lbha$, drops by an order of magnitude. 

\begin{figure*}
\includegraphics{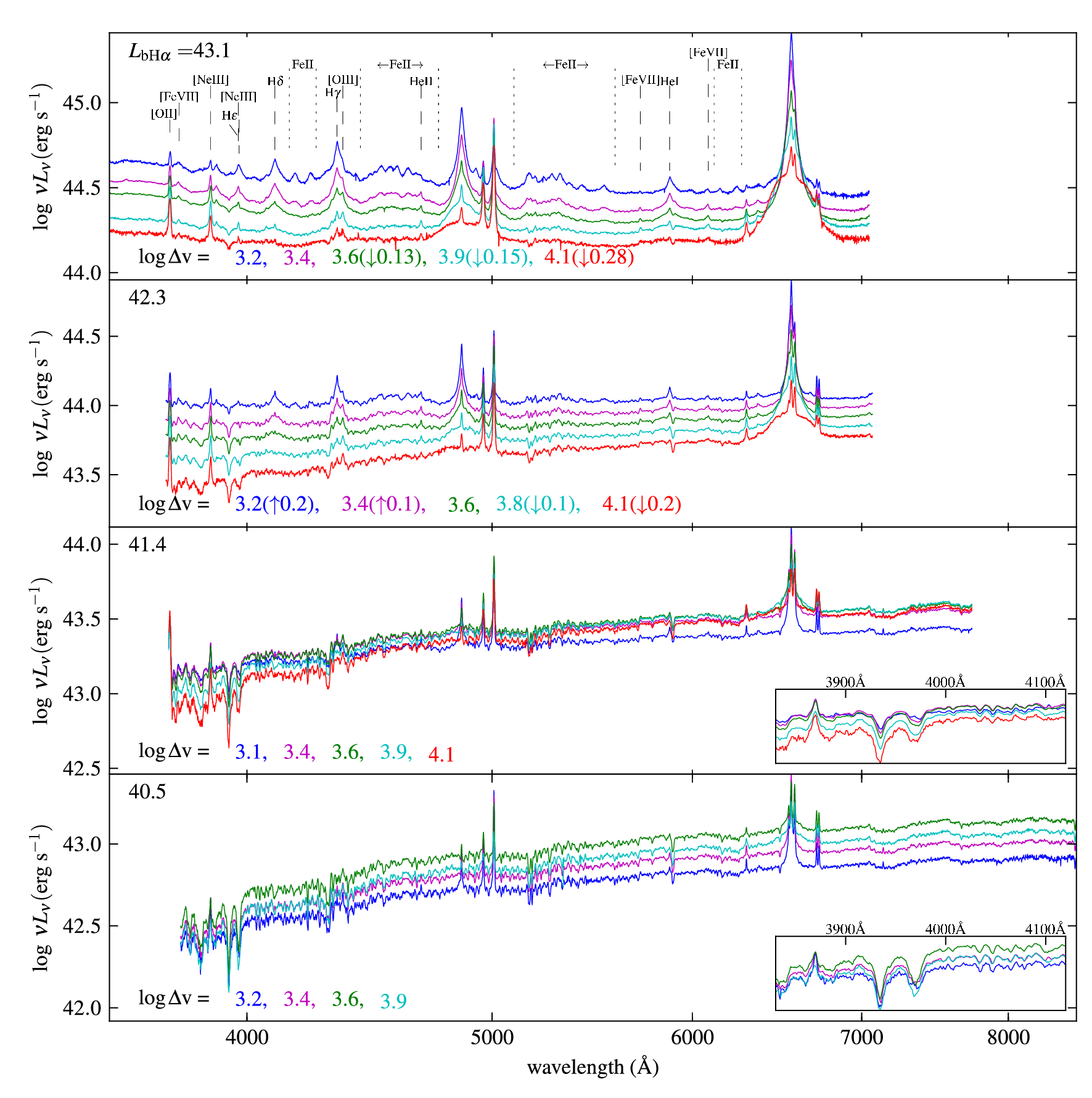}
\caption{
The mean optical spectra as a function of $\lbha$ and $\dva$ (listed in each panel). To increase clarity, some mean spectra were shifted, as noted in the legends. Some prominent emission lines and \feii\ multiplets are marked. The inset zooms on the spectrum near 4000\AA, as in Fig. 16. The highest $\lbha$ bin shows the increasing EV1 emission (BG92) with decreasing $\dva$. The BG92 analysis was near \Hb\, and this plot shows the extension of the EV1 features from 7000\AA\ to below 4000\AA. Note the change in the \feii\ multiplets near 4250\AA\ and 6250\AA, and in the various forbidden lines at $\lambda>5500\AA$ and $\lambda<4500\AA$. 
At $\log\lbha= 42.3$ there is an increase in stellar absorption features with increasing $\dva$, in addition to the EV1 trend. Higher $\dva$ implies a higher $\mbh$ and thus a more massive and more luminous bulge (up to the SDSS aperture effect) 
for the same mean AGN luminosity. At $\log\lbha\le 41.4$, the host dominates the continuum, and 
at $\log\lbha=40.5$ the broad \Hb\ becomes undetectable due to the host contribution. Such objects will be defined as net type 2 AGN based on \Hb\ only.
}
\label{fig: fwhm binned L2}
\end{figure*}

\subsection{The dependence of the SED on $\lledd$ and $\mbh$} \label{subsec: sed by lledd and mbh}

\begin{figure}
\includegraphics{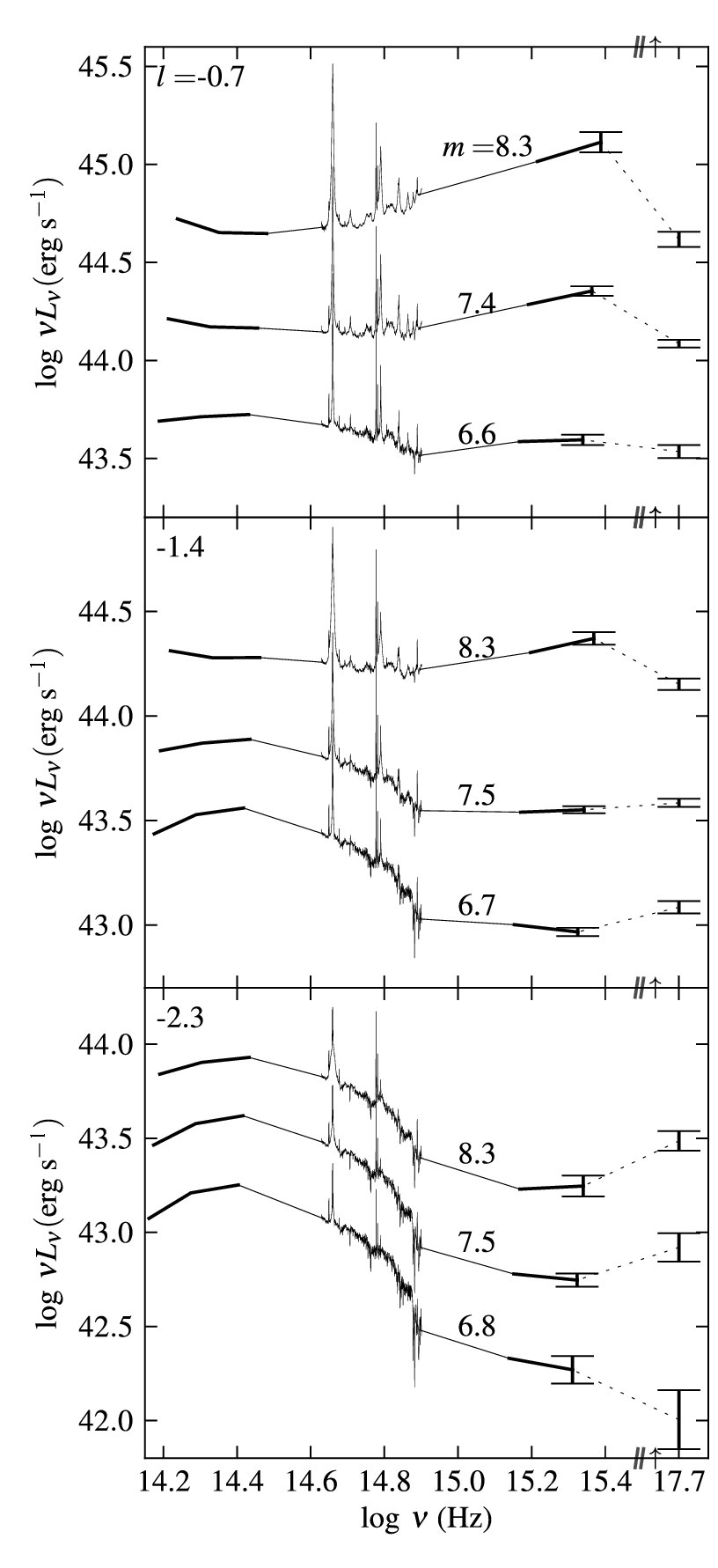}
\caption{The mean observed SED as a function of $\lledd$ and $\mbh$.  The bins are a decade wide in both $m$ and $l$, in the ranges $-3<l<0$ and $6<m<9$. Each panel is for a given $l$ bin (mean values in the upper left corner), subdivided by $m$ (mean values near each curve). The values of $L_{\rm X}$ are shifted in the y-axis (+1 dec) for presentation. The error bars denote the error in the mean $L_{\rm X}$ and $\nln$(FUV). Note the increase in the relative host contribution (the bump near $\log \nu=14.4$) with decreasing $l$ (top to bottom panel), as expected since $L_{\rm AGN}/\mbh$ corresponds to $L_{\rm AGN}/L_{\rm bulge}$ and the T1 sample is bulge dominated (\S3.5). 
Note also the increase in the relative host contribution with decreasing $m$, at a given $l$, in the middle and upper panels. This likely reflects the increase in $L_{\rm host}/L_{\rm bulge}$ with decreasing $m$, as smaller bulges are often found in disc galaxies. At $l<-2$ (lowest panel), the optical-NUV SED does not change with $\mbh$, as the host dominates for all $\mbh$. 
The fraction of X-ray detections is $>50\%$ only in the four top curves, so other plotted $L_{\rm X}$ values can be significantly biased. 
}
\label{fig: L and MBH binned}
\end{figure}

Figure 19, presents the SED binned by $\lledd$, for each panel, and by $\mbh$, within each panel. 
The bins are a decade wide, with $\lledd$ in the range $-3<l<0$ and $\mbh$ in the range $6<m<9$. 
With decreasing $\lledd$ the host contribution becomes more dominant, as expected since $\lledd$ provides a measure of $L/L_{\rm bulge}$. However, we also see a trend where the host contribution becomes more dominant (the NIR region of the SED turning from a dip into a peak) with decreasing $\mbh$ at a given $\lledd$, in particular for $l \ge -1.4$ (the two upper panels). 
This likely arises from the fact that $L_{\rm host}/L_{\rm bulge}$ rises with decreasing $L_{\rm bulge}$, as expected in a flux limited sample (SDSS), where the low luminosity bulges are more commonly found in more luminous disc galaxies, rather than in lower luminosity bulge-only galaxies (i.e. dwarf ellipticals), which will be below the detection limit. Thus, Fig. 19 indicates that at a fixed $l$, the host tends to become disc dominated with decreasing $m$. 
In the lowest $L/L_{\rm bulge}$ bin ($l= -2.3$) the host dominates at all $m$, although some AGN contribution can be noticed in the FUV and X-ray, in particular at $m=8.3$, which corresponds to the highest AGN luminosity.

The SED trends seen in Fig. 19 are dominated by the host contribution. To explore trends in the net AGN SED with $\lledd$ and $\mbh$, we use the emission bands least affected by the host contribution, which are the FUV and X-ray bands (e.g. Figs. 10, 13). Following the discussions in \S 3.3 and \S 3.6, in order to avoid host contribution in these bands, we utilize only objects with $\log \lbha > 42$ when analyzing the FUV emission, and objects with $\log \lbha > 41$ when analyzing X-ray emission. Also, to reduce detection biases when analyzing X-ray emission, we use only objects with $\fbha>10^{-13.5}\ \flux$ (Fig. 3, \S 3.6), where the X-ray detection rate is $>50$\%. 

Below we explore which of the parameters, $\lbha$, $\dva$, $\mbh$ and $\lledd$, appears to be the dominant parameter which drives the values of three SED parameters, $L_{\rm X}/\lbha$, $\nln$(FUV)/$\lbha$, and their ratio $\aoxb \equiv 0.42 \times \log\ \l_\nu(\rm{2\kev}) / l_\nu (\rm{FUV})$. Since $\lbha \propto \lbol$ (eq. 6), the ratios of $L_{\rm X}$ and of $\nln$(FUV) to $\lbha$ is a measure of the fraction of the bolometric luminosity emitted at each band. The value of $\aoxb$ is closely related to the more commonly used 2500\AA--2 \kev\ slope ($\aox$), which in eq. 8 was calculated by deriving the 2500\AA\ luminosity from $\lbha$ (eq. 5).

The binning into $\lledd$ and $\mbh$ is performed in the following manner. The objects are sorted by either $m$ or by $l$ and divided into four equal size groups. Each of these groups is then sorted by the other property, and again divided into four equal size groups. This ensures similar statistical errors in all bins. We repeat this processes by sorting based on $\dva$ and $\lbha$.

Figure 20, left column, presents the derived relations of the mean values of the three SED parameters as a function of $l$, at a fixed $m$, for $\nln$(FUV)/$\lbha$, and as a function of $m$ at a fixed $l$ for $L_{\rm X}/\lbha$ and for $\aoxb$. Error bars denote the error in the mean. The upper panels show that $\nln$(FUV)/$\lbha$ is largely set by $l$, and is independent of $m$. The drop of $\nln$(FUV)/$\lbha$ steepens with $l$, with a drop mostly at $l<-1.3$.
The right hand column of Fig. 20 presents $\nln$(FUV)/$\lbha$ as a function of $\dva$ for different bins in $\lbha$. Both parameters contribute, as $\nln$(FUV)/$\lbha$ decreases with increasing $\dva$, at a fixed $\lbha$, and also with decreasing $\lbha$, at a fixed $\dva$. The comparison of the upper two panels indicates that $l$ is the main driver of the $\nln$(FUV)/$\lbha$ distribution, and the trends with  $\dva$ and $\lbha$ are driven by the $l$ dependence. 

As discussed above, the tail towards lower $\nln$(FUV)/$\lbha$ values (Figs. 15, 17) is likely due to reddening. The mean $\nln$(FUV)/$\lbha=18$ at $l=-2$ (Fig. 20) may result from an increased amount of reddening with
decreasing $\lledd$, in particular close to $\lledd=-2$. A trend of $\nln$(FUV)/$\lbha$ with $l$ is consistent with the trend described above of an increasing reddening with the host/AGN luminosity ratio.

Figure 20, middle row panels, explore the driving parameter of $L_{\rm X}/\lbha$. Here there are mixed results. There is a general trend of decreasing $L_{\rm X}/\lbha$ with increasing $m$, but with some $l$ dependence of the amplitude for  $m>7.5$, where a higher $l$ implies a lower $L_{\rm X}/\lbha$. There is also a general trend of decreasing $L_{\rm X}/\lbha$ with increasing $\lbha$, with a weaker trend of decreasing $L_{\rm X}/\lbha$ with increasing $\dva$. None of the four parameters appear to be a dominant parameter.

Figure 20, lower row panels, explore the driving parameter of $\aoxb$. The slope gets steeper with increasing $m$ and with increasing $l$, and both parameters contribute similarly to the trends. The right hand panel shows that $\aoxb$ is mostly driven by $\lbha$. There is only a weak dependence of $\aoxb$ on $\dva$. This indicates that the dependence of $\aoxb$ on $\mbh$ and $\lledd$ found by Kelly et al. (2008) is probably due to the correlation between these parameters and AGN luminosity. 

\begin{figure*}
\includegraphics{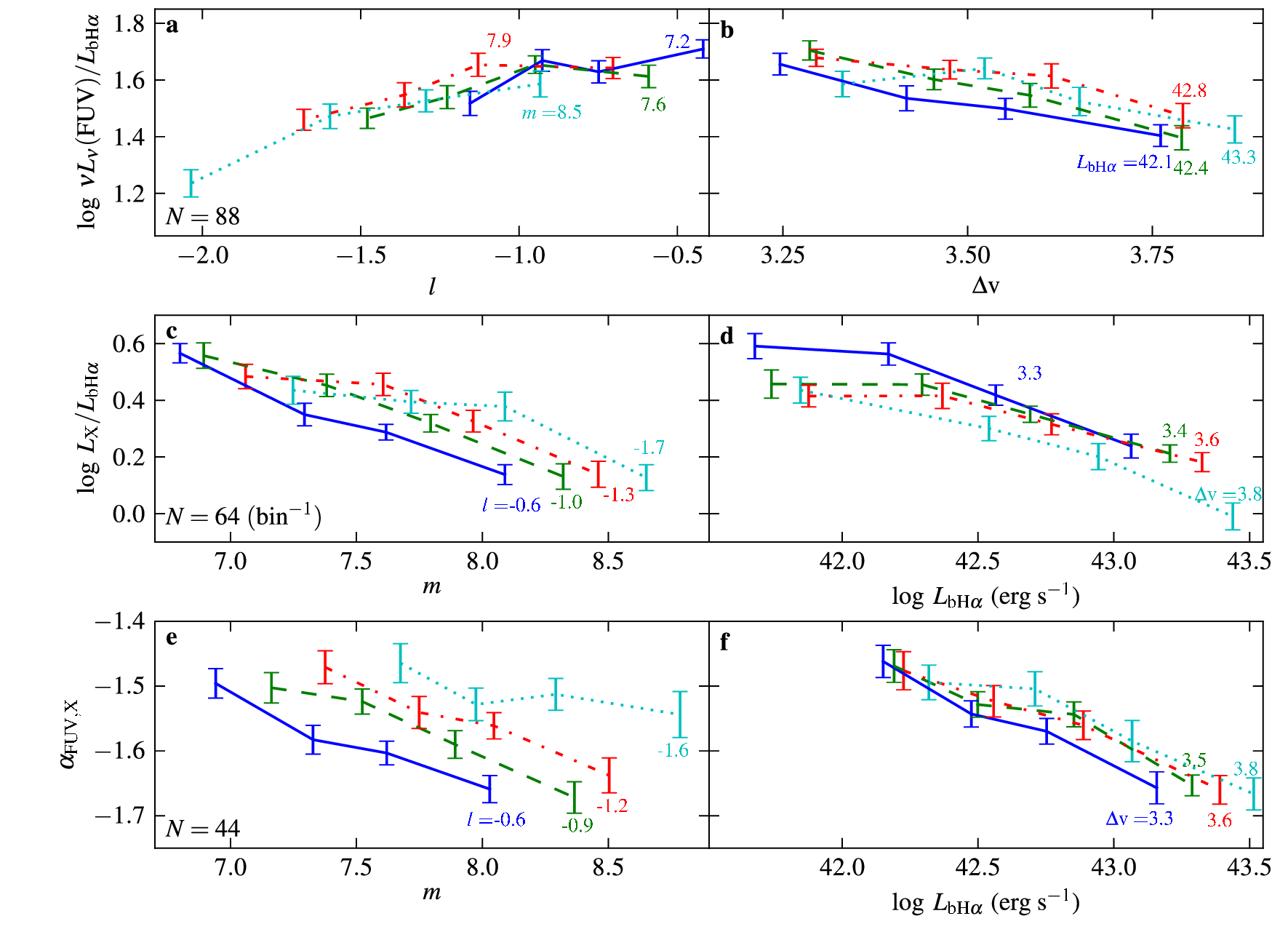}
\caption{The FUV, X-ray, and broad \Ha\ emission ratios as a function of $m$ and $l$ (left column), and as a function of $\lbha$ and $\dva$ (right column), at the $\lbha$ luminosity range where the host contribution is negligible. All bins in each panel have the same number of objects (lower left corner). 
In search of the primary parameter which drives a given ratio, we select for the x-axis either $m$ or $l$ (left column), and $\lbha$ or $\dva$ (right column) to create the smallest spread for the different curves, where each curve includes T1 objects with a specific value of the other parameter. 
Error bars denote the error in the mean value in the bin. To reduce non-detection biases in the X-ray, in the two lower rows only $\fbha>10^{-13.5}\ \flux$ objects are used.
{\bf (a,b)} Note that the primary parameter which sets the mean $\nln$(FUV)/$\lbha$ is $l$. The ratio is independent of $\mbh$. Binning by $\lbha$ and $\dva$ produces a larger dispersion. 
{\bf (c,d)} Here either $m$ or $\lbha$ are the dominant parameters.
{\bf (e,f)} $\aoxb$, equivalent to the ratio of the upper two ratios. Here $\lbha$ appears to be the dominant parameter.
}
\label{fig: mbh and lledd binned}
\end{figure*}

\section{DISCUSSION}

\subsection{The distribution of $\dva$}

Croom (2011) already noted that the mean value of $\dva$ of the \Hb\ line in the SDSS quasar samples is independent of luminosity. Here we extend this result and show that the distribution of $\dva$ values is also independent of luminosity (Fig. 7). Why is this result surprising? The observed value of $\dva$ is determined by the AGN luminosity, which sets the radius of the BLR, and by $\mbh$, which sets the velocity dispersion at that radius. Thus, the distribution of $\dva$ value, $f(\dva)$, is set by the AGN black hole mass function, $f(\mbh)$, and by the AGN luminosity function, $f(\lbol)$, which may vary with $\mbh$. The observed $f(\dva)$ is also set by the detection efficiency of individual broad line AGN.
As shown above, selection effects are not expected to dominate $f(\dva)$ (Fig. 6, upper panel), thus the lack of dependence of $f(\dva)$ on $\lbha$ appears to be an intrinsic property of the SDSS T1 sample. At face value, the physical implications are that the mean $\mbh$ increases roughly as $\lbol^{1/2}$, and similarly the mean $\lledd$ also scales as $\lbol^{1/2}$ (Fig. 6 lower panel). These relations reflect the simple fact that the distribution of the objects along the $\lbol$ axis and along the $\dva$ axis are independent of each other. The only significant deviations are the rise in the lower limit of $\dva$ with $\lbol^{1/4}$, consistent with the physical Eddington limit, i.e. $l=0$, and the drop in the upper limits on $\dva$ with $\lbol^{-1/4}$, consistent with an upper limit to $\mbh$ of a few $10^9\ \msun$. The convergence of these upper and lower limits on $\dva$ at $\lbol >$ a few $10^{45}\ \ergs$ leads to a narrower $f(\dva)$, which produces a small bump in $f(\dva)$ at $\dv \sim 7$ (Fig. 7). A similar convergence of the range of FWHM for the \mgii\ and \civ\ lines at the highest continuum luminosities was noted by Fine et al. (2008, 2010).

Could there be another physical effect which controls the form of $f(\dva)$, other than the the black hole mass function and the AGN luminosity function? Laor \& Davis (2011) noted that the accretion disc continuum becomes colder with increasing $\dva$, and for a large enough value (depending on the black hole spin), the accretion disc can become non-ionizing, and thus form a lineless quasar which is harder to detect. This may explain the rarity of objects with $\dva>10,000-20,000\ \kms$. However, it cannot explain the similarity of $f(\dva)$ at $\dva < 10,000\ \kms$, which therefore remains an open question.

\subsection{The SED}

\subsubsection{A uniform blue optical-UV SED}

As described above, at $\lbha>10^{42}\ \ergs$ the net AGN SED in the 9000\AA--1500\AA\ range is consistent with a fixed shape, and an amplitude which scales linearly with $\lbha$. The SED shape is consistent with the R06 SED of luminous AGN. Furthermore, the linearity of the mean $\lbha$ with $\nln$ in the optical-UV range suggests that the fixed shape SED extends to the ionizing continuum as well. The linearity breaks in the X-ray regime, where the relative X-ray emission drops with increasing luminosity (Fig. 20, and various earlier studies, e.g. Just et al. 2007). Other studies (Maiolino et al. 2007; Treister et al. 2008) indicate there is also a non linearity in the mid-IR emission as well, which also decreases in relative strength with increasing luminosity.

The dispersion in the SED shapes, as indicated by the observed distribution of $\nln(\rm{UV})/\lbha$ and $\an$ values (Fig. 15), is driven at lower luminosities by the range of host contributions to the SED. The luminous AGN ($\log \lbha=43.5$ bin) show a significantly smaller dispersion.
Furthermore, for $-1<\an<0$, within the narrow Gaussian peak of the distribution, there is a correlation of $\an$ with $\nln(\rm{UV})/\lbha$, consistent with reddening (Fig. 17). This suggests that the intrinsic dispersion in $\an$, after correction for reddening, is smaller than observed. These results are consistent with earlier suggestions (Ward et al. 1987; Gaskell et al. 2004) that AGN have a single shape SED over a large luminosity range, which is affected by dust and host contribution.

A rather uniform and blue intrinsic optical-UV SED is expected if AGN are powered by a thin accretion disc, which emits locally close to a blackbody. A locally blackbody Newtonian accretion disc produces the well known $L_{\nu}\propto \nu^{1/3}$ SED at wavelengths emitted from intermediate regions in the disc, away from the inner and outer edges of the disc. Including non-LTE effects in the disc atmospheric emission, and the relativistic corrections for the disc structure and for the observed emission, leads to an optical-UV slopes somewhat steeper than $\nu^{1/3}$, but still bluer than the average observed slope (e.g. Bonning et al. 2007).  Davis et al. (2007) found that a small amount of reddening of $\ebv \sim 0.03-0.055$ of SMC-like dust, is sufficient to make the observed UV SED consistent with the accretion disc model predictions. These values are consistent with a mean $\ebv \sim 0.04$ for SMC-like dust, deduced above (\S3.7.3) by comparing the slope of the 10 percentile of flattest luminous AGN $\an=-0.13$, with the peak of the narrow distribution at $\an=-0.3$, where 60\% of the AGN reside (Table 4).

\subsubsection{The reddening dust}

What is the fraction of AGN affected by reddening? 
As noted above, in the most luminous AGN bin ($\log \lbha=43.5$), about 60\% of the AGN may be slightly affected by reddening with $E(B-V) \sim 0.04$, as indicated by the correlation of $\an$ vs. $\nln(\rm{UV})/\lbha$, consistent with dust extinction affecting the SED at $-1<\an<0$  (Fig. 17). An additional 30\% of these AGN, have a mean $\an=-0.7$ (Table 4), which may correspond to a reddening of $E(B-V) \sim 0.1$ (\S3.7.3), and the remaining 10\% which reside in the $\an=-1.1$ bin may be affected by a reddening of $E(B-V) \sim 0.2$ (note that 6 more objects with $\an \sim 0$ and 4 with $\an < -1.25$ are not included in Table 4, due to our $N\geq10$ requirement, \S3.7.2). The dust suppresses $\nln$(FUV) by a factor of two and five in the $\an=-0.7$ and $\an=-1.1$ bins, indicating significant reprocessing of $\lbol$ by the dust. 

The distribution of extinction values appears independent of the AGN luminosity for $\log \lbol = 43-46$, 
as the tail of the $\nln$(UV)/$\lbha$ distribution remains largely unchanged with $\lbha$ (fig. 15, left panel). 
However, for the quasars sample, Richards et al. (2003) find that 6\% of the quasars appear to have $\ebv \sim 0.13$, while Hopkins et al. (2004) find that only 2\% of the SDSS quasars have $\ebv > 0.1$ and  1\% have $\ebv > 0.2$, significantly lower fractions than found here (but note that some of it may be due to different procedures used to derive $\ebv$).
This suggests a decreasing reddening with increasing $\lbol$, which may be related to the observed
drop in the mid-IR to $\lbol$ with increasing $\lbol$ in AGN (Maiolino et al. 2007; Treister et al. 2008) suggesting a drop in the
covering factor of dusty gas with increasing luminosity. However, the mid-IR emission may be associated
with the `receding torus' scenario (Lawrence \& Elvis 1982) where the associated dust is X-ray absorbing and thus optically very thick. The dust associated with the obscuring torus will therefore not be the reddening dust. Also, as noted above, the T1 sample does not show a luminosity dependence for the reddened fraction
of the objects.

An alternative interpretation of the
lower fraction of reddened AGN in the Richard et al. and Hopkins et al. quasar samples, is a relation of reddening with $\lledd$. As discussed above, quasars have a higher $\lledd$ by selection (Fig. 8) compared to emission line selected T1 sample, and higher
$\lledd$ AGN tend to have lower reddening (Fig. 20). 

What is the nature of the dust?  Richards et al. (2003) and Hopkins et al. (2004) find that the SED of their red quasars is well reproduced, down to $\sim 1500$\AA, by extinction of SMC-like dust. A similar conclusion is reached here based on $\an$ (Fig. 17). The local UV slope argues against SMC-like dust, but given the large scatter and the lack of spectroscopy in the UV, the discrepancy may not be significant.

Where is the dust located? The location of the reddening dust can be explored by comparing the near, mid and far-IR emission of reddened and unreddened objects, which can tell how far from the center the bulk of the reddening dust resides. The possible relation of the reddening with the host/AGN luminosity ratio, or (equivalently) with $\lledd$, quantities which are not inclination dependent, suggests the reddening dust does not reside on the smallest possible scales.

\subsection{The host galaxy}

\subsubsection{$\lhost$ vs. $\lagn$}

In \S3.5.1, we use the $z$ band host galaxy luminosity as a measure of the host stellar mass (within the SDSS aperture).
We find that the correlation of the T1 host luminosity with the AGN luminosity ($\lbha$) is induced by the correlation of both quantities with $z$. Partial correlation analysis of the host luminosity versus $\lbha$, for a fixed $z$, yields no significant correlation. A similar result was obtained by Hao et al. (2005b) for emission line selected type 1 and type 2 AGN from the SDSS. It reflects, as noted by Hao et al., the large range of $\lledd$ in emission line selected AGN samples, in contrast with colour and quasi-stellar
selected AGN samples, which by selection have a high $\lledd$ (e.g. \S 3.1.2), which induces a correlation between the AGN and the host luminosity.

\subsubsection{The host type}

Kauffmann et al. (2003a, hereafter K03) compared the host properties of type 2 AGN to inactive galaxies in the SDSS survey, and found that type 2 AGN reside almost exclusively in massive galaxies. 
Here, we verify this effect, and find that at low $z$
the mean $z$ band host luminosity of type 1 and type 2 AGN, are a factor of three to five higher than in inactive galaxies. We further separated the inactive galaxies to NEG and SFG, and find that at the lowest $z$, the $z$ band luminosity of T1 AGN hosts follow very closely the mean NEG luminosity. So, it appears that the preference of AGN for more massive hosts may just reflect the preference for NEG, and the fact that NEG tend to be more massive.

Further support for NEG as the host of the T1 AGN comes from a comparison of the host luminosity distributions. The NEG and the T1 hosts follow similar (fibre $r$-band) luminosity distribution. The relative normalization of T1 AGN is $\sim 3$\% of the NEG. This result is consistent with a scenario where $\sim 3$\% of NEG host a T1 AGN, irrespective of the NEG luminosity. The similar stellar mass distribution of NEG and of type 2 AGN is also noted in Salim et al. (2007, Fig. 17 there). In contrast, if T1 occur in SFG, then the fraction of SFG hosting a T1 AGN increases from $\sim 10^{-3}$ to $\sim 1$ over the observed range of luminosity (Fig. 9). Such a drastic change in the AGN fraction with the host luminosity appears implausible. Furthermore, it contradicts the fact that the T1 hosts are more massive than SFG (Fig. 12, see also Salim et al. 2007).

With increasing luminosity we find that the hosts of AGN become bluer, as was first noted for quasars (Hutchings 1987; Kirhakos et al. 1999; Jahnke et al. 2004). This was verified for type 2 AGN by K03 who found an increasing fraction of younger stars with increasing luminosity. A similar trend was noted by Vanden Berk et al. (2006) for type 1 SDSS AGN, although the non orthogonality of the host and AGN eigenvectors leads to a non-unique solution in the method used in that study (see \S 2.2). Here we find that the T1 and type 2 hosts are similar, and with increasing $\lbha$, the mean host $u-z$ colour for both types gets bluer (Fig. 12). For $\lbha < 10^{41}\ \ergs$, the T1 host colour is red, identical to that of NEG, and for $\lbha > 10^{41}\ \ergs$, it becomes bluer, becoming intermediate between NEG and SFG at $\lbha\gtrsim 3\times 10^{42}\ \ergs$. This colour effect is also characterized by the tendency of AGN (depending on the sample selection criteria, e.g. Hickox et al. 2009) to reside in the `green valley' in the colour-luminosity diagrams (e.g. Fig. 1 in Salim et al. 2007). The implied specific SFR ratios of $10^{-2.9}<b_{300}<10^{-2.25}$ (Table 3) are significantly lower than for `starburst' galaxies, i.e. galaxies forming stars faster than their mean SFR over the Hubble time ($\log b_{300} > -1.5$). The highest luminosity T1 AGN are intermediate between NEG and SFG,
as found in type 2 AGN (Heckman et al. 2004; Wild et al. 2007). However, their concentration index and bulge/total light ratio (K03) remains similar to NEG, indicating early type hosts.

Why do low luminosity AGN avoid SFG? Is it due to suppression of the host SFR? This scenario is unlikely given the low AGN luminosity, which is well below the host luminosity, and thus unlikely to affect the host significantly. In fact, as $\lbol$ increases and the AGN dominates the host emission, the host SFR actually increases, the reverse of what is expected from the AGN SFR suppression scenario. Furthermore, the suppression scenario is inconsistent with the various AGN host galaxy parameters (stellar mass, concentration index, bulge/total light ratio) which indicate AGN occur in bulge dominated galaxies (see K03), in contrast with the average galaxy parameters of SFG which characterize disc dominated galaxies (Kauffmann et al. 2003b). 
A similar preference for earlier type hosts was also found by Ho et al. (1997b) for the lowest luminosity AGN ($\lbha\sim 10^{39}\ \ergs$), for luminous quasars (e.g. McLeod \& Rieke 1995), and for X-ray selected AGN (e.g. Schade et al. 2000). 
The early type host galaxy parameters also argue against a delay scenario, where the host SF episode shuts off by the time the infalling gas reaches the AGN. Massive stars will indeed die off within $\sim 100$ Myr, but the structure of the galaxy will take longer to evolve. Since low luminosity AGN and SFG appear to have different host mass distribution properties, 
they are not likely to be different phases in the same host population.

An alternative explanation is that AGN avoid SFG because SFG tend to have lower mass bulges. If SFG at low $z$ typically have $\mbh \ll 10^6\ \msun$, yet rather luminous discs, the AGN may be optically swamped by the host, even if it shines close to $L_{\rm{Edd}}$. Another option is obscuration by dust, as high SFR is generally associated with excess dust extinction. For example, Sim{\~o}es Lopes et al. (2007) find excess dust even in early type galaxies with AGN activity. 
To test whether AGN avoid SFG because of their high SFR and the associated dust, or because of their low bulge mass, one needs to compare the fraction of SFG and the fraction of NEG galaxies which host an AGN, at the same bulge mass, based on X-ray and mid-IR surveys which are less affected by dust extinction. If the fraction is similar, then the SF activity has no effect on the AGN activity. 

One should note that the B04 definition of an inactive galaxy as SFG or NEG is not physically motivated, but is just based on the ability to detect and measure a set of emission lines. It will clearly be useful to compare the T1 host properties with inactive galaxies separated based on a physical property, such as the absolute or specific SFR.

\subsubsection{The SFR values}

The $u$ band luminosity is mostly set by the SFR in the recent $\sim$ 100 Myr. At $\lbol\sim 5\times 10^{44}\ \ergs$ the T1 hosts $u$ band luminosity is intermediate between SFG and NEG (Fig. 12), which suggests an intermediate SFR. A similar intermediate SFR was found by Salim et al. (2007, Fig. 3 there) for the SDSS type 2 AGN, based on the UV luminosity from GALEX (see also Heckman et al. 2004). The value of the SFR can be estimated from the $u$ band luminosity (Moustakas et al. 2006, eq. 11). The mean host $u$ band luminosity of the T1 AGN then implies a SFR which increases with the AGN $\lbol$. The SFR increases from 0.11~$\msun$ yr$^{-1}$ at $\log\ \lbol=42.2$ to 3.5~$\msun$ yr$^{-1}$ at $\log\ \lbol=44.6$, with a mean SFR of $\sim 1.1$ $\msun$ yr$^{-1}$ for the most abundant $\log\ \lbol$ of 43.7 (derived from $\lbha=41.6$, the most abundant T1 AGN, Fig. 5). The rise in the host SFR with $\lbol$ is qualitatively consistent with earlier results for type 2 AGN (K03, Silverman et al. 2009, Netzer 2009). The values derived above are comparable to the SFR vs. $\lbol$ values presented by Trakhtenbrot \& Netzer (2010, Fig. 1 there) for SDSS type 2 AGN, where the SFR is derived from the $\dbreak$ (B04), and $\lbol$ is derived from the \oiii\ and \oi\ luminosities. Thus, we conclude that inferring $\lbol$ from either the NLR (\oiii) or the BLR (\Ha), and inferring the SFR from either the $u$ band luminosity or from $\dbreak$, leads to similar $\lbol$ versus SFR relations for type 2 and type 1 AGN.

An estimate of the SFR in type 1 AGN was made by Kim et al. (2006), based on the \oii\ luminosity, assuming 1/3 comes from star formation, and 2/3 is powered by the AGN. They used the Kewley et al. (2004) \oii\ calibration of the SFR and derived a SFR for the most abundant $L_{\oii}=5\times10^{40}\ \ergs$ of 0.5 $\msun$ yr$^{-1}$, for an SDSS based sample of $z<0.3$ type 1 AGN, which should overlap well the T1 sample.
Kim et al. note the \oii\  emission may all be powered by the AGN, so the actual SFR could be even smaller. 
Note that the \oii\ based SFR values for the SDSS type 2 AGN, presented by Silverman et al. (2009, Fig. 10 there), is typically in the range of $0.1-1\ \msun$ yr$^{-1}$, which is significantly below the UV and the \Ha\ based estimate for type 2 SDSS AGN, of $1-10\ \msun$ yr$^{-1}$ (Salim et al. 2007, Fig. 3 there). So, apparently there is a discrepancy between the $u/UV$ and the \oii\ derived SFR in type 2 AGN (see discussion in Ho 2005), which may also be present in the T1 sample, if \oii\ is mostly powered by the AGN, and the implied SFR should be lower. Since \oii\ measures the most recent SFR compared to the UV/$u$ continuum, the discrepancy
may imply a post starburst.

\section{CONCLUSION}

We present and analyze a new sample (T1) of \smpsz\ broad \Ha\ selected AGN from the SDSS DR7, with $\log \lbha=40-44$, which spans $m=6-9$ and $l=-3-0$. We add UV (GALEX), IR (2MASS), and X-ray (ROSAT) luminosities to form the mean SED. The main results are:

\begin{enumerate}
\item The \Ha\ FWHM velocity distribution ${\rm d} N/{\rm d}\log \dva$ is independent of luminosity and falls exponentially with $\dva$. 
The origin of this distribution remains to be understood.
\item The observed mean 9000\AA--1500\AA\ SED, as a function of $\lbha$, is consistent with a sum of the mean SED of luminous quasars, which scales linearly with $\lbha$, and a host galaxy contribution.
\item The host galaxy $r$-band luminosity function of T1 objects with an extended morphology 
follows the NEG luminosity function, with a relative normalization of $\sim 3$\%, suggesting that the host of broad line AGN are NEG, and the AGN probability of occurrence is independent of the host mass. 
\item The mean $z$ band luminosity and the $u-z$ band colour of the lowest luminosity T1 host is identical to NEG. The host colour becomes progressively bluer with increasing luminosity. The implied mean SFR versus $\lbol$ is similar to that found in type 2 AGN.
\item The dispersion in the optical-UV SED in luminous AGN ($\log \lbha \ge 43$), is consistent with reddening. 
This indicates the intrinsic SED of AGN is blue, with a small dispersion, as predicted from thermal thin accretion disc models.
\item Reddening by dust along the line of sight to T1 AGN is common (40\% with $\ebv \gtrsim 0.1$).
\item The $\lbha$ versus $L_{\rm{X}}$ correlation provides a useful probe for unobscured narrow line AGN. It can be used to test if the absence of a broad \Ha, in X-ray detected AGN, is significant.
\item The primary parameter which drives $\alpha_{\rm FUV, X}$ is the luminosity, rather than $\mbh$ or $\lledd$. 
\item The primary parameter which drives $\log \nlnl$(FUV)$/\log\ \lbha$ is $\lledd$, which may indicate
that lower $\lledd$ AGN are more likely to be reddened.

\end{enumerate}

This publication makes use of data products from the SDSS project, funded by the Alfred P. Sloan Foundation, data from GALEX supported by NASA, data from the Two Micron All Sky Survey, funded by the NASA and the NSF, and from the ROSAT Data Archive of the Max-Planck-Institut für extraterrestrische Physik (MPE) at Garching, Germany. A.L. acknowledges support by the Israel Science Foundation grant 407/08.

\appendix
\appendixpage

\section{SAMPLE CREATION DETAILS}\label{append: pipeline details}
\subsection{Bad pixels}\label{append: bad pixels}
During the initial filtering of the SDSS database (\S 2.1), we require each spectrum to comply with the following criteria, concerning `bad' pixels. A pixel is considered as bad if it has any of the flags listed in table 11 of Stoughton et al. (2002), except `MANYBADCOLUMNS' and `NEARBADPIXEL'. Also, due to SDSS oversampling, each pixel within two pixels of a flagged pixel is considered bad.
\begin{itemize}
 \item $<20\%$ bad pixels near the broad \Ha\ (6250\AA--6880\AA)
 \item 0 bad pixels near the narrow \Ha\ (6530\AA--6600\AA)
 \item $<50\%$ bad pixels in the continuum fitting ranges (6125\AA--6250\AA\ \& 6880\AA--7000\AA)
\end{itemize}

\subsection{Host subtraction}\label{append: high chisq subtraction}
In \S 2.2, each spectrum is fit with three galaxy eigenspectra (Yip et al. 2004) and a power-law component.
The fit is performed on the following wavelength ranges, chosen since they are devoid of the strong AGN and stellar emission lines: 3455\AA--3490\AA, 3500\AA--3550\AA, 3600\AA--3700\AA, 4200\AA--4250\AA, 4400\AA--4460\AA, 4490\AA--4600\AA, 5130\AA--5680\AA, 5750\AA--5830\AA, 5920\AA--6050\AA, and 6130\AA--6200\AA. 

An example of a spectrum before and after subtracting the fit of the three eigenspectra is shown in Figure A1. 

\begin{figure}
\includegraphics{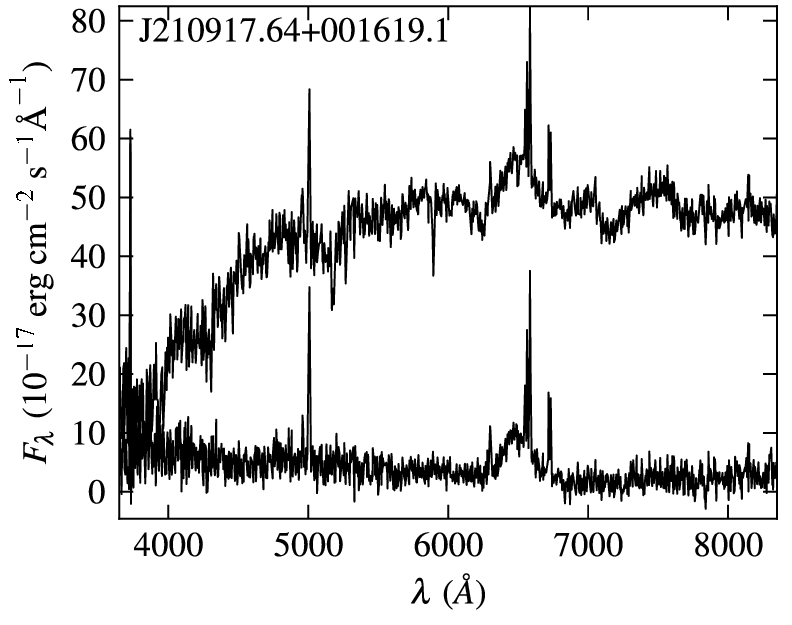}
\caption{An example of host subtraction using the Yip04 eigenspectra (\S 2.2). The top line is the SDSS spectrum, after correcting for Galactic extinction. The bottom line is the eigenspectra-subtracted spectrum. Note the high contrast of the broad \Ha\ in the residual.
}
\label{fig: subtraction low AGN to galaxy ratio}
\end{figure}

\subsection{Broad line flux}\label{append: excess flux}

A depiction of the excess flux identification process presented in \S 2.3 is shown in Figure A2.
\begin{figure}
\includegraphics{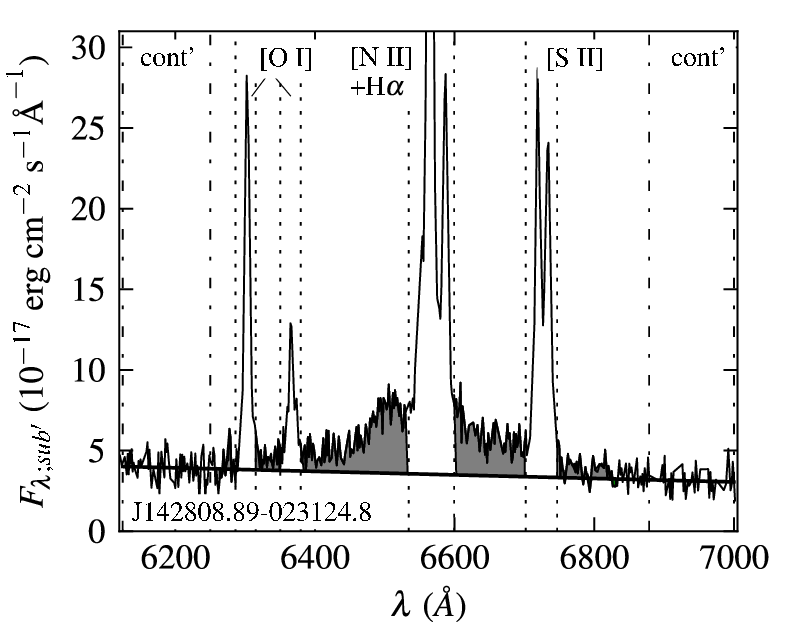}
\caption{Identifying broad \Ha\ emission (\S 2.3), shown on an example host-subtracted spectrum. The continuum (solid straight line) is interpolated from the mean of the regions bordered by dash-dot lines. The potential BLR flux ($\Delta F$) is filled with gray. Flux near the strong narrow lines (delimited by dotted lines) does not enter the calculation of $\Delta F$.}
\label{fig: excess search example}
\end{figure}

Figure A3 shows the $\normdf$ histogram of the parent sample (\S 2.1). Near $\normdf \approx 0$, the distribution fits a Gaussian, as expected from the central limit theorem. The objects in the positive wing are potential T1 objects. We choose a criterion of $\normdf>2.5$, since objects with lower $\normdf$ have a $<1\%$ chance to pass the additional criteria described in \S 2.5. The $\normdf$ distribution of the T1 sample, after the criteria in \S 2.5, is also shown. It can be seen that a criterion of $\normdf \gtrsim 12$ would have been roughly sufficient to avoid false identifications. 

\begin{figure}
\includegraphics{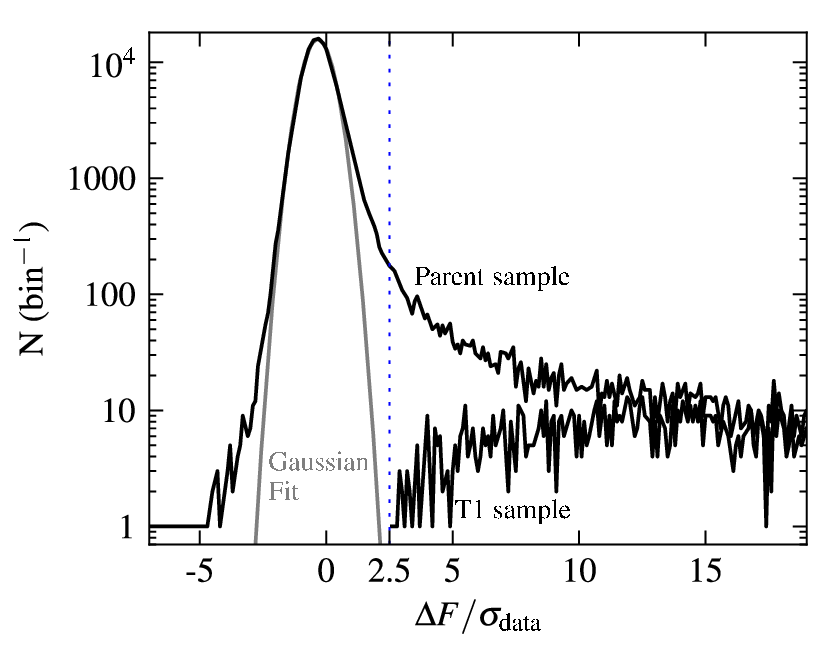}
\caption{The distribution of the normalized excess flux (\S 2.3) in the parent sample (\psmpsz\ objects). Normalization is by the $F_\lambda$ dispersion  ($\sone$) in the continuum windows (Fig. A2). Bin size is 0.1. To avoid objects with excess flux due to the normal dispersion (see Gaussian fit to the distribution), only the \midsmpsz\ objects with $\normdf>\minfluxinwingsA$ passed to the next stage of the algorithm. Also shown is the final T1 sample, the \smpsz\ objects that pass the final criteria in \S 2.5. 
}
\label{fig: excess flux distribution}
\end{figure}

\subsection{Broad \Ha\ and narrow lines fit}\label{append: high chisq line fit}
The emission lines are fit with Gauss-Hermites functions (GHs) in three stages (\S 2.4). The details of this process are described below. We use the conventions and notations of van der Marel \& Franx (1993): $\gamma, V, \sigma$ and $h^i$ are the flux, mean velocity, dispersion, and i\th\ Hermite coefficient, respectively. Fourth-order GHs have finite $\hc$ and $\hd$, while the other $h^i$ are zero. A line name in subscript denotes a property of a specific line. The SDSS flux density error at $\lambda$ is denoted by $\epsilon_\lambda$.

\subsubsection{[O~{III}]}\label{subsubsec: oiii fit} 
If there are $<20\%$ bad pixels at 4967\AA--5250\AA, the range is fit with three components: 
\begin{itemize}
\item An iron template derived from observations of I Zw 1 (kindly provided by T. Boroson). We fit its flux, velocity, and width (by convolving with a Gaussian), denoted $\gamma_{\feii}$, $V_{\feii}$ and $\sigma_{\feii}$, respectively. The value of $\gamma_{\feii}$ is initialized from the power-law coefficient found in \S 2.2, assuming a \feii\ equivalent width of 7\AA\ in the fit range. During the fit we keep $|V_{\feii}|< 550\ \kms$ and $\sigma_{\feii}<20\,000\ \kms$.
\item A broken power law for the underlying continuum and the red wing of the broad \Hb. Initialization is as follows: the wavelength and flux at the break -- 5082\AA, $F_\lambda(5082\AA)$; shorter $\lambda$ slope -- interpolation between $F_\lambda(4967\AA)$ and $F_\lambda(5082\AA)$; longer $\lambda$ slope -- 0.
\item A 4\th-order GH for \oiii. The $\gamma_{\oiii}$, $V_{\oiii}$ and $\sigma_{\oiii}$ are initialized from the SDSS fit. The values of $\hc_{\oiii}$ and $\hd_{\oiii}$ are initialized to 0. During the fit we keep $|V_{\oiii}|<550\ \kms$ and FWHM$_\oiii < 1\,400\ \kms$. The fit is run once with $\hc_{\oiii}$ and $\hd_{\oiii}$ fixed to 0 (a Gaussian), and once when they are allowed to vary. The final fit is chosen according to an F-test ($p < 0.05$).
\end{itemize}

The fit is disregarded if there are bad pixels $<6\AA$ from the fit \oiii\ peak. The \oiii\ is considered `detected' if it has $F_\lambda > 3.5 \times \epsilon_\oiii$, and $V_{\oiii}$ has not reached the limit mentioned above.

\subsubsection{\Ha\ region}\label{subsubsec: Ha fit}
We use nine components to fit the 6250\AA--6880\AA\ region:
\begin{itemize}
 \item A continuum. 
 \item An up to 10\th -order GH for the broad \Ha\ (b\Ha). 
 \item Seven up to 4\th -order GHs for the narrow lines (NLs). If no good pixel is found within 5\AA\ of the rest wavelength of a NL, it is not fit. Below, a subscript `NL' refers to all seven NLs simultaneously.
\end{itemize}
Initially, the broad \Ha\ is fit by a 4\th -order GH, in order to avoid fitting the narrow lines by high-order Hermite polynomials. The mean velocities of NLs in a doublet are always kept equal, and the flux ratios in the \nii\ and \oi\ doublets are fixed to 3. We limit $-550 < V_{\rm{NL}} < 550\ \kms$. The Levenberg-Marquardt best-fitting algorithm (Press et al. 1992) is used.

Several fit attempts are made, differing by the initialization and attainable values of the parameters:
\begin{enumerate}[\bf A{.}]
 \item Initialization of the $\gamma_{b\Ha}$, $\sigma_{b\Ha}$ and continuum is based on the continuum and excess flux found in \S 2.3. The $V_{\bHa}$ and all $h^i_{\bHa}$ are initialized to 0. The $V_{\rm{NL}}, \sigma_{\rm{NL}}$, $\hc_{\rm{NL}}$ and $\hd_{\rm{NL}}$ are initialized by the fit properties of \oiii. The $\gamma_{\rm{NL}}$ are initialized using NL flux ratios of typical Seyferts\footnote{$\gamma_{\nii\ 6583\AA}, \gamma_{\rm{n\Ha}} ,\gamma_{\oi 6300\AA}, \gamma_{\sii 6731\AA}, \gamma_{\sii 6716\AA}$ are initialized to $1/3,1/3,1/20,1/9,1/9$ of $\gamma_{\oiii}$, respectively.}$^{,}$
 \footnote{If \oiii\ is not detected (\insignificantoiiiprcnt\% of T1 sample), $\gamma_{\rm{NL}}$ initialization is derived from the upper limit on $\gamma_{\oiii}$. If bad pixels prevented fitting \oiii\ (\badpixeloiiiprcnt\% of T1 objects), $\gamma_{\rm{NL}}$ and $V_{\rm{NL}}$ are initialized from the SDSS fits. In both cases $\sigma_{\rm{NL}}, \hc_{\rm{NL}}$ and $\hd_{\rm{NL}}$ are initialized to a net Gaussian with FWHM = $300\ \kms$.}. During the fit, $\sigma_{\rm{NL}},\ \hc_{\rm{NL}}$ and $\hd_{\rm{NL}}$ are kept fixed to the \oiii\ values.
 \item Similar to {\bf A}, but starting from the result of {\bf A} and allowing all $\sigma_{\rm{NL}}$ to vary together (one additional degree of freedom). We limit $\sigma_{\rm{NL}} < 1.5 \times \sigma_{\oiii}$\footnote{If \oiii\ is not detected, {\bf B} is not run. Instead, {\bf A} is run with $\sigma_{\rm{NL}}$ allowed to vary, keeping FWHM$_{\rm{NL}} < 650\ \kms$.}.
 \item In this attempt we avoid implausibly large troughs in the broad \Ha\ profile, which can be created by the \nii\ and \nHa\ components. The maximum $F_\lambda$ of these 3 NLs are limited such that the broad profile will not have a trough deeper than 20\% of its local $F_\lambda$. Also, the error entering the fit near these NLs ($\pm 10$\AA) is reduced to a 1/3 of its initial value, effectively improving the fit of this region on the expense of regions farther from \Ha. The b\Ha\ and continuum are initialized as in {\bf A}. The $\gamma_{\rm{NL}}$, $V_{\rm{NL}}$ and $\sigma_{\rm{NL}}$ are initialized from the result of {\bf B}, keeping $\sigma_{\rm{NL}} < 1.5 \times \sigma_{\oiii}$ during the fit. The $\hc_{\rm{NL}}, \hd_{\rm{NL}}$ are initialized to 0, and their absolute value is kept below $\hc_{\oiii}$ and $\hd_{\oiii}$, respectively. During the fit, $\sigma_{\rm{NL}}$, $\hc_{\rm{NL}}$ and $\hd_{\rm{NL}}$ of different NLs are kept equal.
 \item The b\Ha\ and continuum are initialized as in {\bf A}. The $\gamma_{\rm{NL}}$ and $V_{\rm{NL}}$ are initialized from the highest-$F_\lambda$ pixel within 5\AA\ of the rest wavelength of the line. The $\sigma_{\rm{NL}}$ is initialized so that FWHM$_{\rm{NL}} = 300\ \kms$ and limited to $< 1.5 \times \sigma_{\oiii}$. As before, all $\sigma_{\rm{NL}}$ vary together. The $\hc_{\rm{NL}}$ and $\hd_{\rm{NL}}$ are fixed to 0.
\end{enumerate}

The $\chisq$ of the {\bf B}, {\bf C} and {\bf D} fits are compared, using the sum of the SDSS flux density error and 10\% of the fit flux density as the error, and the best result is chosen. Objects in which $\sigma_{\rm{nHa}}$ reached the limit ($1.5 \times \sigma_{\oiii}$) are found to have non- or barely-detectable NLs near \Ha. When the narrow \Ha\ component is not restricted by other NLs, it erroneously fits the top of the broad \Ha. In such cases, we use result {\bf A}, in which $\sigma_{\rm{NL}} = \sigma_{\oiii}$. In the \Ntotalprofilehighfwhm\ of these objects that were fit with $F_{\rm{n\Ha}} / F_{\oiii}>2$, the \oiii\ is weak as well, and the problem remains. Therefore, in these \Ntotalprofilehighfwhm\ objects we assume $F_{\rm{n\Ha}},F_{\nii} = 0$.

Finally, in order to fully account for the diversity of the broad \Ha\ profile, we run fits with a high-order GH for the broad \Ha. The initial values of the parameters are taken from the chosen fit above, while the new $h^i$ are initialized to 0. Only the parameters of the broad \Ha\ and the $\gamma_{\rm{NL}}$'s vary during the fit. First, a fit is attempted with a 6\th-order GH, using $h_1$ \& $h_2$\footnote{When $V$ and $\gamma$ of a GH are allowed to vary, $h_1$ \& $h_2$ hold higher-order information than $\hc$ \& $\hd$ (van der Marel \& Franx 1993).}. Then we add $h_5$ \& $h_6$, and lastly $h_7$ \& $h_8$. Additional Hermite coefficients are retained only if they pass an F-test ($p<0.05$). To avoid having negative or distinct features in the broad \Ha\ profile, the broad \Ha\ flux density is set to zero at velocities beyond a velocity in which it is $<10\%$ of the flux density error. 

After this process, \NhighChisq\ objects (\NprcntHighchisq\%) still have a reduced $\chisq > 2$. By eye-inspection, \Npasshighscore\ of them have a broad \Ha\ and a reasonable fit. They were added to the T1 sample. Also, \NbadlyfitBLRs\ show a BLR which was improperly fit. Since their number is relatively small, we do not attempt any further improvement of their fit, and they do not enter the T1 sample. The rest of the high $\chisq$ objects do not show a clear broad \Ha. 

\subsubsection{\Hb}\label{subsubsec: Hb fit}
Finally, we fit the region $\pm 6\ \sigma_{\rm{n}\Ha}$ from the \Hb\ peak, using:
\begin{itemize}
 \item A 4\th -order GH for the narrow \Hb. Only $\gamma_{\rm{n}\Hb}$ is fit, initialized to $\gamma_{\rm{n}\Ha} / 3$. Other GH parameters equal those fit to \Ha.
 \item A parabola for the continuum + top of the broad \Hb. Initialization is by the edges and middle of the fit range. The non-linear coefficient of the parabola is non-zero only if necessary (F-test, $p < 0.05$).
\end{itemize}
The fit is not performed if there are bad pixels less than $4\times \sigma_{\rm{n}\Ha}$ from the peak.

\section{REJECTED OBJECTS}

Table B1 lists objects which passed the selection criteria described in \S2.1 and \S2.3, but failed one of the selection criteria described in \S2.5. 

\begin{table}
\begin{center}
\begin{tabular}{l|c|l}
Object Name & $\Delta F / \sigma_{\rm data}$ & Rejection reason \\ 
\hline
J000202.71-010508.8 &	2.5	& $\dv > 25$ \\
J000500.03+002055.1 &	2.5	& $\dv > 25$ \\
J000605.59-092007.0 &	2.9	& ${\Delta F' / \sone} < \minfluxinwingsB$ \\
J000834.72+003156.1 &	11.3	& ${\Delta F' / \sone} < \minfluxinwingsB$ \\
J000911.58-003654.7 &   14.2	& $\normhieght < \minHafluxdensity$ \\
\end{tabular}
\caption{Objects with excess flux near \Ha\ that were rejected from the T1 sample. Col. 2 lists the normalized excess flux found in \S2.3. The criterion which the object failed (\S2.5) is listed in Col. 3. The electronic version of the paper includes all \nRejected\ objects.}
\label{table: sample}
\end{center}
\end{table}

\label{lastpage}


\begin{thebibliography}{}\addcontentsline{toc}{section}{References}

\bibitem[Abazajian et al.(2004)]{2004AJ....128..502A} Abazajian, K., Adelman-McCarthy, J.~K., Ag{\"u}eros, M.~A., et al.\ 2004, AJ, 128, 502 

\bibitem[Abazajian et al.(2009)]{Abazajian09} Abazajian, K.~N., et al.\ 2009, ApJS, 182, 543 

\bibitem[Adelman-McCarthy et al.(2008)]{2008ApJS..175..297A} Adelman-McCarthy, J.~K., Ag{\"u}eros, M.~A., Allam, S.~S., et al.\ 2008, ApJS, 175, 297 

\bibitem[Baldwin, Phillips \& Terlevich (1981)]{BPT} Baldwin, J.~A., Phillips, M.~M., \& Terlevich, R.\ 1981, PASP, 93, 5 

\bibitem[Balogh et al.(1999)]{Balogh99} Balogh, M.~L., Morris, S.~L., Yee, H.~K.~C., Carlberg, R.~G., \& Ellingson, E.\ 1999, ApJ, 527, 54 

\bibitem[Baskin \& Laor(2005)]{BaskinLaor05} Baskin, A., \& Laor, A.\ 2005, MNRAS, 356, 1029 

\bibitem[Bauer et al.(2004)]{Bauer04} Bauer, F.~E., Alexander, D.~M., Brandt, W.~N., Schneider, D.~P., Treister, E., Hornschemeier, A.~E., \& Garmire, G.~P.\ 2004, AJ, 128, 2048 

\bibitem[Becker et al.(1995)]{Becker95} Becker, R.~H., White, R.~L., \& Helfand, D.~J.\ 1995, ApJ, 450, 559 

\bibitem[Bentz et al.(2009)]{Bentz09} Bentz, M.~C., Peterson, B.~M., Netzer, H., Pogge, R.~W., \& Vestergaard, M.\ 2009, ApJ, 697, 160 

\bibitem[Bica \& Alloin(1986)]{BicaAlloin86} Bica, E., \& Alloin, D.\ 1986, A\&A, 162, 21 

\bibitem[Blanton \& Roweis(2007)]{2007AJ....133..734B} Blanton, M.~R., \& Roweis, S.\ 2007, AJ, 133, 734 

\bibitem[Bonning et al.(2007)]{2007ApJ...659..211B} Bonning, E.~W., Cheng, L., Shields, G.~A., Salviander, S., \& Gebhardt, K.\ 2007, ApJ, 659, 211 

\bibitem[Boroson \& Green(1992)]{BG92} Boroson, T.~A., \& Green, R.~F.\ 1992, ApJS, 80, 109 (BG92)

\bibitem[Brinchmann et al.(2004)]{Brinchman04} Brinchmann, J., Charlot, S., White, S.~D.~M., Tremonti, C.,Kauffmann, G., Heckman, T., \& Brinkmann, J.\ 2004, MNRAS, 351, 1151 (B04)

\bibitem[Bruzual \& Charlot(2003)]{2003MNRAS.344.1000B} Bruzual, G., \& Charlot, S.\ 2003, MNRAS, 344, 1000 

\bibitem[Caccianiga \& Severgnini(2011)]{2011MNRAS.415.1928C} Caccianiga, A., \& Severgnini, P.\ 2011, MNRAS, 415, 1928 

\bibitem[Cardelli et al.(1989)]{Cardelli89} Cardelli, J.~A., Clayton, G.~C., \& Mathis, J.~S.\ 1989, ApJ, 345, 245 

\bibitem[Cutri et al.(2003)]{2MASSwebsite} Cutri, R. M., et al. 2003, Explanatory Supplement to the 2MASS All Sky Data
Release (Washington: NASA), http://www.ipac.caltech.edu/2mass/releases/
allsky/doc/explsup.html

\bibitem[Croom et al.(2002)]{Croom02} Croom, S.~M., et al.\ 2002, MNRAS, 337, 275 

\bibitem[Croom et al.(2004)]{2004MNRAS.349.1397C} Croom, S.~M., Smith, R.~J., Boyle, B.~J., et al.\ 2004, MNRAS, 349, 1397 

\bibitem[Croom(2011)]{2011ApJ...736..161C} Croom, S.~M.\ 2011, ApJ, 736, 161 

\bibitem[Davis et al.(2007)]{2007ApJ...668..682D} Davis, S.~W., Woo, J.-H., \& Blaes, O.~M.\ 2007, ApJ, 668, 682 

\bibitem[Davis \& Laor(2011)]{2011ApJ...728...98D} Davis, S.~W., \& Laor, A.\ 2011, ApJ, 728, 98 

\bibitem[Dietrich et al.(2002)]{Dietrich02} Dietrich, M., Hamann, F., Shields, J.~C., Constantin, A., Vestergaard, M., Chaffee, F., Foltz, C.~B., \& Junkkarinen, V.~T.\ 2002, ApJ, 581, 912 

\bibitem[Elitzur \& Shlosman(2006)]{2006ApJ...648L.101E} Elitzur, M., \& Shlosman, I.\ 2006, ApJ, 648, L101 

\bibitem[Filippenko(1997)]{Filippenko97} Filippenko, A.~V.\ 1997, ARA\&A, 35, 309 

\bibitem[Fine et al.(2008)]{2008MNRAS.390.1413F} Fine, S., Croom, S.~M., Hopkins, P.~F., et al.\ 2008, MNRAS, 390, 1413

\bibitem[Fine et al.(2010)]{2010MNRAS.409..591F} Fine, S., Croom, S.~M., Bland-Hawthorn, J., et al.\ 2010, MNRAS, 409, 591 

\bibitem[Fukugita et al.(1996)]{Fukugita96} Fukugita, M., Ichikawa, T., Gunn, J.~E., Doi, M., Shimasaku, K., \& Schneider, D.~P.\ 1996, AJ, 111, 1748 

\bibitem[Gaskell et al.(2004)]{Gaskell04} Gaskell, C.~M., Goosmann, R.~W., Antonucci, R.~R.~J., 
\& Whysong, D.~H.\ 2004, ApJ, 616, 147 

\bibitem[Gaskell \& Benker(2007)]{Gaskell07} Gaskell, C.~M., \& Benker, A.~J.\ 2007, arXiv:0711.1013 

\bibitem[Green et al.(1986)]{1986ApJS...61..305G} Green, R.~F., Schmidt, M., \& Liebert, J.\ 1986, ApJS, 61, 305 

\bibitem[Greene \& Ho(2004)]{2004ApJ...610..722G} Greene, J.~E., \& Ho, L.~C.\ 2004, ApJ, 610, 722 

\bibitem[Greene \& Ho(2005)]{GH05} Greene, J.~E., \& Ho, L.~C.\ 2005, ApJ, 630, 122 

\bibitem[Greene \& Ho(2007)]{GH07} Greene, J.~E., \& Ho, L.~C.\ 2007, ApJ, 667, 131 (GH07)

\bibitem[Grupe et al.(2010)]{2010ApJS..187...64G} Grupe, D., Komossa, S., Leighly, K.~M., \& Page, K.~L.\ 2010, ApJS, 187, 64 

\bibitem[Gunn et al.(1998)]{Gunn98} Gunn, J.~E., et al.\ 1998, AJ, 116, 3040 

\bibitem[Hao et al.(2005)]{Hao05} Hao, L., et al.\ 2005a, AJ, 129, 1783 

\bibitem[Hao et al.(2005)]{2005AJ....129.1795H} Hao, L., Strauss, M.~A., Fan, X., et al.\ 2005b, AJ, 129, 1795 

\bibitem[Heckman et al.(2004)]{2004ApJ...613..109H} Heckman, T.~M., Kauffmann, G., Brinchmann, J., et al.\ 2004, ApJ, 613, 109 

\bibitem[Hewett et al.(1995)]{1995AJ....109.1498H} Hewett, P.~C., Foltz, C.~B., \& Chaffee, F.~H.\ 1995, AJ, 109, 1498 

\bibitem[Hickox et al.(2009)]{Hickox09} Hickox, R.~C., et al.\ 2009, ApJ, 696, 891 

\bibitem[Ho et al.(1997a)]{Ho97} Ho, L.~C., Filippenko, A.~V., Sargent, W.~L.~W., \& Peng, C.~Y.\ 1997a, ApJS, 112, 391 

\bibitem[Ho et al.(1997b)]{Ho97b} Ho, L.~C., Filippenko, A.~V., \& Sargent, W.~L.~W.\ 1997b, ApJ, 487, 568 

\bibitem[Ho et al.(2003)]{2003ApJ...583..159H} Ho, L.~C., Filippenko, A.~V., \& Sargent, W.~L.~W.\ 2003, ApJ, 583, 159 

\bibitem[Ho(2005)]{2005ApJ...629..680H} Ho, L.~C.\ 2005, ApJ, 629, 680 

\bibitem[Hopkins et al.(2004)]{Hopkins04} Hopkins, P.~F., et al.\ 2004, AJ, 128, 1112 

\bibitem[Hutchings(1987)]{1987ApJ...320..122H} Hutchings, J.~B.\ 1987, ApJ, 320, 122 

\bibitem[Jahnke et al.(2004)]{2004MNRAS.352..399J} Jahnke, K., Kuhlbrodt, B., \& Wisotzki, L.\ 2004, MNRAS, 352, 399 

\bibitem[Just et al.(2007)]{2007ApJ...665.1004J} Just, D.~W., Brandt, W.~N., Shemmer, O., et al.\ 2007, ApJ, 665, 1004 

\bibitem[Kaspi et al.(2005)]{Kaspi05} Kaspi, S., Maoz, D., Netzer, H., Peterson, B.~M., Vestergaard, M., \& Jannuzi, B.~T.\ 2005, ApJ, 629, 61 

\bibitem[Kauffmann et al.(2003a)]{Kauffmann03} Kauffmann, G., et al.\ 2003a, MNRAS, 346, 1055 (K03)

\bibitem[Kauffmann et al. (2003b)]{2003MNRAS.341...54K} Kauffmann, G., Heckman, T.~M., White, S.~D.~M., et al.\ 2003b, MNRAS, 341, 54 

\bibitem[Kelly et al.(2008)]{Kelly08} Kelly, B.~C., Bechtold, J., Trump, J.~R., Vestergaard, M., \& Siemiginowska, A.\ 2008, ApJs, 176, 355 

\bibitem[Kewley et al.(2004)]{2004AJ....127.2002K} Kewley, L.~J., Geller, M.~J., \& Jansen, R.~A.\ 2004, AJ, 127, 2002 

\bibitem[Kim et al.(2006)]{2006ApJ...642..702K} Kim, M., Ho, L.~C., \& Im, M.\ 2006, ApJ, 642, 702 

\bibitem[Kirhakos et al.(1999)]{1999ApJ...520...67K} Kirhakos, S., Bahcall, J.~N., Schneider, D.~P., \& Kristian, J.\ 1999, ApJ, 520, 67 

\bibitem[Kollmeier et al.(2006)]{Kollmeier06} Kollmeier, J.~A., et al.\ 2006, ApJ, 648, 128 (K06)

\bibitem[Laor et al.(1994)]{Laor94} Laor, A., Fiore, F., Elvis, M., Wilkes, B.~J., \& McDowell, J.~C.\ 1994, ApJ, 435, 611 

\bibitem[Laor et al.(1997)]{Laor97} Laor, A., Fiore, F., Elvis, M., Wilkes, B.~J., \& McDowell, J.~C.\ 1997, ApJ, 477, 93 

\bibitem[Laor (2003)]{Laor03} Laor, A.\ 2003, ApJ, 590, 86

\bibitem[Laor \& Davis(2011)]{2011MNRAS.417..681L} Laor, A., \& Davis, S.~W.\ 2011, MNRAS, 417, 681 

\bibitem[Magorrian et al.(1998)]{Magorrian98} Magorrian, J., et al.\ 1998, AJ, 115, 2285 

\bibitem[Lawrence \& Elvis(1982)]{1982ApJ...256..410L} Lawrence, A., \& Elvis, M.\ 1982, ApJ, 256, 410 

\bibitem[Maiolino et al.(2007)]{Maiolino07} Maiolino, R., Shemmer, O., Imanishi, M., Netzer, H., Oliva, E., Lutz, D., \& Sturm, E.\ 2007, A\&A, 468, 979 

\bibitem[Maoz et al.(2005)]{2005ApJ...625..699M} Maoz, D., Nagar, N.~M., Falcke, H., \& Wilson, A.~S.\ 2005, ApJ, 625, 699 

\bibitem[Martin et al.(2005)]{Martin05} Martin, D.~C., et al.\ 2005, ApJ, 619, L1 

\bibitem[McLeod \& Rieke(1995)]{1995ApJ...454L..77M} McLeod, K.~K., \& Rieke, G.~H.\ 1995, ApJ, 454, L77 

\bibitem[Morrissey et al.(2007)]{Morrissey07} Morrissey, P., et al.\ 2007, ApJS, 173, 682 

\bibitem[Moustakas et al.(2006)]{2006ApJ...642..775M} Moustakas, J., Kennicutt, R.~C., Jr., \& Tremonti, C.~A.\ 2006, ApJ, 642, 775 

\bibitem[Netzer et al.(2004)]{2004ApJ...614..558N} Netzer, H., Shemmer, O., Maiolino, R., et al.\ 2004, ApJ, 614, 558 

\bibitem[Netzer(2009)]{2009MNRAS.399.1907N} Netzer, H.\ 2009, MNRAS, 399, 1907 

\bibitem[Neugebauer et al.(1987)]{Neugebauer87} Neugebauer, G., Green, R.~F., Matthews, K., Schmidt, M., Soifer, B.~T., \& Bennett, J.\ 1987, ApJS, 63, 615 

\bibitem[Nicastro(2000)]{Nicastro00} Nicastro, F.\ 2000, ApJ, 530, L65 

\bibitem[Osterbrock(1981)]{1981ApJ...249..462O} Osterbrock, D.~E.\ 1981, ApJ, 249, 462 

\bibitem[Pei(1992)]{Pei92} Pei, Y.~C.\ 1992, ApJ, 395, 130 

\bibitem[Pfeffermann et al.(1987)]{Pfeffermann87} Pfeffermann, E., et al.\ 1987, Proc. SPIE, 733, 519 

\bibitem[Phillips(1978)]{Phillips78} Phillips, M.~M.\ 1978, ApJS, 38, 187 

\bibitem[Press et al.(1992)]{Numerical Recipies} Press, W.~H., Teukolsky, S.~A., Vetterling, W.~T., \& Flannery, B.~P.\ 1992, Cambridge: University Press, |c1992, 2nd ed.,  

\bibitem[Richards et al.(2003)]{Richards03} Richards, G.~T., et al.\ 2003, AJ, 126, 1131 

\bibitem[Richards et al.(2006)]{Richards06} Richards, G.~T., et al.\ 2006, ApJS, 166, 470 (R06)

\bibitem[Rigby et al.(2006)]{Rigby06} Rigby, J.~R., Rieke, G.~H., Donley, J.~L., Alonso-Herrero, A., \& P{\'e}rez-Gonz{\'a}lez, P.~G.\ 2006, ApJ, 645, 115 

\bibitem[Salim et al.(2007)]{2007ApJS..173..267S} Salim, S., Rich, R.~M., Charlot, S., et al.\ 2007, ApJS, 173, 267 

\bibitem[Sanders et al.(1989)]{1989ApJ...347...29S} Sanders, D.~B., Phinney, E.~S., Neugebauer, G., Soifer, B.~T., \& Matthews, K.\ 1989, ApJ, 347, 29 

\bibitem[Schade et al.(2000)]{2000MNRAS.315..498S} Schade, D.~J., Boyle, B.~J., \& Letawsky, M.\ 2000, MNRAS, 315, 498 

\bibitem[Schartel et al.(1996)]{Schartel96} Schartel, N., et al.\ 1996, MNRAS, 283, 1015 

\bibitem[Schlegel et al.(1998)]{Schlegel98} Schlegel, D.~J., Finkbeiner, D.~P., \& Davis, M.\ 1998, ApJ, 500, 525 

\bibitem[Schneider et al.(2002)]{2002AJ....123..567S} Schneider, D.~P., Richards, G.~T., Fan, X., et al.\ 2002, AJ, 123, 567 

\bibitem[Schneider et al.(2005)]{2005AJ....130..367S} Schneider, D.~P., Hall, P.~B., Richards, G.~T., et al.\ 2005, AJ, 130, 367 

\bibitem[Schneider et al.(2010)]{Schneider10} Schneider, D.~P., et al.\ 2010, AJ, 139, 2360 

\bibitem[Shen et al.(2008)]{2008ApJ...680..169S} Shen, Y., Greene, J.~E., Strauss, M.~A., Richards, G.~T., \& Schneider, D.~P.\ 2008, ApJ, 680, 169 

\bibitem[Shen et al.(2011)]{2011ApJS..194...45S} Shen, Y., et al.\ 2011, ApJS, 194, 45 

\bibitem[Shi et al.(2010)]{2010ApJ...714..115S} Shi, Y., Rieke, G.~H., Smith, P., Rigby, J., Hines, D., Donley, J., Schmidt, G., \& Diamond-Stanic, A.~M.\ 2010, ApJ, 714, 115 

\bibitem[Silverman et al.(2009)]{2009ApJ...696..396S} Silverman, J.~D., Lamareille, F., Maier, C., et al.\ 2009, ApJ, 696, 396 

\bibitem[Sim{\~o}es Lopes et al.(2007)]{2007ApJ...655..718S} Sim{\~o}es Lopes, R.~D., Storchi-Bergmann, T., de F{\'a}tima Saraiva, M., \& Martini, P.\ 2007, ApJ, 655, 718 

\bibitem[Skrutskie et al.(2006)]{Skrutskie06} Skrutskie, M.~F., et al.\ 2006, AJ, 131, 1163 

\bibitem[Soltan(1982)]{1982MNRAS.200..115S} Soltan, A.\ 1982, MNRAS, 200, 115 

\bibitem[Stark et al.(1992)]{Stark92} Stark, A.~A., Gammie, C.~F., Wilson, R.~W., Bally, J., Linke, R.~A., Heiles, C., \& Hurwitz, M.\ 1992, ApJS, 79, 77 

\bibitem[Steffen et al.(2006)]{Steffen06} Steffen, A.~T., Strateva, I., Brandt, W.~N., Alexander, D.~M., Koekemoer, A.~M., Lehmer, B.~D., Schneider, D.~P., \& Vignali, C.\ 2006, AJ, 131, 2826 

\bibitem[Stoughton et al.(2002)]{EDR} Stoughton, C., et al.\ 2002, AJ, 123, 485 

\bibitem[Strauss et al.(2002)]{Strauss02} Strauss, M.~A., et al.\ 2002, AJ, 124, 1810 

\bibitem[Trammell et al.(2007)]{Trammell07} Trammell, G.~B., Vanden Berk, D.~E., Schneider, D.~P., Richards, G.~T., Hall, P.~B., Anderson, S.~F., \& Brinkmann, J.\ 2007, AJ, 133, 1780 

\bibitem[Tran et al.(2011)]{Tran11} Tran, H.~D., Lyke, J.~E., \& Mader, J.~A.\ 2011, ApJ, 726, L21 

\bibitem[Trakhtenbrot \& Netzer(2010)]{2010MNRAS.406L..35T} Trakhtenbrot, B., \& Netzer, H.\ 2010, MNRAS, 406, L35 

\bibitem[Treister et al.(2008)]{Treister08} Treister, E., Krolik, 
J.~H., \& Dullemond, C.\ 2008, ApJ, 679, 140 

\bibitem[Trump et al.(2009)]{Trump09} Trump, J.~R., et al.\ 2009, ApJ, 706, 797 

\bibitem[Vanden Berk et al.(2001)]{VB01} Vanden Berk, D.~E., et al.\ 2001, AJ, 122, 549 

\bibitem[Vanden Berk et al.(2006)]{VB06} Vanden Berk, D.~E., et al.\ 2006, AJ, 131, 84 

\bibitem[van der Marel \& Franx(1993)]{vanderMarel93} van der Marel, R.~P., \& Franx, M.\ 1993, ApJ, 407, 525 

\bibitem[Vasudevan \& Fabian(2009)]{2009MNRAS.392.1124V} Vasudevan, R.~V., \& Fabian, A.~C.\ 2009, MNRAS, 392, 1124 

\bibitem[Veilleux \& Osterbrock(1987)]{1987ApJS...63..295V} Veilleux, S., \& Osterbrock, D.~E.\ 1987, ApJS, 63, 295 

\bibitem[Voges et al.(1999)]{Voges99} Voges, W., et al.\ 1999, A\&A, 349, 389 

\bibitem[Voges et al.(2000)]{Voges00} Voges, W., et al.\ 2000, IAU Circ. 7432, 3 

\bibitem[Wang et al.(2005)]{Wang05} Wang, T.-G., Dong, X.-B., Zhang, X.-G., Zhou, H.-Y., Wang, J.-X., \& Lu, Y.-J.\ 2005, ApJ, 625, L35 

\bibitem[Ward et al.(1987)]{1987ApJ...315...74W} Ward, M., Elvis, M., Fabbiano, G., et al.\ 1987, ApJ, 315, 74 

\bibitem[Wild et al.(2007)]{2007MNRAS.381..543W} Wild, V., Kauffmann, G., Heckman, T., et al.\ 2007, MNRAS, 381, 543 

\bibitem[Wyder et al.(2007)]{Wyder07} Wyder, T.~K., et al.\ 2007, ApJS, 173, 293 

\bibitem[Yip et al.(2004)]{Yip04} Yip, C.~W., et al.\ 2004, AJ, 128, 585 (Yip04)

\bibitem[Zheng et al.(1997)]{1997ApJ...475..469Z} Zheng, W., Kriss, G.~A., Telfer, R.~C., Grimes, J.~P., \& Davidsen, A.~F.\ 1997, ApJ, 475, 469 

\end{thebibliography}
\end{document}